\documentclass[article, shortnames]{jss}

\usepackage{orcidlink,thumbpdf,lmodern}

\usepackage{framed}
\usepackage{amsmath, amssymb, amsthm}
\newtheorem{remark}{Remark}
\usepackage{subcaption} 
\usepackage{rotating, threeparttable, booktabs, caption, dcolumn, pdflscape}
\usepackage{dsfont, pifont}
\usepackage{multirow}
\usepackage{soul} 
\usepackage{placeins}

\usepackage{graphicx}
\graphicspath{{../code/figure/}{code/figure/}}

\newcommand{\class}[1]{`\code{#1}'}
\newcommand{\fct}[1]{\code{#1()}}
\renewcommand{\vec}[1]{\mbox{\boldmath ${#1}$}}

\def\argmin{\mbox{\rm argmin}}
\def\argmax{\mbox{\rm argmax}}

\author{Jiangyan Zhao\\East China Normal University
   \And Kunhai Qing\\East China Normal University
   \AND Jin Xu\\East China Normal University}
\Plainauthor{Jiangyan Zhao, Kunhai Qing, Jin Xu}

\title{\pkg{BKP}: An \proglang{R} Package for Beta Kernel Process Modeling}
\Plaintitle{BKP: An R Package for Beta Kernel Process Modeling}
\Shorttitle{Beta Kernel Process Modeling in \proglang{R}}

\Abstract{
	Estimating input-dependent probability surfaces from binary, binomial, categorical, or multinomial response data is a common task in statistics and machine learning. 
	Latent Gaussian process classifiers provide flexible nonparametric models for such problems, but posterior inference with discrete responses typically requires approximation or simulation. 
	We discuss an implementation of probability-scale beta and Dirichlet kernel models in the \pkg{BKP} package for \proglang{R}. 
	The package implements the Beta Kernel Process (BKP), which uses kernel-weighted pseudo-count aggregation and beta-binomial conjugacy to obtain closed-form conjugate posterior summaries and posterior predictive distributions for binomial probabilities. 
	It also implements the Dirichlet Kernel Process (DKP) for multi-class responses, together with TwinBKP and TwinDKP, scalable twinning-based global-local approximations for larger datasets. 
	The resulting workflow supports transparent kernel-weighted evidence borrowing, several kernel families, fixed and data-adaptive priors, effective-sample-size calibration, loss-based hyperparameter tuning, and standard S3 methods for fitting, prediction, simulation, visualization, and extraction of posterior summaries. 
	Reproducible examples demonstrate probability-surface estimation, binary and multi-class classification, computational comparison, and real-data applications to \emph{Loa loa} infection prevalence mapping and Mourning Warbler distribution modeling.
}

\Plainkeywords{Beta kernel process, Dirichlet kernel process, conjugate updating, kernel smoothing, scalable computation, kernel-weighted evidence borrowing, R}

\Address{
	Jiangyan Zhao, Kunhai Qing {\it (equal contribution)}\\
	School of Statistics\\
	East China Normal University\\
	
	Jin Xu {\it (corresponding author)}\\
	School of Statistics\\
	\emph{and}\\
	Key Laboratory of Advanced Theory and Application in Statistics and Data Science - MOE\\
	East China Normal University\\
	3663 North Zhongshan Road\\
	Shanghai 200062, China\\
	E-mail: \email{jxu@stat.ecnu.edu.cn}\\
	URL: \url{https://faculty.ecnu.edu.cn/_s35/xj2_en/main.psp}
}

\begin{document}

\newpage

\section{Introduction } \label{sec:intro}

Estimating an input-dependent probability function from binary or binomial observations is a fundamental task in modern statistics and machine learning \citep{Hastie2009StatLearning, Rolland2019BKP, Murphy2022PML}. Such problems arise when the goal is to infer an unknown probability surface from individual Bernoulli outcomes or aggregated binomial counts over a continuous input space. Representative applications include binary classification \citep{MacKenzie2014, Wen2025KLR}, probability calibration \citep{Sung2020binaryCalibration, Dimitriadis2023}, relative abundance modeling \citep{Martin2020}, longitudinal analysis of patient-reported outcomes \citep{Najera2018HRQoL, Najera2019PRO}, infectious disease risk mapping \citep{Diggle2007Loa, Diggle2016Prevalence}, and adaptive clinical trial design for dose exploration and dose optimization \citep{Benest2022CoBe, Zhao2026SKBD}.

Classical approaches such as logistic regression and generalized additive models provide computationally efficient tools for binomial regression, but they usually impose either a fixed parametric form or an additive smoothing structure, which may limit their ability to capture complex nonlinear probability surfaces \citep{Hastie2009StatLearning}. Latent Gaussian process (GP) classifiers provide a flexible nonparametric alternative for probabilistic classification \citep{Rasmussen2006GPML}. However, binary and categorical responses lead to non-Gaussian likelihoods, so posterior inference typically requires approximation or simulation, such as Laplace approximation, expectation propagation, variational inference, or Markov chain Monte Carlo methods \citep{Nickisch2008BinaryGP}. These approximations can increase computational cost and implementation complexity, motivating probability-scale methods that avoid latent-variable inference and retain closed-form conjugate posterior summaries.

An alternative route is to model the probability surface directly on the probability scale. Rather than placing a GP prior on a latent real-valued function and mapping it to probabilities through a link function, beta-kernel methods borrow information across nearby inputs through kernel weights and produce beta or Dirichlet conjugate updates by aggregating fractional pseudo-counts. Related ideas have appeared under different names in several research communities, including smoothed beta-distribution models for robotic grasping \citep{Montesano2009, Montesano2012}, continuous correlated beta processes for sharing evidence among Bernoulli experiments \citep{Goetschalckx:2011}, Bayesian beta kernel models for binary classification and online learning \citep{MacKenzie2014}, and smooth beta-process methods for learning probability functions from Bernoulli tests \citep{Rolland2019BKP}. Recent work on kernelized bandits with Bernoulli rewards further reinforces this distinction: probability-scale kernel learning is not a routine Gaussian-to-Bernoulli substitution, because GP-based estimators are not naturally constrained to the unit interval and Bernoulli-specific theoretical guarantees in kernelized settings remain an open problem \citep{Mussi2024KMAB}.

The \pkg{BKP} package \citep{Zhao2026BKP}, developed in \proglang{R} \citep{R} and available from the Comprehensive R Archive Network (CRAN), implements this probability-scale modeling strategy through the \emph{Beta Kernel Process} (BKP). It combines kernel-weighted local likelihoods with beta-binomial conjugate updating. For each prediction input, neighboring observations contribute fractional success and failure counts through kernel weights, yielding closed-form conjugate posterior summaries for the local binomial probability. The package supports both individual binary responses and aggregated binomial counts. Its multinomial extension, the \emph{Dirichlet Kernel Process} (DKP), uses the analogous Dirichlet-multinomial conjugate update to model categorical and multinomial responses. 
For larger datasets, the package also provides TwinBKP and TwinDKP, twinning-based global-local approximations to the full BKP and DKP updates \citep{Vakayil2022Twinning,Vakayil2024TwinGP}.

The term ``Beta Kernel Process'' is used descriptively in this article. It is distinct from the beta process in Bayesian nonparametrics \citep{Hjort1990BP}, and is not to be confused with the kernel beta process for covariate-dependent feature learning \citep{Ren2011KBP}. 
Here, ``Beta'' refers to the beta distribution used for local binomial probability summaries, ``Kernel'' refers to kernel-weighted sharing of evidence across inputs, and ``Process'' refers to the input-indexed collection of such beta distributions over the input space. 
The name is intended to emphasize a probability-scale analogue of kernel smoothing with an input-indexed family of probability distributions: BKP constructs an input-indexed family of conjugate beta posteriors directly on the probability scale, rather than placing a Gaussian-process prior on a latent real-valued function.
In this article, BKP is therefore interpreted as a Bayesian-inspired, probability-scale kernel smoothing framework based on local-likelihood conjugate updating, rather than as a fully specified Bayesian stochastic-process model in the Gaussian-process sense.

Mature software libraries are available for GP modeling, including \pkg{DiceKriging} \citep{Roustant2012DiceKriging}, \pkg{GPfit} \citep{MacDonald2015GPfit}, and \pkg{gplite} \citep{Piironen2022gplite} in \proglang{R}, and \pkg{GPyTorch} \citep{Gardner2018GPyTorch}, \pkg{GPflow} \citep{Matthews2017GPflow}, and \pkg{GPy} \citep{gpy2014} in \proglang{Python}. 
In contrast, publicly available software dedicated to probability-scale beta and Dirichlet kernel modeling of binomial and multinomial responses remains limited. 
To our knowledge, \pkg{BKP} is the first publicly available \proglang{R} package dedicated to BKP and DKP methodology, providing a unified and extensible implementation for BKP, DKP, TwinBKP, and TwinDKP modeling. 
It includes model fitting, prediction, posterior simulation, visualization, prior specification, multiple kernel functions, loss-based hyperparameter tuning, effective-sample-size calibration, efficient C++ routines implemented through \pkg{Rcpp} \citep{Eddelbuettel2011Rcpp} and \pkg{RcppArmadillo} \citep{Eddelbuettel2014RcppArmadillo}, and standard S3 methods. 
These features make \pkg{BKP} suitable for probability-surface modeling tasks in biomedicine, ecology, social science, and industrial statistics.

The contribution of this article is twofold. 
Methodologically, it consolidates BKP and DKP under a common local-likelihood conjugate updating formulation, clarifies their Bayesian-inspired interpretation, and extends the framework with fixed and data-adaptive prior specifications, effective-sample-size calibration, compactly supported kernels, and twinning-based global-local approximations. 
Computationally, it provides a reproducible \proglang{R} implementation that integrates model fitting, prediction, simulation from fitted conjugate posteriors, visualization, hyperparameter tuning, and standard S3 methods for BKP, DKP, TwinBKP, and TwinDKP. The package provides a probability-scale workflow when closed-form conjugate posterior summaries, transparent local evidence borrowing, and direct binomial or multinomial updating are desirable.

The remainder of the paper is organized as follows. 
Section~\ref{sec:model} presents the statistical foundation of BKP, DKP, and the TwinBKP/TwinDKP approximations. 
Section~\ref{sec:package} describes the package interface, including fitting, prediction, simulation, visualization, and S3 methods. 
Section~\ref{sec:example} provides illustrative examples for binomial and multinomial responses, together with GP-based comparisons and timing experiments. 
Section~\ref{sec:real_data} demonstrates the workflow in two real-data applications. 
Section~\ref{sec:summary} concludes with limitations and future directions.
Appendix~\ref{app:coverage} reports a simulation-based assessment of empirical pointwise interval coverage.

\section{Statistical Foundation} \label{sec:model}
 
\subsection{Beta Kernel Process}\label{sec:review-bkp}

Let $\vec{x} = (x_1, x_2, \ldots, x_d) \in \mathcal{X} \subset \mathbb{R}^d$ denote a $d$-dimensional input. Suppose the success probability surface $\pi(\vec{x}) \in [0,1]$ is unknown. 
At each location $\vec{x}$, the observed response is modeled as
\[
	y(\vec{x}) \sim \mathrm{Binomial}(m(\vec{x}), \pi(\vec{x})),
\]
where $y(\vec{x})$ is the number of successes out of $m(\vec{x})$ independent trials. 
The full dataset comprises $n$ observations $\mathcal{D}_n = \{(\vec{x}_i, y_i, m_i)\}_{i=1}^n$, where we write $y_i= y(\vec{x}_i)$ and $m_i= m(\vec{x}_i)$ for brevity.

At each input location $\vec{x}$, the success probability is assigned a Beta prior,
\[
	\pi(\vec{x}) \sim \mathrm{Beta}(\alpha_0(\vec{x}), \beta_0(\vec{x})),
\]
where $\alpha_0(\vec{x}) > 0$ and $\beta_0(\vec{x}) > 0$ are input-dependent shape parameters. Details on prior specification are discussed in Section~\ref{sec:prior}.

Let $k: \mathcal{X} \times \mathcal{X} \to [0,1]$ denote a user-defined kernel function measuring the similarity between input locations. 
Using the kernel-weighted local likelihood update, the BKP model constructs a closed-form conjugate posterior for $\pi(\vec{x})$ as
\begin{align}\label{eq:BKP_model}
	\pi(\vec{x}) \mid \mathcal{D}_n &\sim \mathrm{Beta}\left(\alpha_n(\vec{x}), \beta_n(\vec{x})\right), \notag \\
	\alpha_n(\vec{x}) &= \alpha_0(\vec{x}) + \sum_{i=1}^{n} k(\vec{x}, \vec{x}_i) y_i = \alpha_0(\vec{x}) + \vec{k}^\top(\vec{x}) \vec{y}, \\
	\beta_n(\vec{x})  &= \beta_0(\vec{x}) + \sum_{i=1}^{n} k(\vec{x}, \vec{x}_i) (m_i - y_i) = \beta_0(\vec{x}) + \vec{k}^\top(\vec{x}) (\vec{m} - \vec{y}), \notag
\end{align}
where $\vec{k}(\vec{x}) = [k(\vec{x}, \vec{x}_1), \ldots, k(\vec{x}, \vec{x}_n)]^\top$ is the vector of kernel weights, $\vec{y}=(y_1,\ldots,y_n)^\top$,  and $\vec{m}=(m_1,\ldots,m_n)^\top$ \citep{Goetschalckx:2011, Rolland2019BKP}.

The corresponding posterior predictive distribution for a future binomial response at $\vec{x}$ with known trial size $m(\vec{x})$ is
\begin{equation}\label{eq:BKP_predictive_count}
	y(\vec{x}) \mid \mathcal{D}_n, m(\vec{x})
	\sim \mathrm{Beta\mbox{-}Binomial}
	\left(m(\vec{x}), \alpha_n(\vec{x}), \beta_n(\vec{x})\right).
\end{equation}

\begin{remark}[Bayesian interpretation]
For a fixed input location $\vec{x}$ and fixed kernel hyperparameters, the kernel-weighted local likelihood is
\[
	\widetilde{L}(\pi(\vec{x}); \mathcal{D}_n) 
	\propto \prod_{i=1}^{n} \left\{\pi(\vec{x})^{y_i}(1-\pi(\vec{x}))^{m_i - y_i} \right\}^{k(\vec{x}, \vec{x}_i)}
	= \pi(\vec{x})^{\vec{k}^\top(\vec{x}) \vec{y}} \{1 - \pi(\vec{x})\}^{\vec{k}^\top(\vec{x}) (\vec{m}-\vec{y})}.
\]
Combining this local likelihood with the beta prior yields the conjugate beta posterior in \eqref{eq:BKP_model}. 
This construction is analogous to local likelihood estimation \citep{Fan1998LocalMLE}, where observations are reweighted according to their distance from the target input. 
Thus, although the BKP update leverages beta-binomial conjugacy, it differs from a fully specified Bayesian stochastic-process model. 
Moreover, kernel parameters are selected by empirical risk minimization rather than by posterior inference. 
BKP is therefore best interpreted as a Bayesian-inspired, probability-scale smoothing framework based on local-likelihood conjugate updating.
\end{remark}

\begin{remark}[Pointwise prediction cost]
	For fixed kernel hyperparameters, computing the BKP conjugate posterior at a new input location requires $\mathcal{O}(n)$ kernel evaluations and weighted aggregations, because the update in \eqref{eq:BKP_model} is available in closed form. 
	The overall cost of model fitting, including LOOCV-based hyperparameter tuning, is discussed in Section~\ref{sec:model-selection}.
\end{remark}

Based on the resulting conjugate posterior in \eqref{eq:BKP_model}, the posterior mean
\begin{equation}\label{eq:BKPmean}
	\widehat{\pi}_n(\vec{x}) = \mathbb{E}[\pi(\vec{x}) \mid \mathcal{D}_n] = \frac{\alpha_n(\vec{x})}{\alpha_n(\vec{x}) + \beta_n(\vec{x})}
\end{equation}
serves as a smooth estimator of the latent success probability. The corresponding posterior variance  
\begin{equation}\label{eq:BKPvar}
	\sigma^2_n(\vec{x})=\mathrm{Var}[\pi(\vec{x}) \mid \mathcal{D}_n] = \frac{\widehat{\pi}_n(\vec{x})\{1 - \widehat{\pi}_n(\vec{x})\}}{\alpha_n(\vec{x}) + \beta_n(\vec{x}) + 1}
\end{equation}
provides a local posterior measure of uncertainty under the fitted BKP update.
These posterior summaries can be used to visualize the fitted probability surface and its local uncertainty across the input space, particularly in regions with sparse data coverage. See Section~\ref{sec:example} for illustrations.

When the inferential target is the future binomial count rather than the
latent success probability, the posterior predictive distribution in
\eqref{eq:BKP_predictive_count} gives
\[
	\mu_n(\vec{x})	 = \mathbb{E}\{y(\vec{x}) \mid \mathcal{D}_n, m(\vec{x})\} = m(\vec{x})\widehat{\pi}_n(\vec{x}).
\]
The corresponding posterior predictive variance is
\[
	v^2_n(\vec{x}) = \mathrm{Var}\{y(\vec{x}) \mid \mathcal{D}_n, m(\vec{x})\} = m(\vec{x})	\{\alpha_n(\vec{x})+\beta_n(\vec{x})+m(\vec{x})\}\sigma^2_n(\vec{x}).
\]
This predictive variance accounts for both uncertainty in the latent success
probability and binomial sampling variability.

For binary classification, the posterior mean can be thresholded to produce hard prediction through
\begin{equation}\label{eq:class_cri}
	\widehat{y}(\vec{x}) =
	\begin{cases}
		1 & \text{if } \widehat{\pi}_n(\vec{x}) > \pi_0, \\
		0 & \text{otherwise},
	\end{cases}
\end{equation}
where $\pi_0 \in (0,1)$ is a user-specified threshold, typically set to be 0.5.

\subsection{Effective-sample-size calibration}\label{sec:ESS}

For a beta distribution, the sum of the two shape parameters controls the concentration of the distribution and is commonly interpreted as a posterior precision or an effective sample size (ESS) \citep{Morita2008ESS}. 
For the BKP update in \eqref{eq:BKP_model}, the data-driven contribution to this precision at input location $\vec{x}$ is
\[
m_K(\vec{x}) = \vec{k}(\vec{x})^\top \vec{m}.
\]
When the binomial trial sizes $m_i$ are highly heterogeneous, this kernel-weighted trial size may either overstate or understate the effective information available near $\vec{x}$. 
Consequently, posterior uncertainty summaries may be overly concentrated or overly diffuse relative to the intended local amount of information.
The purpose of ESS calibration is to rescale the kernel-weighted successes and failures by a common factor so that their total pseudo-count matches a target ESS, while leaving the kernel-weighted empirical success proportion unchanged.
Here, calibration refers to matching the scale of the kernel-weighted pseudo-counts to a target effective sample size, rather than to a formal frequentist coverage guarantee.

To construct the target ESS, let $m_{\mathrm S}(\vec{x})$ denote a continuous Shepard interpolation \citep{Shepard1968} of the observed trial sizes $m_1,\ldots,m_n$. 
We then attenuate this interpolated trial size according to the proximity of $\vec{x}$ to the observed design points by defining
\[
m_{\mathrm{tar}}(\vec{x}) = \rho(\vec{x})m_{\mathrm S}(\vec{x}),
\qquad
\rho(\vec{x}) = \max_{1\le i\le n} k(\vec{x},\vec{x}_i).
\]
At the observed design points, the Shepard interpolation is exact for distinct inputs. Together with $k(\vec{x}_i,\vec{x}_i)=1$, this gives
\[
	m_{\mathrm{tar}}(\vec{x}_i)=m_i,\qquad i =1,\ldots,n.
\]
Away from the observed inputs, $\rho(\vec{x})$ decreases as the maximum kernel similarity decreases, so that the target ESS is reduced in regions with weak support from the training data.

The ESS-calibrated BKP posterior is obtained by replacing the data contribution in \eqref{eq:BKP_model} with
\[
	\alpha_n^{\mathrm{ESS}}(\vec{x}) 
	= \alpha_0(\vec{x}) + c(\vec{x})\vec{k}(\vec{x})^\top\vec{y},
	\qquad
	\beta_n^{\mathrm{ESS}}(\vec{x}) 
	= \beta_0(\vec{x}) + c(\vec{x})\vec{k}(\vec{x})^\top(\vec{m}-\vec{y}),
\]
where $c(\vec{x}) = m_{\mathrm{tar}}(\vec{x})/m_K(\vec{x})$ whenever $m_K(\vec{x})>0$. 
If $m_K(\vec{x})=0$, no training observation contributes to the update and the posterior is taken to be the baseline beta prior at $\vec{x}$. 
For $m_K(\vec{x})>0$, the adjusted data contribution satisfies
\[
	c(\vec{x})\vec{k}(\vec{x})^\top\vec{m}=m_{\mathrm{tar}}(\vec{x}).
\]
Thus, the calibration changes only the total pseudo-count contributed by the data, while preserving the kernel-weighted empirical success proportion.
The empirical pointwise coverage of the resulting intervals is examined in Appendix~\ref{app:coverage}.

\subsection{Prior Specification}\label{sec:prior}

The \pkg{BKP} package implements three prior specification strategies. 
These options differ in how the prior mean and prior precision are specified across the input space.

\begin{itemize}
	\item \textbf{Non-informative prior:}  
	The default choice is the uniform beta prior, with $\alpha_0(\vec{x})=\beta_0(\vec{x})\equiv 1$ for all $\vec{x}$. 
	This option is appropriate when no external prior information is available and a weak baseline prior is desired.
	
	\item \textbf{Informative prior with fixed mean:}  
	When prior information about the overall success probability $p_0\in(0,1)$ is available, a fixed informative prior can be specified by
	\[
	\alpha_0(\vec{x}) = r_0p_0,
	\qquad
	\beta_0(\vec{x}) = r_0(1-p_0),
	\]
	for all $\vec{x}$, where $r_0>0$ controls the prior precision. 
	Larger values of $r_0$ imply a more concentrated prior distribution around $p_0$. 
	The uniform prior is recovered as the special case $r_0=2$ and $p_0=0.5$. 
	When no external value of $p_0$ is available, a practical default is the empirical mean of the observed proportions, $\bar{p}=n^{-1}\sum_{i=1}^n y_i/m_i$.
	
	\item \textbf{Data-adaptive prior:}  
	To accommodate variation over the input space, \pkg{BKP} also supports a data-adaptive prior in which both the prior mean and prior precision vary with $\vec{x}$. 
	Let $S_K(\vec{x})=\sum_{i=1}^n k(\vec{x},\vec{x}_i)$ denote the local kernel mass. 
	When $S_K(\vec{x})>0$, define normalized kernel weights by $w_i(\vec{x})=k(\vec{x},\vec{x}_i)/S_K(\vec{x})$. 
	The adaptive prior uses
	\[
	p(\vec{x}) = \sum_{i=1}^n w_i(\vec{x})\frac{y_i}{m_i},
	\qquad
	r(\vec{x}) = r_0 S_K(\vec{x}),
	\]
	and sets
	\[
	\alpha_0(\vec{x}) = r(\vec{x})p(\vec{x}),
	\qquad
	\beta_0(\vec{x}) = r(\vec{x})\{1-p(\vec{x})\}.
	\]
	This empirical-Bayes-type construction uses local kernel-weighted information to define a location-dependent baseline prior. 
	In regions with stronger support from nearby observations, the prior becomes more concentrated; in regions with weaker support, it remains more diffuse. 
	The goal is to stabilize the conjugate update under heterogeneous sampling density, rather than to impose a fixed prior strength uniformly over the input space. 
	For numerical stability, the implementation uses small positive lower bounds for the local kernel mass and for the resulting beta shape parameters, ensuring valid prior parameters when the kernel mass is zero or nearly zero, as may occur with compactly supported kernels. 
	This construction is in the spirit of local likelihood modeling \citep{Fan1996LocalModel}.
\end{itemize}

The choice of the global precision parameter $r_0$ depends on the response format and the intended role of the prior. 
For binary classification problems, where each input location typically contributes a single Bernoulli observation, a relatively small value such as $0.01 \leq r_0 \leq 0.1$ is often useful to avoid prior domination and keep the fitted probabilities primarily data-driven. 
This behavior is illustrated in Example~1 of Section~\ref{sec:example_bkp}. 
For aggregated binomial responses, a reasonable starting value is the mean number of trials per input location, which makes the adaptive prior strength comparable to the typical information content of one observed binomial count. 
In applications without reliable prior information, sensitivity analysis over a grid such as $r_0 \in \{0.001, 0.01, 0.1, 1, 5, 10\}$ can be used to assess the stability of fitted probability surfaces and decision boundaries.

\subsection{Model Selection via Kernel Hyperparameter Tuning}\label{sec:model-selection}

\subsubsection{Kernel Functions}

Let $h(\vec{x}, \vec{x}'; \vec{\theta})$ denote the scaled Euclidean distance
\[
	h(\vec{x}, \vec{x}'; \vec{\theta}) = \sqrt{\sum_{j=1}^{d} \left(\frac{x_{j} - x_{j}'}{\theta_{j}}\right)^{2}},
\]
where $\vec{\theta} = (\theta_1, \theta_2, \ldots, \theta_d)$ contains positive kernel length-scale parameters. 
Smaller values of $\theta_j$ make the kernel more sensitive to differences in the $j$th input component, whereas larger values imply smoother borrowing along that dimension. 
The \pkg{BKP} package supports anisotropic kernels, where each input dimension has its own length-scale parameter, and isotropic kernels, where a single common length-scale is used, that is, $\theta_1=\cdots=\theta_d=\theta$.

Based on this metric, the kernel function is written as $k(\vec{x},\vec{x}')=k(h)$, where the functional form of $k(\cdot)$ determines the kernel type \citep{Rasmussen2006GPML}. 
The \pkg{BKP} package implements four kernel functions: Gaussian, Mat{\'e}rn $5/2$, Mat{\'e}rn $3/2$, and a compactly supported Wendland kernel \citep{Wendland1995}, as summarized in Table~\ref{tab:kernel-functions}.

\begin{table}[!t]
	\renewcommand{\arraystretch}{1.5}
	\centering
	\begin{tabular}{ll}
		\toprule
		\textbf{Kernel Type} & \textbf{Function $k(h)$}\\
		\midrule
		Gaussian & $k(h) = \exp(-h^2)$ \\
		Mat\'{e}rn $\nu = 5/2$ & $k(h) = \left(1 + \sqrt{5}h + \frac{5}{3}h^2 \right)\exp(-\sqrt{5}h)$ \\
		Mat\'{e}rn $\nu = 3/2$ & $k(h) = \left(1 + \sqrt{3}h\right)\exp(-\sqrt{3}h)$ \\
		Wendland & $k(h)=\left(\zeta h+1\right)\max\{0,1-h\}^\zeta,\quad \zeta=\lfloor d/2\rfloor+3$ \\
		\bottomrule
	\end{tabular}
	\caption{Kernel functions implemented in the \pkg{BKP} package.}
	\label{tab:kernel-functions}
\end{table}

\subsubsection{Loss Functions}

Kernel hyperparameters are selected by minimizing a user-specified loss function evaluated by leave-one-out cross-validation (LOOCV) \citep{Montesano2012}. 
For each data point $\vec{x}_i$, the prediction $\widehat{\pi}_n^{-i}(\vec{x}_i;\vec{\theta})$ is computed with the $i$th observation excluded. 
In the implementation, this is achieved by setting the diagonal entries of the kernel matrix to zero, so that all leave-one-out predictions are obtained from a single kernel matrix rather than by refitting the model $n$ times. 
The LOOCV procedure is known to mitigate overfitting and produce better generalization performance than marginal likelihood-based approaches \citep{Rasmussen2006GPML, Vehtari2017LOOCV}.

The \pkg{BKP} package supports two loss functions: the Brier score and the log-loss. 
Let $\widetilde{\pi}_i = y_i/m_i$ denote the empirical success proportion at input location $\vec{x}_i$.

\paragraph{Brier Score.}
The LOOCV Brier score is defined as
\begin{equation}\label{eq:Brier}
	\mathrm{BS}(\vec{\theta}; \mathcal{D}_n) = \frac{1}{n} \sum_{i=1}^{n} \left\{ \widehat{\pi}_n^{-i}(\vec{x}_i;\vec{\theta}) - \widetilde{\pi}_i \right\}^2.
\end{equation}
It penalizes the squared deviation between the leave-one-out predicted probability and the observed empirical proportion.

\paragraph{Log-Loss.}
The LOOCV log-loss is defined as
\begin{equation}\label{eq:log-loss}
	\mathrm{LL}(\vec{\theta}; \mathcal{D}_n) = -\frac{1}{n} \sum_{i=1}^{n} \left[ \widetilde{\pi}_i \log \widehat{\pi}_n^{-i}(\vec{x}_i;\vec{\theta}) + (1 - \widetilde{\pi}_i) \log \left\{1 - \widehat{\pi}_n^{-i}(\vec{x}_i;\vec{\theta})\right\} \right].
\end{equation}
This is the cross-entropy loss evaluated using the empirical proportions and the LOOCV predictive probabilities. 

\begin{remark}[Choice of loss function]
	Although both criteria are evaluated through LOOCV, they emphasize different aspects of predictive performance \citep{Gneiting2007ProperScoring, Flores2026ClassifierEvaluation}. 
	The Brier score penalizes squared deviations from empirical proportions and is relatively less sensitive to isolated extreme probability errors, making it a convenient choice for smooth probability-surface estimation with aggregated binomial data. 
	In contrast, the log-loss corresponds to a cross-entropy criterion based on empirical proportions and penalizes overconfident mispredictions more strongly, especially when predicted probabilities are close to 0 or 1. 
	In practice, the choice between the two depends on whether smoother probability estimation or classification-oriented probabilistic prediction is the primary objective.
\end{remark}

\subsubsection{Hyperparameter Optimization}\label{sec:HPO}

To enforce positivity and improve numerical stability, kernel length-scale parameters are optimized on the logarithmic scale. 
Following the reparameterization strategy used in \citet{MacDonald2015GPfit}, we define
$\gamma_j=\log_{10}(\theta_j)$ for $j=1,\ldots,d_\gamma$, where $d_\gamma=1$ for isotropic kernels and $d_\gamma=d$ for anisotropic kernels. 
Let $\vec{\gamma}=(\gamma_1,\ldots,\gamma_{d_\gamma})$ denote the transformed parameter vector. 
For example, in the anisotropic Gaussian case,
\begin{equation}\label{eq:reparameterization}
	k(\vec{x}, \vec{x}'; \vec\gamma) = \exp\left\{ -\sum_{j=1}^{d}\left(\frac{x_j - x_j'}{10^{\gamma_j}} \right)^2 \right\}.
\end{equation}

The LOOCV loss functions in \eqref{eq:Brier} and \eqref{eq:log-loss} can be non-convex in the kernel hyperparameters. 
To reduce dependence on a single initialization and seek a good approximation to the global minimum, \pkg{BKP} uses a multi-start derivative-free local optimization strategy. 
By default, the initial values are generated from a Latin hypercube design (LHD) with $n_0=10d_\gamma$ starting points \citep{Loeppky:2009}. 
Each starting point is then refined using the SBPLX algorithm \citep{Rowan1990SBPLX}, as implemented in \pkg{nloptr} \citep{Johnson2007NLopt}.

Following \citet{MacDonald2015GPfit}, the initial search region for $\vec{\gamma}$ is
\begin{equation}\label{eq:gamma-bound}
	\Omega_0 =	\left[	\frac{\log_{10} d_\gamma-\log_{10} 500}{2},
	\frac{\log_{10}d_\gamma+2}{2}	\right]^{d_\gamma}.
\end{equation}
For all implemented kernels, including the Gaussian, Mat{\'e}rn, and Wendland kernels, the same initial search region is used to generate starting values. 
The subsequent local optimization is performed over the broader box $\Omega=[-3,3]^{d_\gamma}$.

For each initial value, the optimizer solves
\[
	\widehat{\vec{\gamma}}^{(i)} = \underset{\vec{\gamma} \in \Omega}{\argmin}\; L(\vec{\gamma};\mathcal{D}_n), \quad i=1,\ldots, n_0,
\]
where $L(\cdot)$ denotes either the Brier score in \eqref{eq:Brier} or the log-loss in \eqref{eq:log-loss}. 
The final estimate is selected as the best local solution:
\[
	i^\star = \underset{1\leq i\leq n_0}{\argmin}\; L(\widehat{\vec{\gamma}}^{(i)};\mathcal{D}_n),
	\qquad
	\widehat{\vec{\gamma}}=\widehat{\vec{\gamma}}^{(i^\star)}.
\]
The fitted kernel length-scale parameters are then obtained by the inverse transformation $\widehat{\theta}_j=10^{\widehat{\gamma}_j}$.

The optimization procedure is summarized as follows:
\begin{enumerate}
	\item Generate $10d_\gamma$ initial values in $\Omega_0$ using a space-filling LHD.
	\item Run the SBPLX algorithm from each initial value over $\Omega$.
	\item Select the solution with the smallest LOOCV loss among all starts.
\end{enumerate}

\begin{remark}[Full-BKP fitting cost]
	Compared with full GP inference, full BKP avoids the $\mathcal{O}(n^3)$ dense covariance-matrix factorization step by using kernel-weighted beta-binomial aggregation. 
	For fixed input dimension, evaluating the BKP posterior at all training locations requires $\mathcal{O}(n^2)$ kernel evaluations and weighted aggregations. 
	In the full-kernel implementation, this computation is carried out by constructing an $n\times n$ kernel matrix, which requires $\mathcal{O}(n^2)$ memory. 
	Under LOOCV-based tuning, each loss evaluation uses one zero-diagonal kernel matrix to obtain all leave-one-out predictions simultaneously. 
	If $T$ denotes the total number of loss evaluations across all starts in the multi-start optimization, the overall computational cost of hyperparameter tuning is $\mathcal{O}(Tn^2)$, while the memory requirement remains $\mathcal{O}(n^2)$. 
	Thus, full BKP avoids cubic GP-type matrix factorization but remains a quadratic full-kernel method.
\end{remark}

\subsection{Scalable BKP via a Twinning-based Global-Local Approximation}\label{sec:TwinBKP}

For large datasets, the full BKP update in \eqref{eq:BKP_model} may still be computationally demanding, as model fitting requires repeated construction of kernel matrices and each posterior prediction requires kernel evaluations against all $n$ observations.
To improve scalability, \pkg{BKP} implements a twinning-based global-local approximation, referred to as TwinBKP, inspired by the global-local design of TwinGP \citep{Vakayil2024TwinGP}. 
Unlike TwinGP, which introduces the global-local structure through Gaussian process covariance matrices, TwinBKP adapts the idea to the conjugate beta-binomial updating structure of BKP.

\subsubsection{Global-local subset construction}

TwinBKP first selects a global subset
\[
G \subset \{1,\ldots,n\}, \qquad |G|=g,
\]
using the twinning algorithm \citep{Vakayil2022Twinning}. 
This subset is shared across all prediction locations and is intended to provide a representative summary of the observed training data. 
For a prediction location $\vec{x}$, TwinBKP further selects a local subset
\[
L(\vec{x}) \subset \{1,\ldots,n\}\setminus G, \qquad |L(\vec{x})|=\ell,
\]
consisting of the $\ell$ nearest neighbours of $\vec{x}$ among the remaining observations. 
The prediction-specific subset is then
\[
U(\vec{x}) = G\cup L(\vec{x}).
\] 
For fixed input dimension, selecting the global subset by twinning has average cost $\mathcal{O}(n\log n)$, while constructing $L(\vec{x})$ requires an $\ell$-nearest-neighbour query, with average cost $\mathcal{O}(\log n+\ell)$ per prediction location using the \code{nanoflann} kd-tree implementation \citep{Blanco2014nanoflann}.
In the default implementation, \pkg{BKP} sets the global subset size to 
\[
g=\min\{n-1,50d,\max(\lfloor\sqrt{n}\rfloor,10d)\}
\]
and uses 
\[
\ell=\min\{n-g,\max(25,3d)\}
\]
nearest neighbours for the local subset at each prediction location. 
This rule allows the global subset to increase with the sample size for moderate $n$, while imposing a dimension-dependent upper cap and keeping the local refinement controlled for fixed input dimension.

\subsubsection{Two-stage conjugate update}

TwinBKP can be interpreted as a two-stage conjugate update. 
First, the global subset defines a global-informed intermediate beta distribution,
\[
\alpha_G(\vec{x}) 
= \alpha_0(\vec{x}) + \sum_{i\in G} k_g(\vec{x},\vec{x}_i;\vec{\theta}_g)y_i,
\qquad
\beta_G(\vec{x}) 
= \beta_0(\vec{x}) + \sum_{i\in G} k_g(\vec{x},\vec{x}_i;\vec{\theta}_g)(m_i-y_i),
\]
where $k_g$ is a global kernel chosen from the implemented kernel family in Table~\ref{tab:kernel-functions}, typically a smooth Gaussian or Mat{\'e}rn kernel. 
Second, this global-informed distribution is updated using the local subset:
\begin{align}
	\pi(\vec{x})\mid \mathcal{D}_{U(\vec{x})} 
	&\sim \mathrm{Beta}(\alpha_U(\vec{x}),\beta_U(\vec{x})), \notag\\
	\alpha_U(\vec{x}) 
	&= \alpha_G(\vec{x}) 
	+ \sum_{i\in L(\vec{x})} k_\ell(\vec{x},\vec{x}_i;\theta_\ell)y_i, \label{eq:twin-bkp}\\
	\beta_U(\vec{x}) 
	&= \beta_G(\vec{x}) 
	+ \sum_{i\in L(\vec{x})} k_\ell(\vec{x},\vec{x}_i;\theta_\ell)(m_i-y_i), \notag
\end{align}
where $k_\ell$ is a local kernel, taken in the default implementation to be a compactly supported Wendland kernel. 
Thus, the global subset supplies broad background information shared across prediction locations, while the local subset refines the update near the target location. 
The baseline prior parameters $\alpha_0(\vec{x})$ and $\beta_0(\vec{x})$ follow the same specification strategies as in Section~\ref{sec:prior}. 
Posterior summaries and predictive probabilities are then computed from the resulting beta posterior and the corresponding beta-binomial predictive distribution, as described in Section~\ref{sec:review-bkp}.

\subsubsection{Kernel parameter specification}

Kernel parameters in TwinBKP are handled separately for the global and local components. 
The global kernel parameter $\vec{\theta}_g$ is estimated by minimizing the same LOOCV-based loss as in Section~\ref{sec:model-selection}, but only on the global subset
\[
\mathcal{D}_G=\{(\vec{x}_i,y_i,m_i):i\in G\}.
\]
This avoids constructing the full $n\times n$ kernel matrix during tuning. 
Let $T_g$ denote the total number of loss evaluations used in the global-subset optimization. The tuning cost is reduced from $\mathcal{O}(T n^2)$ for full BKP to $\mathcal{O}(T_g g^2)$ for TwinBKP, and the kernel-matrix memory requirement is reduced from $\mathcal{O}(n^2)$ to $\mathcal{O}(g^2)$. 

The scalar local range parameter $\theta_\ell$ is not optimized separately. 
It is set to be the empirical covering radius \citep{Vakayil2024TwinGP} of the global subset on the normalized input scale,
\[
\widehat{\theta}_\ell 
= \max_{1\le i\le n} \min_{j\in G} \|\vec{x}_i-\vec{x}_j\|_2 .
\]
This quantity is the fill distance \citep{Fasshauer2007Meshfree} of the global subset relative to the empirical training design, measuring how well the global subset covers the observed training inputs.
For the compactly supported Wendland kernel, it provides a conservative support radius for local borrowing and avoids an additional local hyperparameter optimization step.
Using a kd-tree built on the global subset, this empirical covering radius can be computed in $\mathcal{O}(n\log g)$ time for fixed input dimension.

\begin{remark}[TwinBKP fitting and prediction cost]
	For fixed input dimension, the overall computational cost of TwinBKP is driven by global subset selection, global-subset tuning, and prediction-specific local updates. 
	With global subset size $g$ and local subset size $\ell$, the fitting cost is 
	$\mathcal{O}(n\log n + T_g g^2)$. 
	The corresponding memory requirement is $\mathcal{O}(n+g^2)$, including storage of the training inputs and the global kernel matrix. 
	For prediction at $t$ locations, the computational cost is $\mathcal{O}\{t(\log n+g+\ell)\}$, combining kd-tree local-neighbour queries with beta-binomial pseudo-count aggregation over the global and local subsets.	
	Under the default scaling rule, $g$ is controlled by both the sample size and the dimension-dependent cap $50d$, while $\ell$ is bounded by $\max(25,3d)$ for fixed $d$. 
	With fixed optimization budgets and fixed input dimension, the fitting cost is therefore dominated by the $\mathcal{O}(n\log n)$ global-subset construction term. 
\end{remark}

\subsection{Extension to Dirichlet Kernel Process}\label{sec:dkp}

The BKP model extends naturally to multi-class responses through the Dirichlet Kernel Process (DKP), which replaces the binomial likelihood with a multinomial likelihood and the beta prior with a Dirichlet prior \citep{MacKenzie2014}. 

Let the response at input $\vec{x}\in\mathcal{X}\subset\mathbb{R}^d$ be 
$\vec{y}(\vec{x})=(y_1(\vec{x}),\ldots,y_q(\vec{x}))$, where $y_s(\vec{x})$ denotes the count of class $s$ and $m(\vec{x})=\sum_{s=1}^q y_s(\vec{x})$ is the total number of trials. 
We assume
\[
\vec{y}(\vec{x}) \sim \mathrm{Multinomial}(m(\vec{x}),\vec{\pi}(\vec{x})),
\]
where $\vec{\pi}(\vec{x})=(\pi_1(\vec{x}),\ldots,\pi_q(\vec{x}))$ is a vector of class probabilities satisfying $\sum_{s=1}^q\pi_s(\vec{x})=1$.

At each input location, a Dirichlet prior is assigned to the class-probability vector,
\[
\vec{\pi}(\vec{x})\sim\mathrm{Dirichlet}(\vec{\alpha}_0(\vec{x})),
\]
where $\vec{\alpha}_0(\vec{x})=(\alpha_{0,1}(\vec{x}),\ldots,\alpha_{0,q}(\vec{x}))$ contains positive prior concentration parameters. 
Given training data $\mathcal{D}_n=\{(\vec{x}_i,\vec{y}_i)\}_{i=1}^n$, let $\mathbf{Y}\in\mathbb{R}^{n\times q}$ denote the response matrix whose $i$th row is $\vec{y}_i^\top$. 
The kernel-smoothed conjugate posterior is
\begin{equation}\label{eq:DKP_model}
	\vec{\pi}(\vec{x}) \mid \mathcal{D}_n 
	\sim 
	\mathrm{Dirichlet}(\vec{\alpha}_n(\vec{x})),
	\qquad
	\vec{\alpha}_n(\vec{x}) 
	= \vec{\alpha}_0(\vec{x}) + \vec{k}^{\top}(\vec{x})\mathbf{Y}.
\end{equation}

The corresponding posterior predictive distribution for a future multinomial response at $\vec{x}$ with known trial size $m(\vec{x})$ is
\begin{equation}\label{eq:DKP_predictive_count}
	\vec{y}(\vec{x}) \mid \mathcal{D}_n, m(\vec{x})
	\sim
	\mathrm{Dirichlet\mbox{-}Multinomial}
	(m(\vec{x}), \vec{\alpha}_n(\vec{x})).
\end{equation}

Let $A_n(\vec{x})=\sum_{s=1}^q\alpha_{n,s}(\vec{x})$. 
The posterior mean and marginal posterior variance of the class probability for class $s$ are
\[
\widehat{\pi}_{n,s}(\vec{x}) 
= \frac{\alpha_{n,s}(\vec{x})}{A_n(\vec{x})},
\qquad
\sigma^2_{n,s}(\vec{x}) 
= \frac{\widehat{\pi}_{n,s}(\vec{x})\{1-\widehat{\pi}_{n,s}(\vec{x})\}}
{A_n(\vec{x})+1},
\qquad s=1,\ldots,q.
\]
For the future count of class $s$, the posterior predictive mean and marginal variance are
\[
\mu_{n,s}(\vec{x}) 
= m(\vec{x})\widehat{\pi}_{n,s}(\vec{x}),
\qquad
v^2_{n,s}(\vec{x}) 
= m(\vec{x})\{A_n(\vec{x})+m(\vec{x})\}\sigma^2_{n,s}(\vec{x}).
\]
For classification tasks, the predicted label is assigned by the posterior predictive MAP rule,
\begin{equation}\label{eq:class_cri_DKP}
	\widehat{y}(\vec{x}) 
	= \underset{s\in\{1,\ldots,q\}}{\argmax}\; 
	\widehat{\pi}_{n,s}(\vec{x}).
\end{equation}

The prior specification strategies in Section~\ref{sec:prior} extend to DKP by replacing the beta shape parameters with Dirichlet concentration parameters. 
For example, a fixed prior can be specified through $\alpha_{0,s}(\vec{x})=r_0p_{0,s}$, where $\vec{p}_0=(p_{0,1},\ldots,p_{0,q})$ is a class-probability vector and $r_0$ controls the total prior concentration. 
The non-informative and data-adaptive priors are constructed analogously by specifying baseline class proportions and a total concentration parameter over the input space.

Kernel hyperparameters are tuned by minimizing LOOCV-based loss functions, as described in Section~\ref{sec:model-selection}. 
For multi-class responses, \pkg{BKP} supports the multi-class Brier score and log-loss,
\begin{align}
	\mathrm{BS}_{\mathrm{multi}}(\vec{\theta}) 
	&= \frac{1}{n}\sum_{i=1}^n\sum_{s=1}^q
	\left\{\widehat{\pi}_{n,s}^{-i}(\vec{x}_i;\vec{\theta})
	-\widetilde{\pi}_{i,s}\right\}^2, \label{eq:dkp-brier}\\
	\mathrm{LL}_{\mathrm{multi}}(\vec{\theta}) 
	&= -\frac{1}{n}\sum_{i=1}^n\sum_{s=1}^q
	\widetilde{\pi}_{i,s}
	\log \widehat{\pi}_{n,s}^{-i}(\vec{x}_i;\vec{\theta}), \label{eq:dkp-logloss}
\end{align}
where $\widetilde{\pi}_{i,s}=y_{i,s}/m_i$ denotes the empirical class proportion and $m_i=\sum_{s=1}^q y_{i,s}$. 
When $q=2$, the log-loss reduces to the binary log-loss in \eqref{eq:log-loss}, while the multi-class Brier score is proportional to the binary Brier score in \eqref{eq:Brier}; this constant factor does not affect hyperparameter selection.

The ESS calibration in Section~\ref{sec:ESS} also extends to DKP by calibrating the total Dirichlet concentration. 
Let $m_K(\vec{x})=\vec{k}(\vec{x})^\top\vec{m}$ and use the same target ESS $m_{\mathrm{tar}}(\vec{x})=\rho(\vec{x})m_{\mathrm S}(\vec{x})$. 
For $m_K(\vec{x})>0$, define $c(\vec{x})=m_{\mathrm{tar}}(\vec{x})/m_K(\vec{x})$. 
The ESS-calibrated DKP posterior is then
\[
\vec{\alpha}_n^{\mathrm{ESS}}(\vec{x}) 
= \vec{\alpha}_0(\vec{x})
+ c(\vec{x})\vec{k}^\top(\vec{x})\mathbf{Y}.
\]
As in the BKP case, if $m_K(\vec{x})=0$, no training observation contributes to the update and the posterior is taken to be the baseline Dirichlet prior at $\vec{x}$.

Finally, the twinning-based global-local approximation extends directly to DKP. 
The resulting TwinDKP model uses the same global-local subset construction as TwinBKP, but replaces the beta pseudo-count updates with Dirichlet pseudo-count vector updates over the global and local subsets.

\section{BKP Package}\label{sec:package}

The \pkg{BKP} package \citep{Zhao2026BKP} implements the Beta Kernel Process (BKP) for binary and binomial response data, together with its multinomial extension, the Dirichlet Kernel Process (DKP). It also provides scalable global-local variants, namely TwinBKP and TwinDKP, for applications where full kernel-matrix construction is computationally demanding. The package follows the standard R modeling workflow, with separate functions for model fitting, posterior prediction, posterior simulation, visualization, and model summarization. It also exposes utilities for kernel construction, prior specification, and loss-based hyperparameter tuning, allowing users to adjust the kernel family, prior structure, loss criterion, and scalability settings according to the data-analysis task.

The main user-facing functions and S3 methods are summarized in Table~\ref{tab:main_functions}. The package can be installed from CRAN using \code{install.packages("BKP")}; the development version is available from GitHub via \code{pak::pak("Jiangyan-Zhao/BKP")}.
An interactive overview of the methodology, package workflow, examples, and related project resources is available from the \href{https://jiangyan-zhao.github.io/BKP-website/}{\pkg{BKP} project website}.

\begin{table}[!t]
	\centering
	\begin{tabular}{@{}lp{0.8\textwidth}@{}}
		\toprule
		& Description \\
		\midrule
		\multicolumn{2}{@{}l}{\textbf{Functions}} \\
		\fct{fit\_BKP} & Fit a Beta Kernel Process (BKP) model for binary/binomial data. \\
		\fct{fit\_DKP} & Fit a Dirichlet Kernel Process (DKP) model for categorical or multinomial data. \\
		\fct{fit\_TwinBKP} & Fit a scalable global-local TwinBKP approximation for binary/binomial data. \\
		\fct{fit\_TwinDKP} & Fit a scalable global-local TwinDKP approximation for categorical or multinomial data. \\
		\midrule
		\multicolumn{2}{@{}l}{\textbf{Core S3 Methods}} \\
		\fct{predict} & Compute posterior predictions at training or new input locations. \\
		\fct{simulate} & Generate posterior simulations from fitted model objects. \\
		\fct{summary} & Summarize model settings and posterior quantities. \\
		\fct{plot} & Visualize fitted probability surfaces and posterior uncertainty. \\
		\fct{print} & Print fitted models, predictions, simulations, and summaries. \\
		\bottomrule
	\end{tabular}
	\caption{Main functions and S3 methods provided in the \pkg{BKP} package. The S3 methods apply to fitted model objects returned by the corresponding fitting functions. Additional accessor methods, such as \fct{fitted}, \fct{parameter}, and \fct{quantile}, are also available.}
	\label{tab:main_functions}
\end{table}

\subsection{BKP Model}\label{sec:BKP_pkg}

We begin with the full BKP interface, which provides the baseline workflow for model fitting, posterior prediction, posterior simulation, visualization, and numerical extraction. The DKP model in Section~\ref{sec:DKP_pkg} follows the same workflow but differs in the response format, posterior distribution, and multi-class prediction summaries. The TwinBKP and TwinDKP models in Section~\ref{sec:Twin_pkg} retain the corresponding BKP and DKP response formats, while replacing full kernel aggregation with the scalable global-local approximation described in Section~\ref{sec:TwinBKP}.

\subsubsection{Model Fitting}

The function \fct{fit\_BKP} fits the BKP model in \eqref{eq:BKP_model} for binary or aggregated binomial response data. Its main arguments are
\begin{Code}
fit_BKP(X, y, m, Xbounds = NULL,
	prior = "noninformative", r0 = 2, p0 = mean(y / m),
	kernel = "gaussian", isotropic = TRUE, theta = NULL,
	loss = "brier", ess = "none", n_multi_start = NULL, n_threads = 1)
\end{Code}

The arguments \code{X}, \code{y}, and \code{m} specify the data. The input \code{X} is an $n \times d$ numeric matrix, \code{y} is a length-$n$ vector of observed successes, and \code{m} is the corresponding vector of binomial trial sizes. The optional argument \code{Xbounds} is a $d \times 2$ matrix containing the lower and upper bounds of each input dimension. When \code{Xbounds} is supplied, the input matrix is rescaled to the unit hypercube $[0,1]^d$ before fitting. When \code{Xbounds = NULL}, \code{X} is assumed to have already been normalized to $[0,1]^d$.

The arguments \code{prior}, \code{r0}, and \code{p0} define the prior specification for the latent success probability surface. The argument \code{prior} selects one of the prior forms described in Section~\ref{sec:prior}: \code{"noninformative"}, \code{"fixed"}, or \code{"adaptive"}. The prior parameters are constructed internally by \fct{get\_prior}. The arguments \code{r0} and \code{p0} specify the prior precision and prior mean when they are required by the selected prior form.

The arguments \code{kernel}, \code{isotropic}, and \code{theta} determine the kernel family and length-scale structure. The argument \code{kernel} selects one of the implemented kernels, namely \code{"gaussian"}, \code{"matern52"}, \code{"matern32"}, or \code{"wendland"}. If \code{isotropic = TRUE}, a single common length-scale is used for all input dimensions; if \code{isotropic = FALSE}, dimension-specific length-scales are used. The argument \code{theta} can be supplied to fix the kernel length-scale parameter directly. In the isotropic case, \code{theta} must be a scalar; in the anisotropic case, it may be either a scalar, which is expanded to all dimensions, or a vector of length $d$. When \code{theta = NULL}, the length-scale parameter is selected by leave-one-out cross-validation.

The remaining arguments control model tuning, effective-sample-size calibration, and computation. 
The argument \code{loss} specifies the LOOCV criterion used for kernel hyperparameter tuning, either \code{"brier"} or \code{"log\_loss"}, as implemented in \fct{loss\_fun}. 
The argument \code{ess} controls optional effective-sample-size calibration. The default \code{ess = "none"} uses the standard BKP posterior update, whereas \code{ess = "shepard"} rescales the kernel-weighted data contribution according to the Shepard-interpolation target described in Section~\ref{sec:ESS}. 
The argument \code{n\_multi\_start} specifies the number of initial values used in multi-start optimization; when it is \code{NULL}, the default is $10d_\gamma$, where $d_\gamma=1$ for isotropic kernels and $d_\gamma=d$ for anisotropic kernels, as in Section~\ref{sec:HPO}.
The argument \code{n_threads} specifies the number of parallel threads used during multi-start optimization when thread-level parallelization is available.

The helper functions \fct{kernel\_matrix}, \fct{get\_prior}, and \fct{loss\_fun} are primarily used internally by \fct{fit\_BKP}. They are exported so that advanced users can inspect the main computational components of the fitted model: \fct{kernel\_matrix} constructs kernel-weight matrices for specified inputs and length-scale parameters, \fct{get\_prior} constructs the prior beta parameters, and \fct{loss\_fun} evaluates the LOOCV loss for a given kernel specification. These functions also support customized workflows built on the same BKP computational structure.

The function \fct{fit\_BKP} returns an object of class \class{BKP}. The object stores the fitted kernel specification, selected length-scale parameter, prior specification, normalized training inputs, observed binomial responses, and posterior beta parameters evaluated at the training inputs. These stored components are used by downstream S3 methods for prediction, simulation, visualization, and model summarization. When ESS calibration is used, the object also records the calibration method and associated diagnostics.

The resulting \class{BKP} object can be passed as the \code{object} argument to \fct{predict}, \fct{simulate}, and \fct{summary}, or as the \code{x} argument to \fct{plot} and \fct{print}.

\subsubsection{Methods for the Fitted BKP Objects}

For a fitted \class{BKP} object \code{BKPmodel}, posterior prediction is obtained by
\begin{Code}
predict(BKPmodel, Xnew = NULL, CI_level = 0.95, threshold = 0.5,
	type = "probability", Mnew = NULL, ...)
\end{Code}
By default, \code{type = "probability"} returns posterior summaries for the latent success probability $\pi(\vec{x})$. For each input location $\vec{x}$ in \code{Xnew}, or at the training locations when \code{Xnew = NULL}, the returned object contains the posterior mean $\widehat{\pi}_n(\vec{x})$, posterior variance $\sigma_n^2(\vec{x})$, lower and upper bounds of the \code{CI_level} credible interval, and the posterior beta parameters $\alpha_n(\vec{x})$ and $\beta_n(\vec{x})$. If the fitted model corresponds to binary classification data, that is, all training trial sizes satisfy \code{m = 1}, the prediction object also includes a binary class label obtained by comparing $\widehat{\pi}_n(\vec{x})$ with the specified \code{threshold}, following the rule in \eqref{eq:class_cri}.

Alternatively, setting \code{type = "count"} returns posterior predictive summaries for a future binomial count. In this case, \code{Mnew} specifies the future trial size at each prediction location, either as a scalar common to all locations or as a vector with length equal to the number of prediction locations. The returned mean, variance, and interval bounds are then computed on the count scale under the beta-binomial posterior predictive distribution in \eqref{eq:BKP_predictive_count}.

The function call
\begin{Code}
simulate(BKPmodel, nsim = 1, seed = NULL, Xnew = NULL, threshold = NULL, ...)
\end{Code}
generates pointwise posterior draws of the latent success probability from a fitted \class{BKP} model. For each input location in \code{Xnew}, or at the training locations when \code{Xnew = NULL}, the method obtains the corresponding posterior beta parameters and independently draws \code{nsim} samples from $\mathrm{Beta}(\alpha_n(\vec{x}), \beta_n(\vec{x}))$. The returned \code{samples} component is a matrix with $n_{\mathrm{new}}$ rows and \code{nsim} columns, where each column contains a collection of independent pointwise draws from the marginal beta posteriors evaluated at the supplied input locations. Thus, these draws should not be interpreted as realizations from a jointly dependent posterior process; in particular, duplicate input locations correspond to independent draws from the same marginal posterior distribution. If \code{threshold} is supplied, the simulated probabilities are additionally converted to binary class labels by thresholding each pointwise draw, and the optional \code{seed} argument sets the random seed for reproducibility. These draws are useful for visualizing marginal posterior uncertainty and for pointwise randomized decision rules. However, because the current BKP construction does not specify cross-location posterior dependence, they do not constitute coherent posterior sample paths or random surfaces. Consequently, downstream procedures requiring function-level posterior samples, such as standard Thompson sampling over a continuous probability surface \citep{Garnett2023BO}, would require an additional joint dependence model.

The function call
\begin{Code}
summary(BKPmodel, ...)
\end{Code}
returns a structured summary of a fitted \class{BKP} object. The summary includes the number of training observations, input dimensionality, kernel family, isotropic setting, selected kernel hyperparameters, loss criterion, minimum achieved loss value, and prior specification. It also reports posterior summaries at the training inputs, including the posterior means and variances of the latent success probabilities. Thus, \fct{summary} provides a compact diagnostic view of both the fitted model configuration and the main posterior quantities used in subsequent prediction and visualization.

The \fct{plot} method provides graphical visualization of a fitted \class{BKP} object, with the display determined by the selected input dimension(s). For one-dimensional displays, it plots the posterior mean of the latent success probability as a curve, together with a shaded credible interval and the observed proportions $\widetilde{\pi}_i = y_i/m_i$. For binary classification data, the classification threshold can also be shown to indicate the induced decision boundary. For two-dimensional displays, the method evaluates the fitted model on a regular prediction grid and produces contour plots of posterior summaries. By default, non-classification plots include the posterior mean, posterior variance, and posterior interval bounds for the latent success probability; for classification data, the posterior mean and variance surfaces of the class probability are displayed. The argument \code{only_mean = TRUE} restricts the display to the posterior mean surface. For inputs with more than two dimensions, users specify one or two coordinates to visualize through the \code{dims} argument.

The \pkg{BKP} package uses S3 classes to organize fitted models and downstream outputs. A fitted \class{BKP} object has a dedicated \fct{print} method for concise console display, while the outputs of \fct{predict}, \fct{simulate}, and \fct{summary} are returned as objects of classes \class{predict\_BKP}, \class{simulate\_BKP}, and \class{summary\_BKP}, respectively. These classes provide method-specific printed summaries and separate the user-facing display from the underlying list structure, making the fitted objects easier to inspect while preserving access to their numerical components.

In addition to the core methods in Table~\ref{tab:main_functions}, the \pkg{BKP} package provides accessor methods for extracting numerical components from fitted objects. For a fitted \class{BKP} object, \fct{fitted} returns the posterior mean success probabilities at the training inputs, \fct{parameter} returns the fitted kernel length-scale parameter together with the posterior beta parameters $\alpha_n(\vec{x})$ and $\beta_n(\vec{x})$, and \fct{quantile} computes posterior quantiles of the latent success probability, with the requested quantile levels specified by \code{probs}. These accessors provide direct programmatic access to the fitted posterior summaries without relying on printed output.

\subsection{DKP Model}\label{sec:DKP_pkg}

The function \fct{fit\_DKP} implements the DKP model in \eqref{eq:DKP_model}, extending the BKP framework from binary or binomial responses to categorical or aggregated multinomial response data. Its main arguments are
\begin{Code}
fit_DKP(X, Y, Xbounds = NULL,
	prior = "noninformative", r0 = 2, p0 = NULL,
	kernel = "gaussian", isotropic = TRUE, theta = NULL,
	loss = "brier", ess = "none", n_multi_start = NULL, n_threads = 1)
\end{Code}
The main difference from \fct{fit\_BKP} is the response format. Instead of separate vectors \code{y} and \code{m}, the DKP interface uses an $n \times q$ count matrix \code{Y}, where each row corresponds to one input location and each column to one response category. The multinomial trial size is defined by the row sum of \code{Y}, so a separate \code{m} argument is not required. When \code{prior = "fixed"}, \code{p0} is a class-probability vector of length $q$; if \code{p0 = NULL}, it is set to the empirical class-proportion vector \code{colMeans(Y / rowSums(Y))}.

The function returns an object of class \class{DKP}, structurally parallel to a \class{BKP} object. It stores the fitted kernel specification, selected length-scale parameter, prior specification, normalized training inputs, observed multinomial counts, and posterior Dirichlet concentration parameters evaluated at the training inputs. The key distinction is that the posterior parameter at an input $\vec{x}$ is a vector $\vec{\alpha}_n(\vec{x})$, rather than a pair of beta shape parameters.

The downstream S3 methods follow the same workflow as for \class{BKP} objects. The \fct{predict} method returns per-class posterior means, marginal variances, and marginal credible intervals for the latent class-probability vector. If each row of \code{Y} sums to one, it also returns predicted class labels using the MAP rule in \eqref{eq:class_cri_DKP}. Setting \code{type = "count"} instead gives per-class marginal posterior predictive summaries for future multinomial counts with trial sizes specified by \code{Mnew}. The \fct{simulate} method draws posterior samples from $\mathrm{Dirichlet}\{\vec{\alpha}_n(\vec{x})\}$ and, for single-label classification data, can also produce sampled class labels. The \fct{plot} method displays class-specific posterior summaries, using curves for one-dimensional inputs and contour plots for two-dimensional displays.

\subsection{TwinBKP and TwinDKP Models}\label{sec:Twin_pkg}

For larger datasets, the package provides the scalable variants \fct{fit\_TwinBKP} and \fct{fit\_TwinDKP}. TwinBKP corresponds to the global-local approximation described in Section~\ref{sec:TwinBKP}, and TwinDKP uses the analogous Dirichlet pseudo-count update described at the end of Section~\ref{sec:dkp}. 
These functions follow the same response conventions as \fct{fit\_BKP} and \fct{fit\_DKP}, respectively: \fct{fit\_TwinBKP} uses \code{X}, \code{y}, and \code{m}, whereas \fct{fit\_TwinDKP} uses \code{X} and the multinomial count matrix \code{Y}. 
Their additional arguments control the global-local approximation and global-subset tuning, including the global and local kernels \code{global_kernel} and \code{local_kernel}, the global and local length-scale parameters \code{theta_g} and \code{theta_l}, the global subset size \code{g}, the local-neighbour size \code{l}, the number of twinning runs \code{twins}, and the diagnostic option \code{store_kernel}. 
In the current implementation, \code{global_kernel} can be selected from the implemented kernel family, whereas \code{local_kernel} is restricted to \code{"wendland"}, corresponding to the compactly supported local kernel used in the TwinBKP and TwinDKP approximations. 
The arguments \code{loss}, \code{n_multi_start}, \code{isotropic}, and \code{n_threads} play the same roles as in the full models, but are applied to global-subset tuning.

The returned objects, of classes \class{TwinBKP} and \class{TwinDKP}, are designed to be used with the same S3 workflow as full BKP and DKP objects. In particular, the methods \fct{predict}, \fct{simulate}, \fct{summary}, \fct{plot}, \fct{print}, \fct{fitted}, \fct{parameter}, and \fct{quantile} are available for the fitted Twin objects. By default, the Twin implementations avoid storing dense $n \times n$ kernel matrices; setting \code{store_kernel = TRUE} stores diagnostic kernel matrices for testing and inspection, at the cost of increased memory usage. Effective-sample-size calibration is currently implemented for the full BKP and DKP models, whereas TwinBKP and TwinDKP use the uncalibrated global-local posterior updates to preserve the scalable approximation.

\section{Examples using BKP}\label{sec:example}

\subsection{BKP Model}\label{sec:example_bkp}

This subsection presents detailed illustrative examples based on the full BKP model. 
Examples for the DKP model and the TwinBKP/TwinDKP scalable approximations are provided in Sections~\ref{sec:example_dkp} and~\ref{sec:example_twin}, respectively.

\paragraph{Example 1} Let $x\in[-2,2]$, and suppose the true Bernoulli probability function is given by
\begin{equation}\label{eq:bkp-1d-generate-function-1}
	\pi_{1}(x) = \frac{1}{1+e^{-3x}},
\end{equation}
which is referred as the function \code{true_pi_fun} in the code below.
We aim to fit the BKP model based on seven input locations that are uniformly distributed over $[-2,2]$, with each location associated with a binomial observation having a maximum trial count of 100. The input locations are generated using the \fct{lhs} function from the \proglang{R} package \pkg{tgp} \citep{Gramacy2010tgp2}. The following \proglang{R} code illustrates how to simulate the data and fit the BKP model using the \fct{fit\_BKP} function.

\begin{CodeChunk}
\begin{CodeInput}    
R> n <- 7
R> Xbounds <- matrix(c(-2,2), nrow = 1) 
R> X <- lhs(n = n, rect = Xbounds) 
R> true_pi <- true_pi_fun(X) 
R> m <- sample(100, n, replace = TRUE)
R> y <- rbinom(n, size = m, prob = true_pi) 
R> BKP_model_1D_1 <- fit_BKP(X, y, m, Xbounds = Xbounds)
\end{CodeInput}
\end{CodeChunk}

The estimates of the parameters of the fitted BKP model can be displayed using the \fct{print} function:

\begin{CodeChunk}
\begin{CodeInput} 
R> print(BKP_model_1D_1)
\end{CodeInput}
\begin{CodeOutput}  
       Beta Kernel Process (BKP) Model    

Number of observations (n):  7
Input dimensionality (d):    1
Kernel type:                 (isotropic) gaussian
Optimized kernel parameters: 0.1748
Minimum achieved loss:       0.01165
Loss function:               brier
Prior type:                  noninformative
\end{CodeOutput}
\end{CodeChunk}

The \code{BKP_model_1D_1} object can be used for visualization and prediction over a grid of input values via the \fct{plot} and \fct{simulate} methods:

\begin{CodeChunk}
\begin{CodeInput}  
R> plot(BKP_model_1D_1)
R> Xnew <- matrix(seq(-2, 2, length = 100), ncol = 1)
R> sim <- simulate(BKP_model_1D_1, nsim = 3, Xnew = Xnew)
\end{CodeInput} 
\end{CodeChunk}

\begin{figure}[!t]
	\centering
	\begin{subfigure}{0.49\textwidth}
		\includegraphics[width=\linewidth]{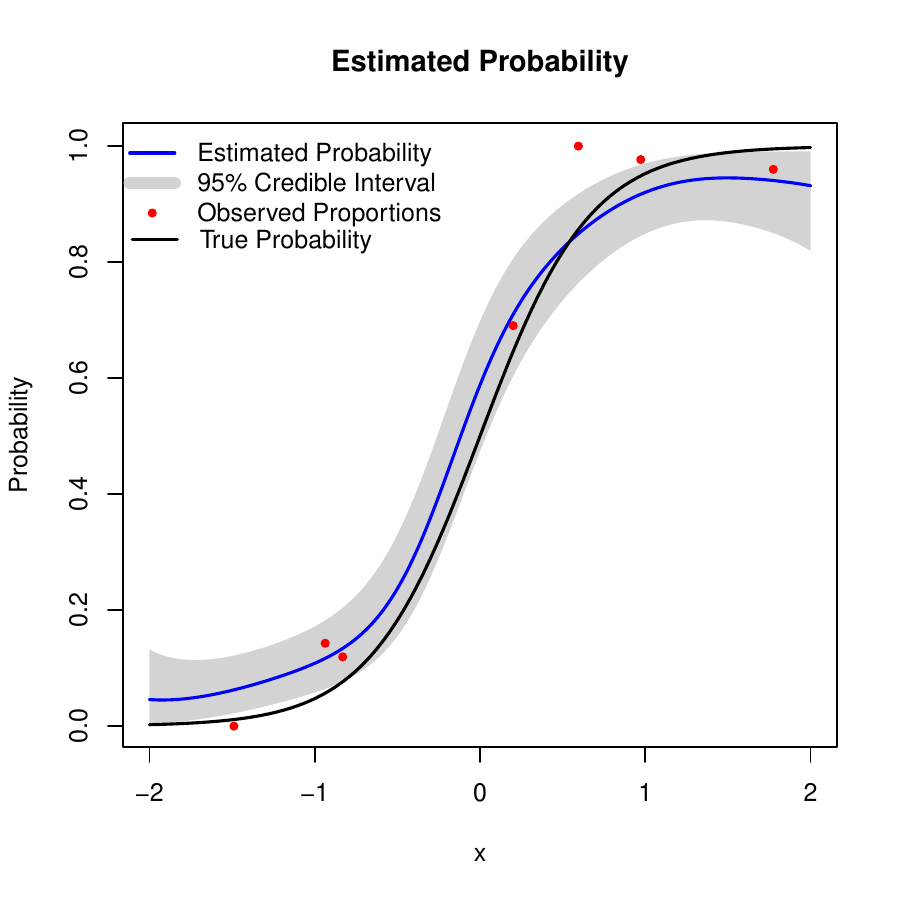}
		\caption{Posterior mean and 95\% CI of Example 1}
		\label{fig:ex1}
	\end{subfigure}
	\hfill 
	\begin{subfigure}{0.49\textwidth}
		\includegraphics[width=\linewidth]{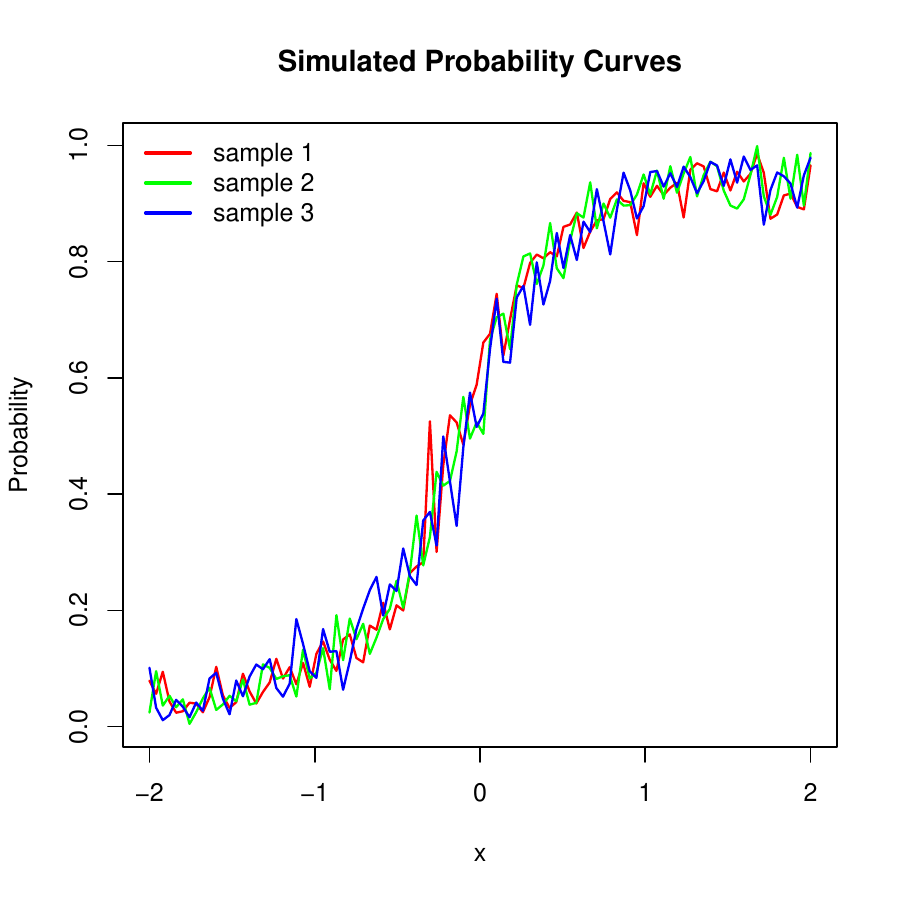}
		\caption{Pointwise posterior draws for Example 1}
		\label{fig:ex1_sim}
	\end{subfigure}
	\medskip
	\begin{subfigure}{0.49\textwidth}
		\includegraphics[width=\linewidth]{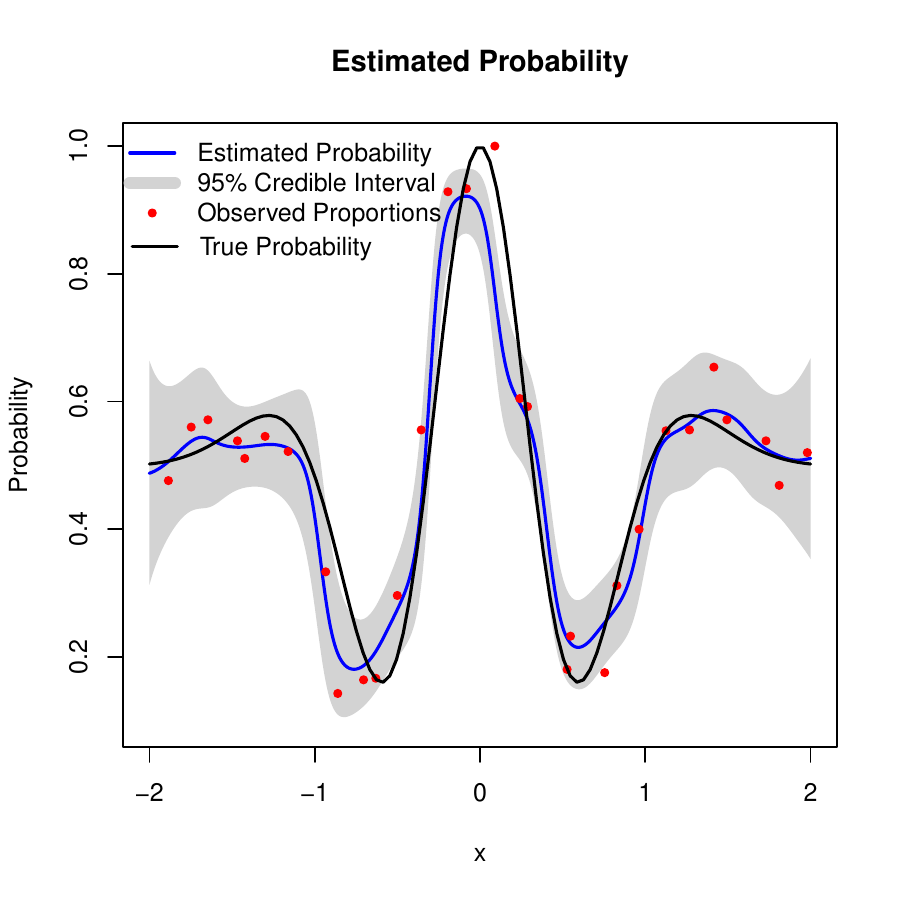}
		\caption{Posterior mean and 95\% CI of Example 2}
		\label{fig:ex2}
	\end{subfigure}
	\hfill 
	\begin{subfigure}{0.49\textwidth}
		\includegraphics[width=\linewidth]{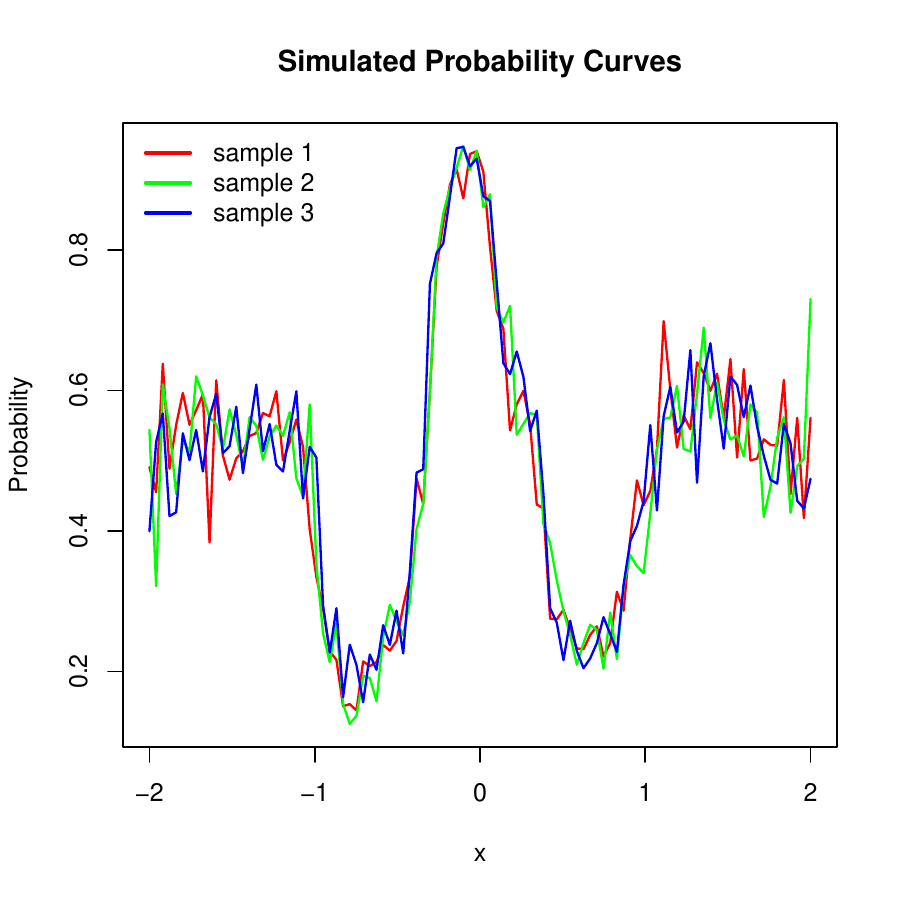}
		\caption{Pointwise posterior draws for Example 2}
		\label{fig:ex2_sim}
	\end{subfigure} 
	
	\caption{Posterior inference and pointwise simulation results from the fitted BKP models. In panels (b) and (d), each displayed curve connects independent draws from the marginal beta posteriors at the plotted input locations for visualization; the connected curves should not be interpreted as realizations from a jointly dependent posterior process.}
	\label{fig:BKP_1D}
\end{figure}

Figure~\ref{fig:BKP_1D} presents two views of the model output. Panel~(a) shows the posterior mean estimate of the probability function $\pi(x)$ (blue), along with a 95\% credible interval (gray band), observed proportions (red dots), and the true underlying probability function (black). Panel~(b) presents three posterior sample curves generated using the \fct{simulate} method, illustrating the variability of the estimated probability surface.

We continue with the same probability function of this example to show the impact of the global precision parameters, $r_0$, in classification tasks. However, the sample size is changed from 7 to 20 for classification. The following \proglang{R} code generates the label of response and fit the BKP model for classification with $r_0$ being 0.01 and 2. 

\begin{CodeChunk}
\begin{CodeInput}  
R> n <- 20
R> Xbounds <- matrix(c(-2, 2), nrow = 1)
R> X <- lhs(n = n, rect = Xbounds)
R> true_pi <- true_pi_fun(X)
R> m <- rep(1, n)
R> y <- as.numeric(true_pi > 0.5)
R> # Fit BKP model with r0 = 0.01
R> BKP_model_1D_1_class_1 <- fit_BKP(
+    X, y, m, Xbounds = Xbounds,
+    prior = "fixed", r0 = 0.01, loss = "log_loss")
R> # Fit BKP model with r0 = 2
R> BKP_model_1D_1_class_2 <- fit_BKP(
+    X, y, m, Xbounds = Xbounds,
+    prior = "fixed", r0 = 2, loss = "log_loss")
\end{CodeInput}
\end{CodeChunk} 
 
\begin{figure}[!t]
	\centering
	\begin{subfigure}{0.49\textwidth}
		\includegraphics[width=\linewidth]{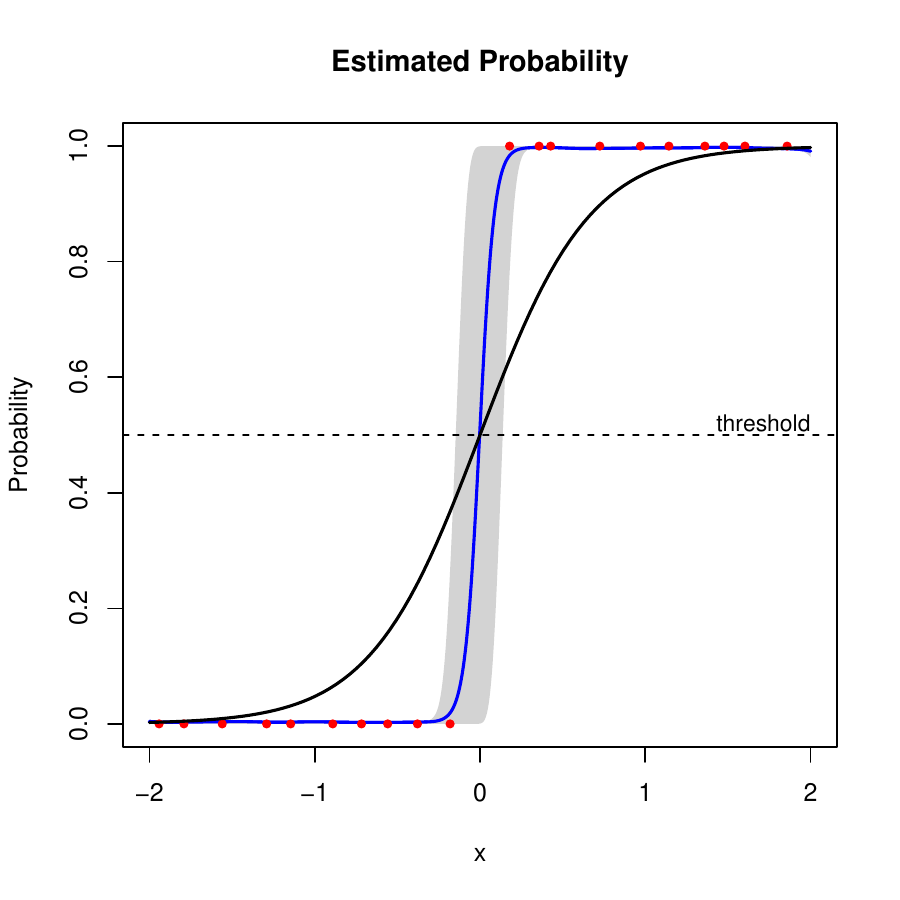}
		\caption{$r_0=0.01$}
		\label{fig:ex1_class_001}
	\end{subfigure}
	\hfill 
	\begin{subfigure}{0.49\textwidth}
		\includegraphics[width=\linewidth]{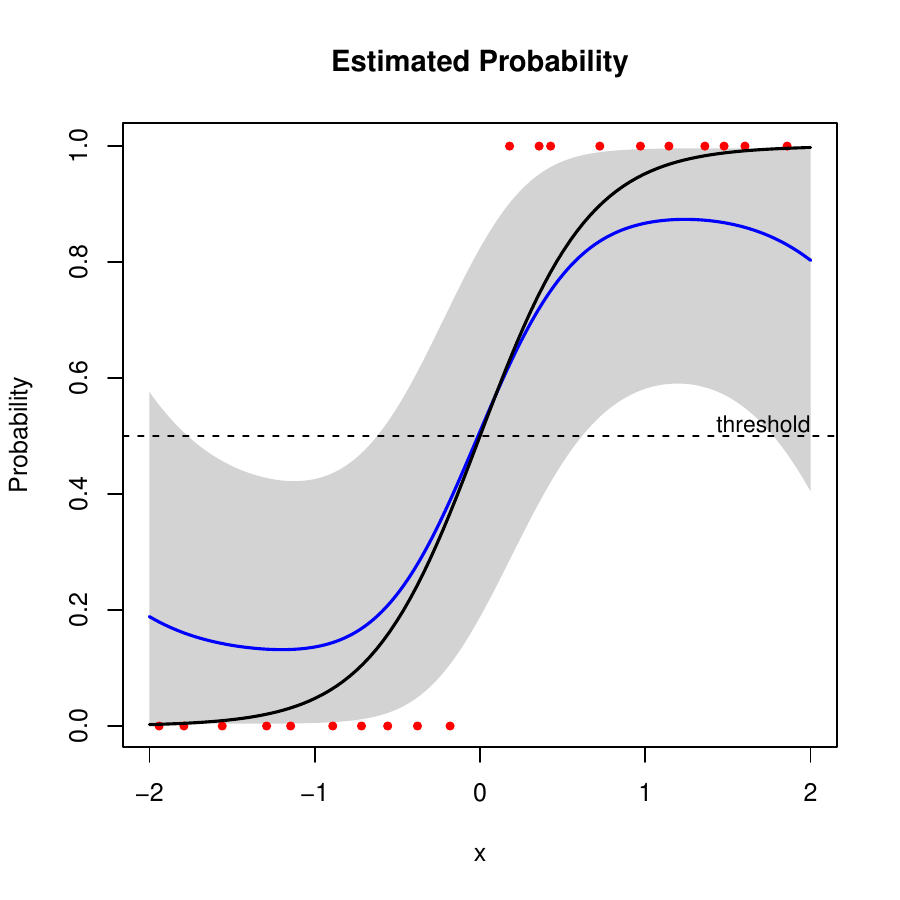}
		\caption{$r_0=2$}
		\label{fig:ex1_class_2}
	\end{subfigure}
	\caption{Posterior mean and 95\% CI of Example 1 (classification task) under $r_0=0.01$ and 2}
	\label{fig:BKP_1D_class}
\end{figure}

Figure~\ref{fig:ex1_class_001} shows a sigmoidal curve with steep slope around zero (in solid line) and a narrow 95\% credible interval (in grey band) under $r_0 = 0.01$, indicating a decisive and confident classification boundary. In contrast, Figure~\ref{fig:ex1_class_2} displays a sine-shape curve with a much wider credible interval, reflecting greater uncertainty and less effective separation. This illustrates that a smaller value of $r_0$ can be preferable in this classification example.
 
\paragraph{Example 2} The first example is essentially a generalized linear model with a smooth logit link, and thus poses limited modeling complexity. To demonstrate the capability of the BKP model in handling more challenging probability surfaces, we consider a second example with a highly nonlinear underlying probability surface. Define the true Bernoulli probability as
\begin{equation}\label{eq:bkp-1d-generate-function-2}
	\pi_{2}(x) = \frac{1}{2}\left[ 1+e^{-x^{2}}\cos \left( 10\frac{1-e^{-x}}{1+e^{-x}} \right) \right],
\end{equation}
where $x \in [-2,2]$  \citep{Goetschalckx:2011}. This example involves rapid local oscillations and strong nonlinearity, making it substantially more difficult to fit than Example~1. Here, we increase the number of locations to 30.

\begin{CodeChunk}
\begin{CodeInput} 
R> n <- 30
R> Xbounds <- matrix(c(-2,2), nrow = 1)
R> X <- lhs(n = n, rect = Xbounds)
R> true_pi <- true_pi_fun(X)
R> m <- sample(100, n, replace = TRUE)
R> y <- rbinom(n, size = m, prob = true_pi) 
R> BKP_model_1D_2 <- fit_BKP(X, y, m, Xbounds = Xbounds)
\end{CodeInput}
\end{CodeChunk}

The results are shown in panels~(c) and (d) of Figure~\ref{fig:BKP_1D}. Panel~(c) shows that the BKP posterior mean follows the main nonlinear features of the probability function, while panel~(d) illustrates posterior variability through three representative realizations of $\pi(x)$.

\paragraph{Example 3} We now consider a two-dimensional test function to further illustrate the modeling capabilities of the \pkg{BKP} package. Let $\vec{x} \in [0,1]^2$, and define the latent surface using a re-scaled version of the Goldstein–Price function \citep{Picheny2013Benchmark}:
\begin{align*}
	f(\vec{x})=& \frac{\log[\{1+a(\vec{x})\}\{30+b(\vec{x})\}]-8.6928}{2.4269}, \quad \text{with}\\
	a(\vec{x}) =& \left(4 x_1 + 4 x_2 - 3 \right)^2  \times  \notag  \\
	& \{75 - 56 \left(x_1 + x_2 \right) + 3\left(4 x_1 - 2 \right)^2 + 6\left(4 x_1 - 2 \right)\left(4 x_2 - 2 \right) + 3\left(4 x_2 - 2 \right)^2\}, \notag \\
	b(\vec{x}) =& \left(8 x_1 - 12 x_2 +2 \right)^2 \times \notag \\
	& \{-14 - 128 x_1 + 12\left(4 x_1 - 2 \right)^2 + 192 x_2 - 36\left(4 x_1 - 2 \right)\left(4 x_2 - 2 \right) + 27\left(4 x_2 - 2 \right)^2 \}.\notag 
\end{align*} 

The true Bernoulli probability surface is then defined by 
\begin{equation}\label{eq:Goldstein-Price}
	\pi_3(\vec{x}) = \Phi\{f(\vec{x})\}, 
\end{equation}
where $\Phi(\cdot)$ is the cumulative distribution function of the standard normal distribution. This formulation produces a smooth yet highly non-linear response surface, providing a challenging test scenario for probabilistic modeling.

To construct the training data, we generate a LHD of size $100$ over $[0,1]^2$. Each location is associated with a binomial observation whose number of trials is randomly drawn from $\{1, \dots, 100\}$. The following \proglang{R} code demonstrates the data simulation and model fitting using the \fct{fit\_BKP} function:

\begin{CodeChunk}
\begin{CodeInput}
R> n <- 100
R> Xbounds <- matrix(c(0, 0, 1, 1), nrow = 2)
R> X <- lhs(n = n, rect = Xbounds)
R> true_pi <- true_pi_fun(X)
R> m <- sample(100, n, replace = TRUE)
R> y <- rbinom(n, size = m, prob = true_pi) 
R> BKP_model_2D <- fit_BKP(X, y, m, Xbounds = Xbounds) 
R> print(BKP_model_2D)
\end{CodeInput}
\begin{CodeOutput} 
       Beta Kernel Process (BKP) Model    

Number of observations (n):  100
Input dimensionality (d):    2
Kernel type:                 (isotropic) gaussian
Optimized kernel parameters: 0.0805
Minimum achieved loss:       0.01125
Loss function:               brier
Prior type:                  noninformative
\end{CodeOutput}
\end{CodeChunk} 

\begin{figure}[!t]
	\centering 
	\begin{subfigure}{0.45\textwidth}
		\includegraphics[width=\linewidth]{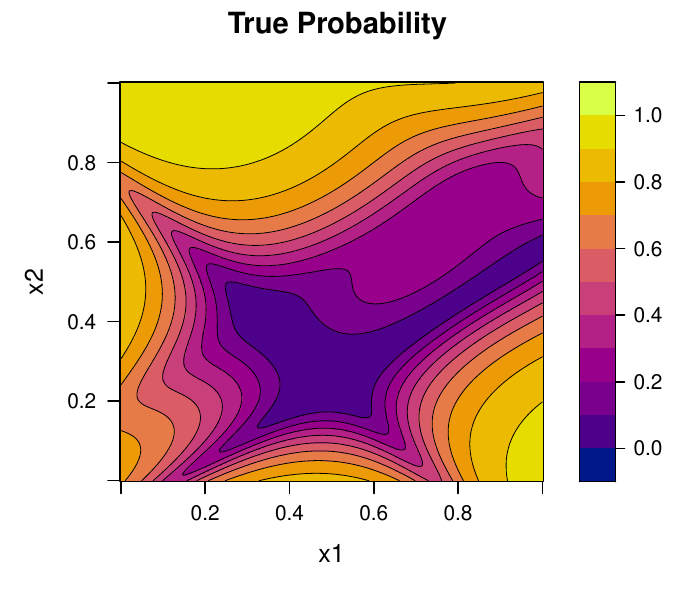} 
	\end{subfigure}   
	\medskip  
	\begin{subfigure}{\textwidth}
		\includegraphics[width=\linewidth]{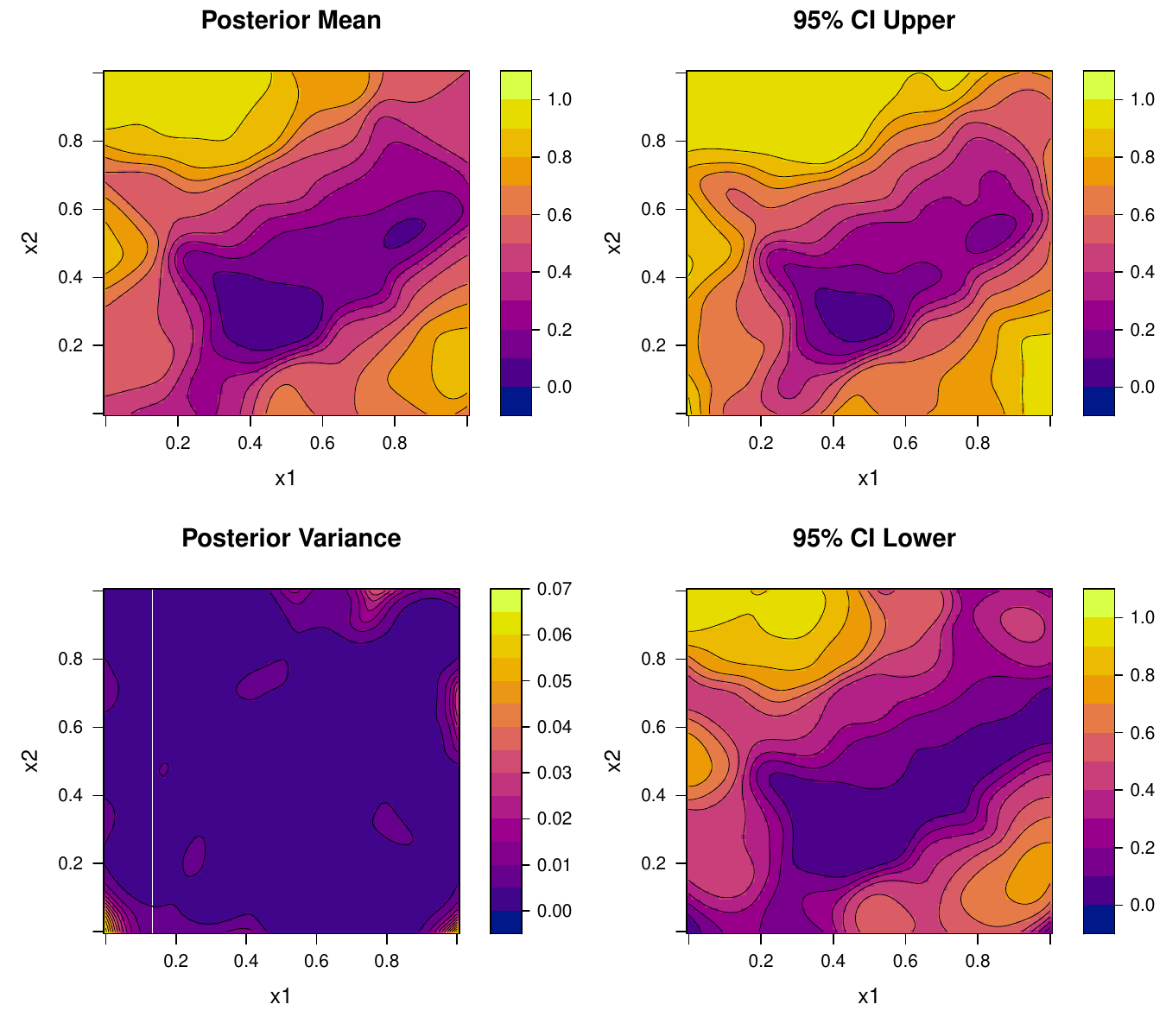}  
	\end{subfigure} 
	\caption{Comparison of the true probability surface and BKP-based posterior summaries of Example~3} 
	\label{fig:BKP_2D}
\end{figure}

Figure~\ref{fig:BKP_2D} is produced by the following call and displays the true probability surface together with the BKP posterior summaries.

\begin{CodeChunk}
\begin{CodeInput}  
R> plot(BKP_model_2D) 
\end{CodeInput} 
\end{CodeChunk}

The results of model prediction are as follows:

\begin{CodeChunk}
\begin{CodeInput}
R> Xnew <- lhs(n = 10, rect = Xbounds)
R> predict(BKP_model_2D, Xnew)
\end{CodeInput}
\begin{CodeOutput}
Prediction results on new data (Xnew).
Total number of prediction points: 10 

Preview of predictions for new data (first 6 of 10 points):
     x1     x2   mean variance 2.5% quantile 97.5% quantile
 0.2184 0.1046 0.3513   0.0042        0.2298         0.4836
 0.9401 0.6108 0.1558   0.0043        0.0512         0.3043
 0.6782 0.8087 0.5144   0.0021        0.4254         0.6031
 0.1591 0.9120 0.9452   0.0006        0.8906         0.9817
 0.7838 0.2556 0.5136   0.0033        0.4005         0.6260
 0.4185 0.5215 0.2639   0.0010        0.2047         0.3277
 ...
\end{CodeOutput}
\end{CodeChunk} 

We continue with Example~3 to examine the computational scaling of the BKP model in comparison with the logistic Gaussian process (LGP) model \citep{Rasmussen2006GPML}. 
The sample size ranges from 200 to 5000. 
Two settings for kernel hyperparameter specification are considered. 
In the first setting, the BKP kernel length-scale parameter is fixed at \code{theta = 1}. 
In the second setting, the kernel parameter is selected by LOOCV-based optimization, using either a single starting value \code{n_multi_start = 1} or the package default \code{n_multi_start = NULL}. 
Under the default rule, the number of starting values is $10d_\gamma$, where $d_\gamma = 1$ for isotropic kernels and $d_\gamma = d$ for anisotropic kernels. 
In the present benchmark, the default isotropic Gaussian kernel is used, so $d_\gamma = 1$ and the multi-start optimization uses 10 starting values.

The LGP model was implemented using the \fct{gp\_fit} and \fct{gp\_optim} functions from the \pkg{gplite} package \citep{Piironen2022gplite}. 
The \pkg{gplite} package uses additional starts only when optimization failure is detected; hence, its timing results are not directly comparable to a fixed multi-start optimization budget.

The timing experiments were conducted on the hardware described in the computational details.
The reported computation times are averages over 20 independent repetitions to reduce variability due to random initialization and computational fluctuations.

\begin{figure}[!t]
	\centering 
	\includegraphics[width=\linewidth]{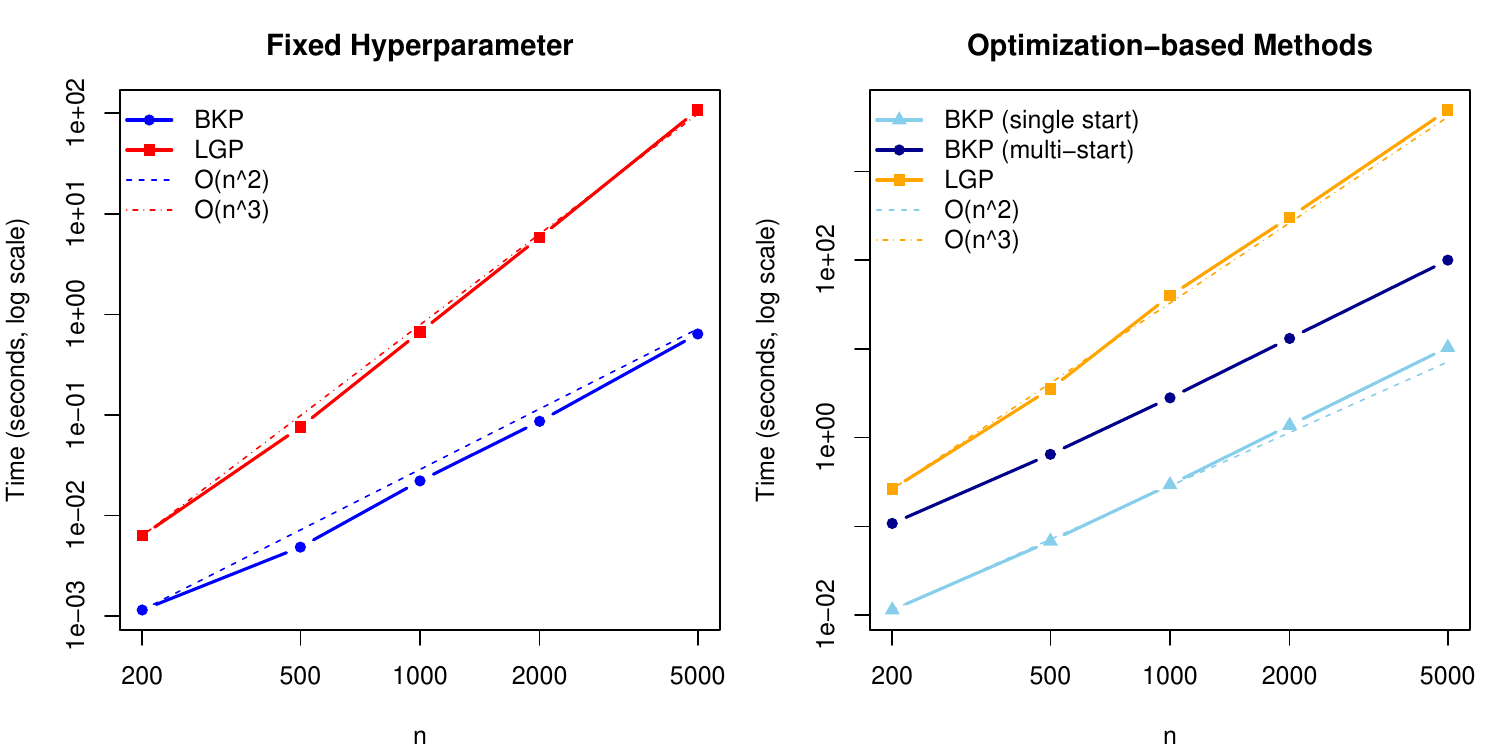}  
	\caption{Comparison of computation times (in log scale) between BKP and LGP methods: (a) fixed hyperparameter; (b) optimization-based methods.}
	\label{fig:elapsed_time}
\end{figure}

The average computation times are shown in Figure~\ref{fig:elapsed_time}. 
The observed trends are consistent with the theoretical scaling discussed in Remark~4: full BKP is a quadratic full-kernel method, whereas full LGP fitting involves cubic matrix-factorization operations. 
For BKP, the cost of multi-start optimization increases approximately in proportion to the number of starting values, although the exact ratio also depends on the number of loss evaluations taken by the optimizer from each start. 
In the present isotropic benchmark, the default multi-start setting uses $10d_\gamma = 10$ starts, so the multi-start timing is expected to be roughly one order of magnitude larger than the single-start timing, up to optimizer and implementation overhead. 
This illustrates the trade-off between increased optimization robustness and additional computation. 

\paragraph{Example 4}  
We next consider a binary classification task using the \emph{Two Spirals} dataset \citep{Chalup2007spirals}, a well-known benchmark consisting of two intertwined spirals in a bounded two-dimensional input space. This dataset is particularly challenging due to the complex, non-linearly separable class structure.

\begin{figure}[!t]
	\centering 
	\begin{subfigure}{\textwidth}
		\includegraphics[width=\linewidth]{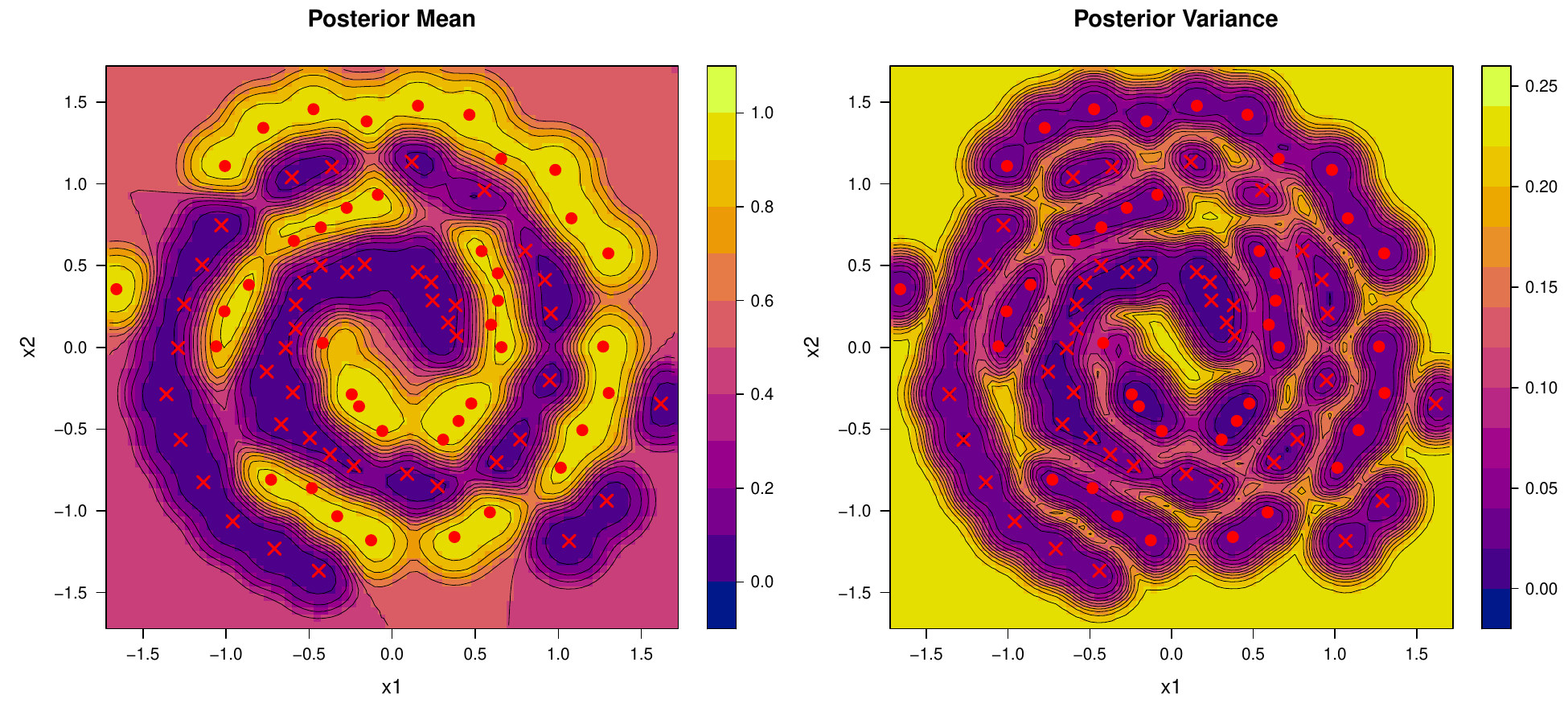}  
		\caption{BKP}
	\end{subfigure} 
	\medspace
	\begin{subfigure}{\textwidth}
		\includegraphics[width=\linewidth]{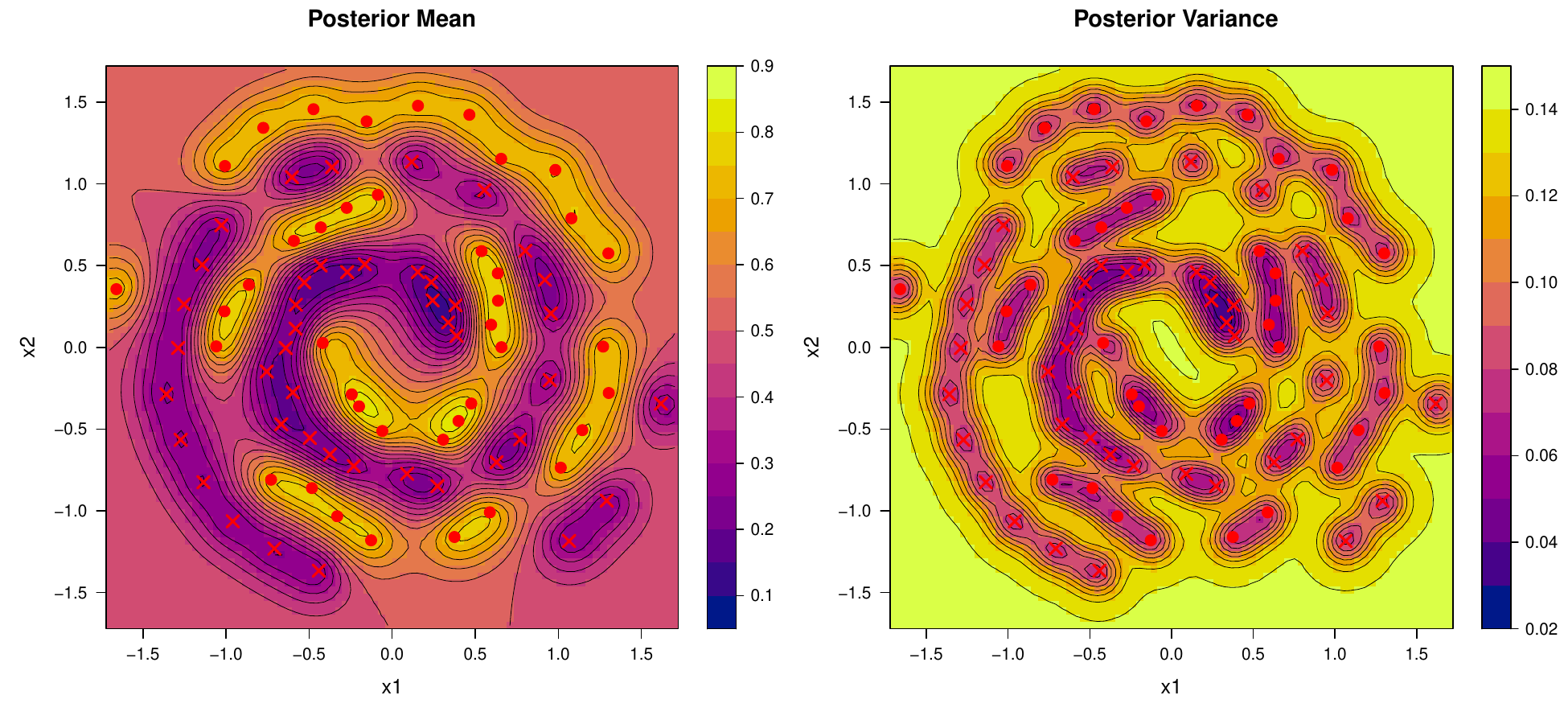}  
		\caption{LGP}
	\end{subfigure}    
	\caption{Probability surfaces for the Two Spirals example fitted by the BKP and LGP models. The left panels show the posterior mean class probabilities, and the right panels show the corresponding model-specific posterior variance summaries. Color shading represents the estimated class probability, while circles and crosses indicate training observations from the two classes.}
	\label{fig:BKP_spiral}
\end{figure}

We generate $n = 120$ observations using the \fct{mlbench.spirals} function from the \proglang{R} package \pkg{mlbench} \citep{Leisch2024mlbench}, with two complete rotations and additive Gaussian noise of standard deviation \code{sd = 0.05}. The inputs $\vec{x}$ are constrained to the domain $[-1.7, 1.7]^2$, and the binary class labels are encoded as 0 and 1. We fit the BKP model using a fixed prior specification with \code{r_0 = 0.1} and \code{p_0 = 0.5}.

\begin{CodeChunk}
\begin{CodeInput}
R> n <- 120
R> data <- mlbench.spirals(n, cycles = 2, sd = 0.05)
R> train_indices <- sample(1:n, 0.7 * n)
R> X_train <- data$x[train_indices, ]
R> y_train <- as.numeric(data$classes[train_indices]) - 1 # Convert to 0/1 for BKP
R> X_test <- data$x[-train_indices, ]
R> y_test <- as.numeric(data$classes[-train_indices]) - 1 
R> m <- rep(1, length(train_indices))
R> Xbounds <- rbind(c(-1.7, 1.7), c(-1.7, 1.7)) 
R> BKP_model_Class <- fit_BKP(
+    X_train, y_train, m, Xbounds = Xbounds,
+    prior = "fixed", r0 = 0.1, p0 = 0.5, loss = "log_loss")
R> prediction <- predict(BKP_model_Class, X_test)
\end{CodeInput}
\end{CodeChunk} 

Figure~\ref{fig:BKP_spiral} presents the posterior mean and variance surfaces of the class probabilities. The upper-left panel shows that the BKP model captures the intricate spiral structure, with smoothly varying posterior mean probabilities that delineate the nonlinear decision boundary. The upper-right panel displays the corresponding posterior variance, which is generally larger near the boundary regions between the spirals. Circles and crosses indicate training observations from the two respective classes. These results illustrate the ability of BKP to model complex classification boundaries while providing model-based posterior uncertainty summaries.

We compare the prediction performance with the LGP model implemented by R package \pkg{gplite}. 
\begin{CodeChunk}
\begin{CodeInput}
R> gp <- gp_init(cf = cf_sexp(), lik = lik_bernoulli())
R> gp <- gp_optim(gp, X_train, y_train, verbose = FALSE) 
R> prediction_gp <- gp_pred(gp, as.matrix(X_test), transform = TRUE)
\end{CodeInput}
\end{CodeChunk} 

\begin{figure}[!t]
	\centering 
	\begin{subfigure}{0.48\textwidth}
		\includegraphics[width=\linewidth]{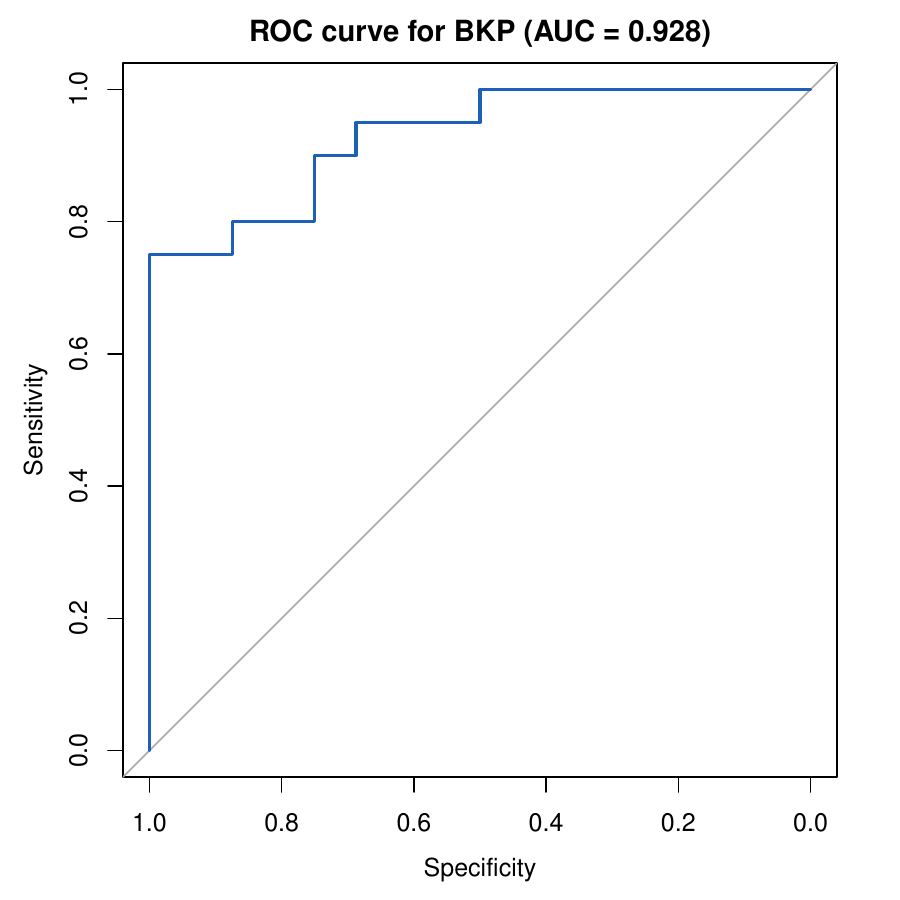}  
		\caption{BKP}
	\end{subfigure} 
	\begin{subfigure}{0.48\textwidth}
		\includegraphics[width=\linewidth]{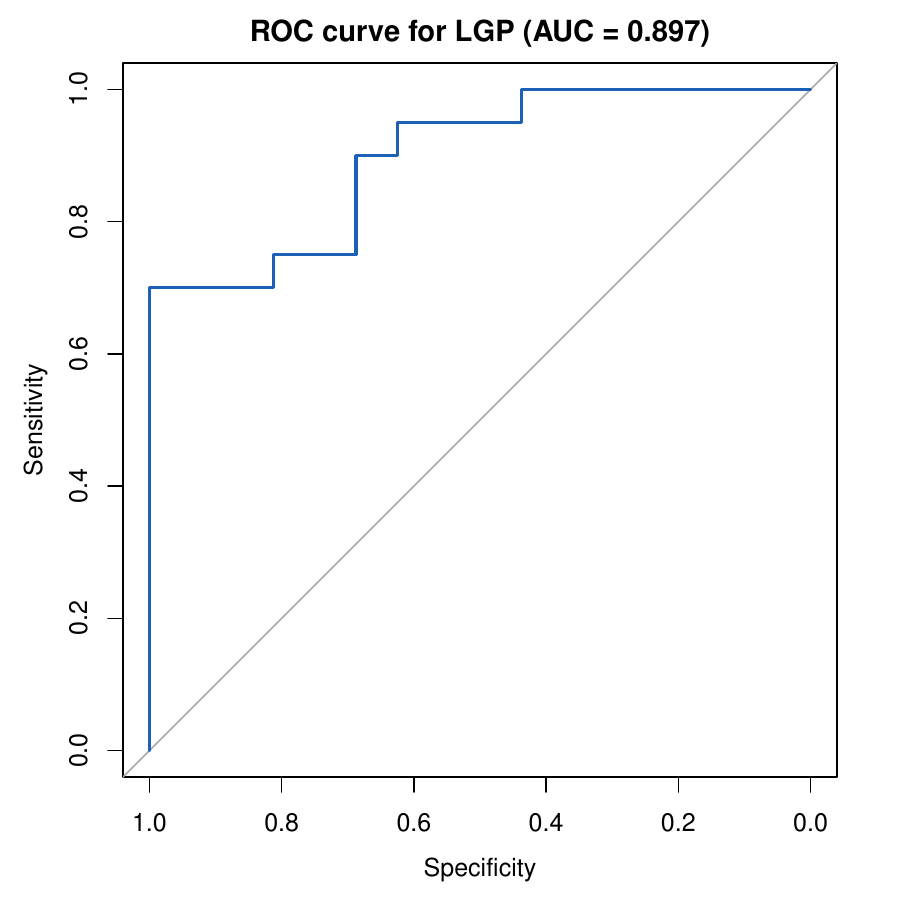}  
		\caption{LGP}
	\end{subfigure}    
	\caption{ROC curves and their respective AUCs for BKP and LGP models}
	\label{fig:ex4-roc}
\end{figure}

The lower panels of Figure~\ref{fig:BKP_spiral} show that the LGP fit produces a different probability surface and a more spatially diffuse variance pattern than BKP. 
Because the two methods use different posterior constructions, their variance surfaces should be interpreted as model-specific uncertainty summaries rather than as direct evidence of comparative calibration. 
The receiver operating characteristic (ROC) curves in Figure~\ref{fig:ex4-roc} provide a predictive comparison on the held-out observations: BKP achieves an AUC of 0.928, compared with 0.897 for LGP, indicating better discrimination for BKP on this particular train--test split.

\subsection{DKP Model}\label{sec:example_dkp}

This subsection presents illustrative examples based on the full DKP model. 
The scalable TwinDKP interface is illustrated together with TwinBKP in Section~\ref{sec:example_twin}.
 
\paragraph{Example 5}
Consider a one-dimensional three-class multinomial response setting. The input is defined on the interval $x \in [-2, 2]$, and the true class probability vector is given by
\begin{equation*} 
	\vec{\pi}(x) = \left[\frac{\pi_{1}(x)}{2}, \frac{\pi_{2}(x)}{2},1-\frac{\pi_{1}(x)}{2}-\frac{\pi_{2}(x)}{2}\right]^\top,
\end{equation*}
where $\pi_{1}(x)$ and $\pi_{2}(x)$ are smooth functions defined in \eqref{eq:bkp-1d-generate-function-1} and \eqref{eq:bkp-1d-generate-function-2}, respectively.

We generate $n = 30$ input locations using Latin hypercube sampling over the interval $[-2, 2]$. At each location, the response is a multinomial vector with probability $\vec{\pi}(x)$ and a random total count sampled from $\{1, \dots, 150\}$. The DKP model is then fitted using the \fct{fit\_DKP} function and visualized with the \fct{plot} method:

\begin{CodeChunk}
\begin{CodeInput}  
R> n <- 30
R> Xbounds <- matrix(c(-2, 2), nrow = 1)
R> X <- lhs(n = n, rect = Xbounds)
R> true_pi <- true_pi_fun(X)
R> m <- sample(150, n, replace = TRUE)
R> Y <- t(sapply(1:n, function(i) rmultinom(1, size = m[i], prob = true_pi[i, ]))) 
R> DKP_model_1D <- fit_DKP(X, Y, Xbounds = Xbounds)
R> plot(DKP_model_1D) 
\end{CodeInput}
\end{CodeChunk} 

\begin{figure}[!t]
	\centering 
	\includegraphics[width=\linewidth]{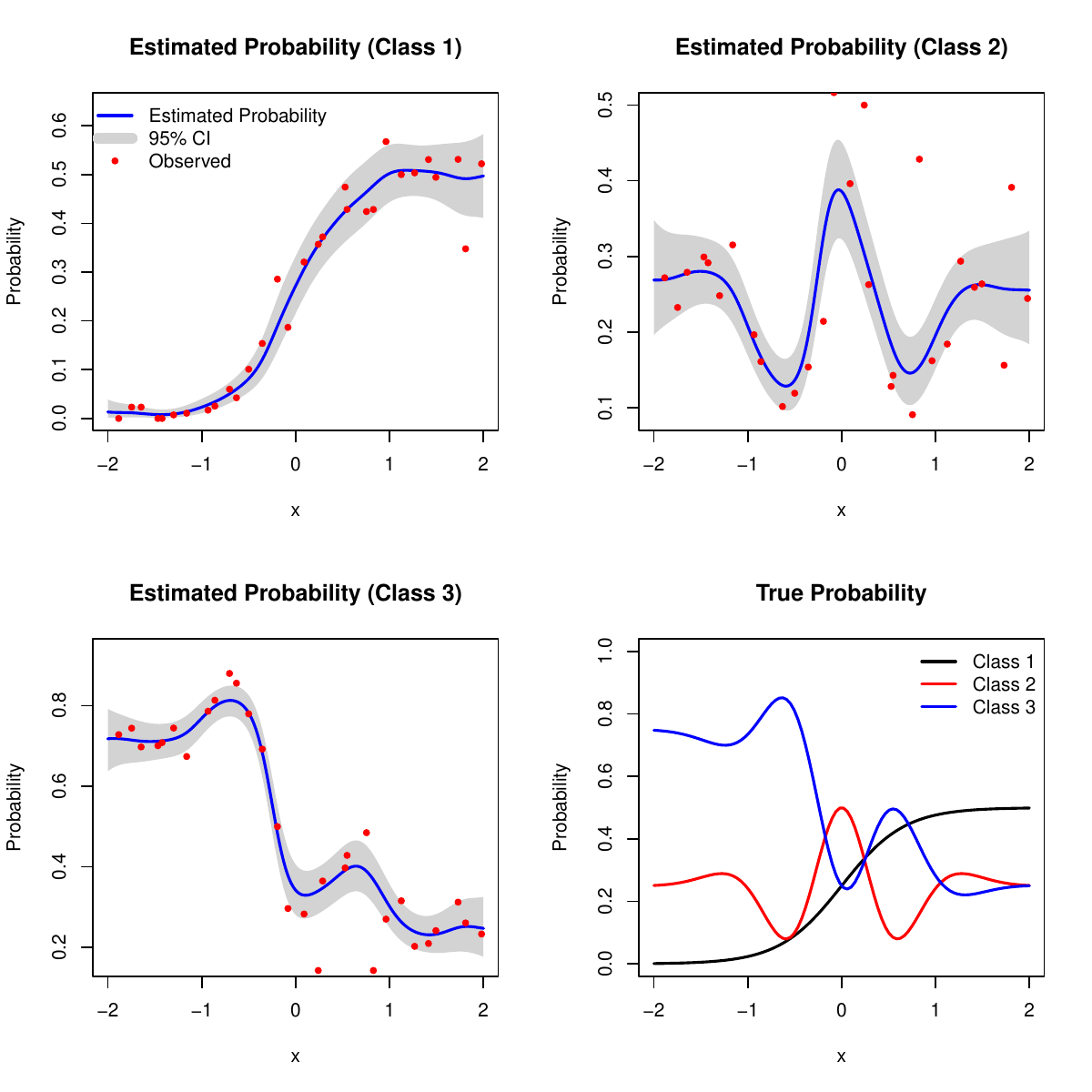}  
	\caption{Posterior inference from the fitted DKP model for a one-dimensional three-class problem} 
	\label{fig:ex5}
\end{figure} 

Figure~\ref{fig:ex5} shows that the DKP fitted mean curves follow the main features of the true class-probability functions, while the shaded bands summarize posterior uncertainty.
The predictive mean curves align with the true underlying structure, and the shaded bands reflect posterior uncertainty.

\paragraph{Example 6} Consider a two-dimensional three-class multinomial response setting. Let $\vec{x} = [x_{1}, x_{2}]^\top \in [0,1]^2$, and define the true class probability function as
\begin{align*} 
	\vec{\pi}(\vec{x})= \left[\frac{\pi_{3}(\vec{x})}{2}, \; \frac{\pi_{4}(\vec{x})}{2}, \; 1-\frac{\pi_{3}(\vec{x})}{2}- \frac{\pi_{4}(\vec{x})}{2}\right]^\top,
\end{align*} 
where  $\pi_{3}(\vec{x})$ is defined in \eqref{eq:Goldstein-Price}, and
\[
	\pi_{4}(\vec{x}) = \sin(\pi x_1) \sin(\pi x_2).
\]

We generate $n = 100$ input points using Latin hypercube sampling over $[0,1]^2$. 
At each location, the response is a multinomial vector with probability $\vec{\pi}(\vec{x})$ and a total count randomly sampled from $\{1, \dots, 150\}$.
We fit the DKP model using \fct{fit\_DKP} and visualize the results with the \fct{plot} method:

\begin{CodeChunk}
\begin{CodeInput} 
R> n <- 100
R> Xbounds <- matrix(c(0, 0, 1, 1), nrow = 2)
R> X <- lhs(n = n, rect = Xbounds)
R> true_pi <- true_pi_fun(X)
R> m <- sample(150, n, replace = TRUE)
R> Y <- t(sapply(1:n, function(i) rmultinom(1, size = m[i], prob = true_pi[i, ]))) 
R> DKP_model_2D <- fit_DKP(X, Y, Xbounds = Xbounds) 
\end{CodeInput}
\end{CodeChunk} 

\begin{figure}[!t]
	\centering 
	\begin{subfigure}{0.5\textwidth}
		\includegraphics[width=\linewidth]{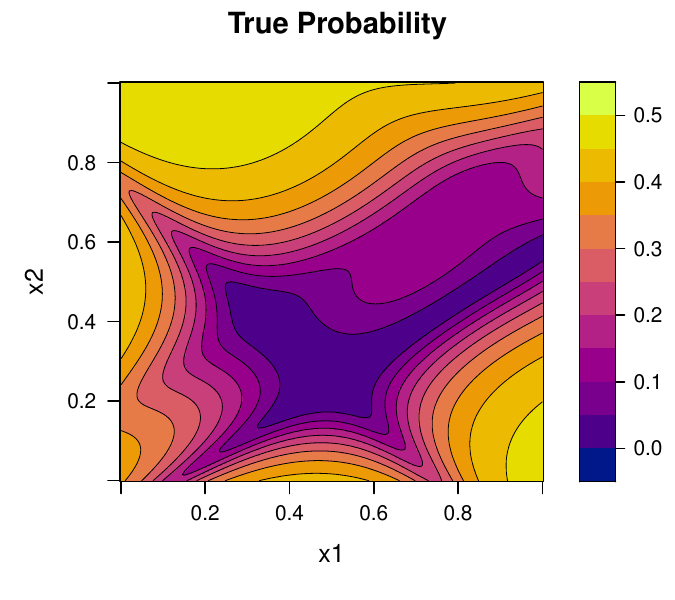} 
	\end{subfigure}   
	\medspace 
	\begin{subfigure}{\textwidth}
		\includegraphics[width=\linewidth]{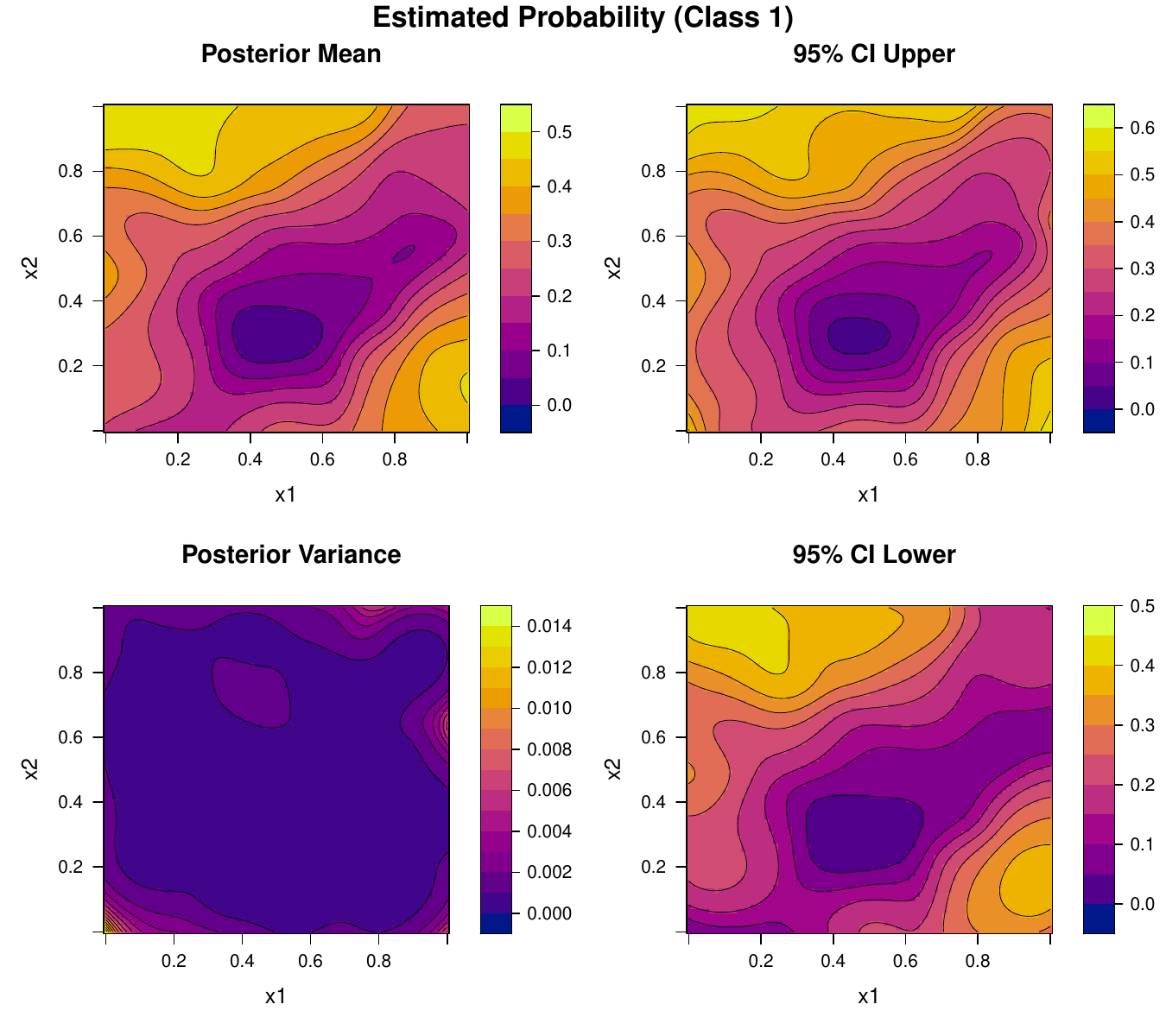}  
	\end{subfigure} 
	\caption{Class 1: true distribution (top) and DKP model posterior summaries (bottom)} 
	\label{fig:ex6-class1}
\end{figure} 

\begin{figure}[!t]
	\centering 
	\begin{subfigure}{0.5\textwidth}
		\includegraphics[width=\linewidth]{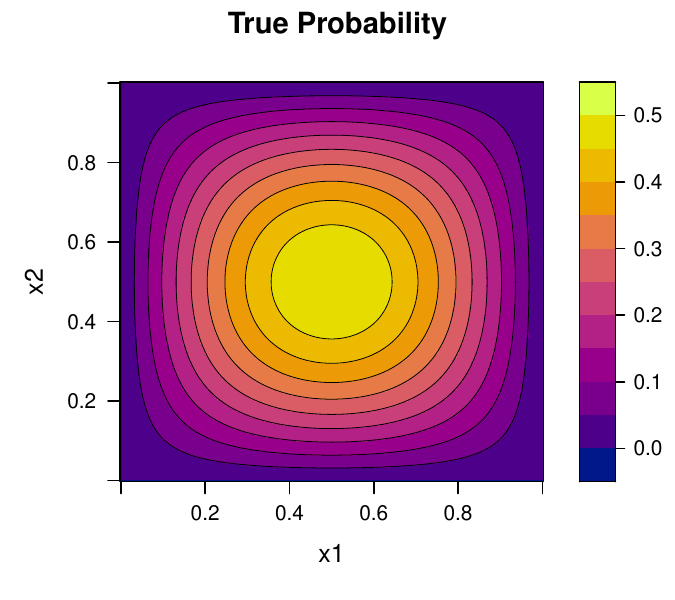} 
	\end{subfigure}   
	\medspace   
	\begin{subfigure}{\textwidth}
		\includegraphics[width=\linewidth]{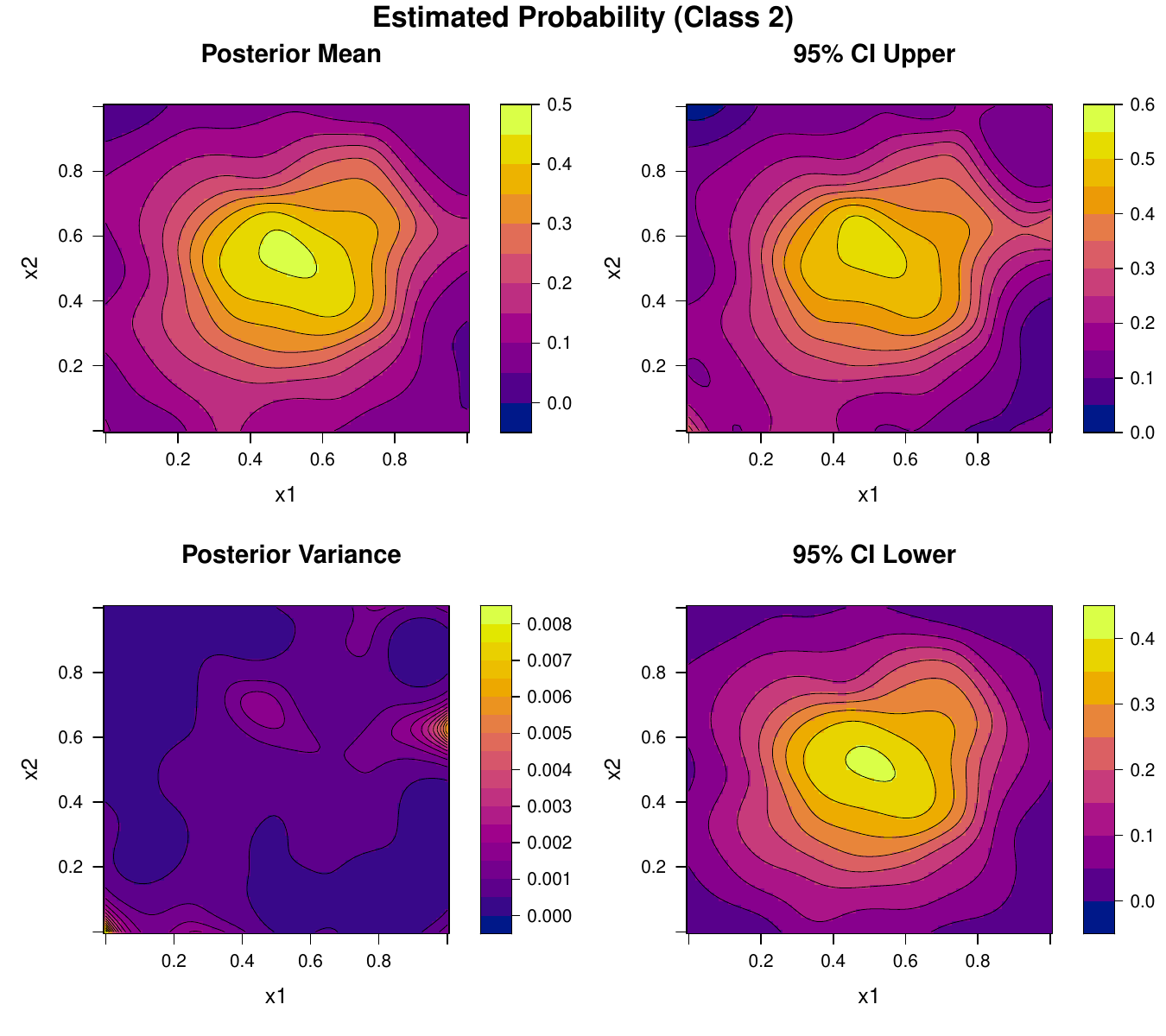}  
	\end{subfigure} 
	\caption{Class 2: true distribution (top) and DKP model posterior summaries (bottom)} 
	\label{fig:ex6-class2}
\end{figure} 

\begin{figure}[!t]
	\centering 
	\begin{subfigure}{0.5\textwidth}
		\includegraphics[width=\linewidth]{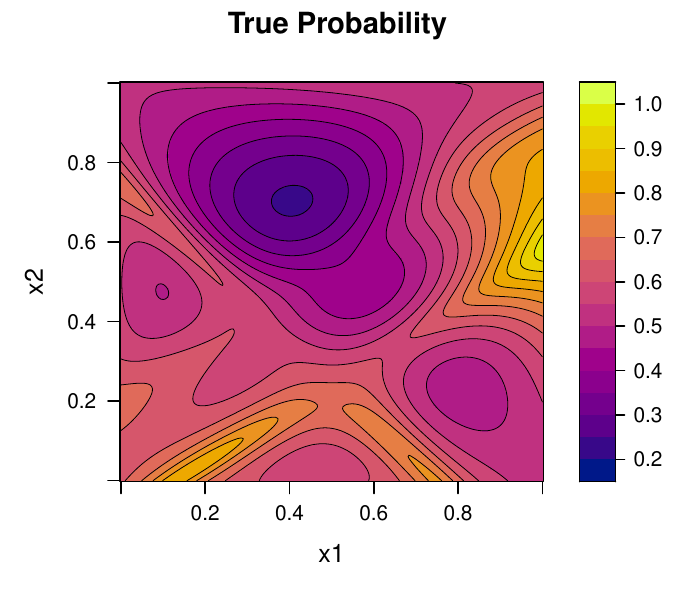} 
	\end{subfigure}   
	\medspace 
	\begin{subfigure}{\textwidth}
		\includegraphics[width=\linewidth]{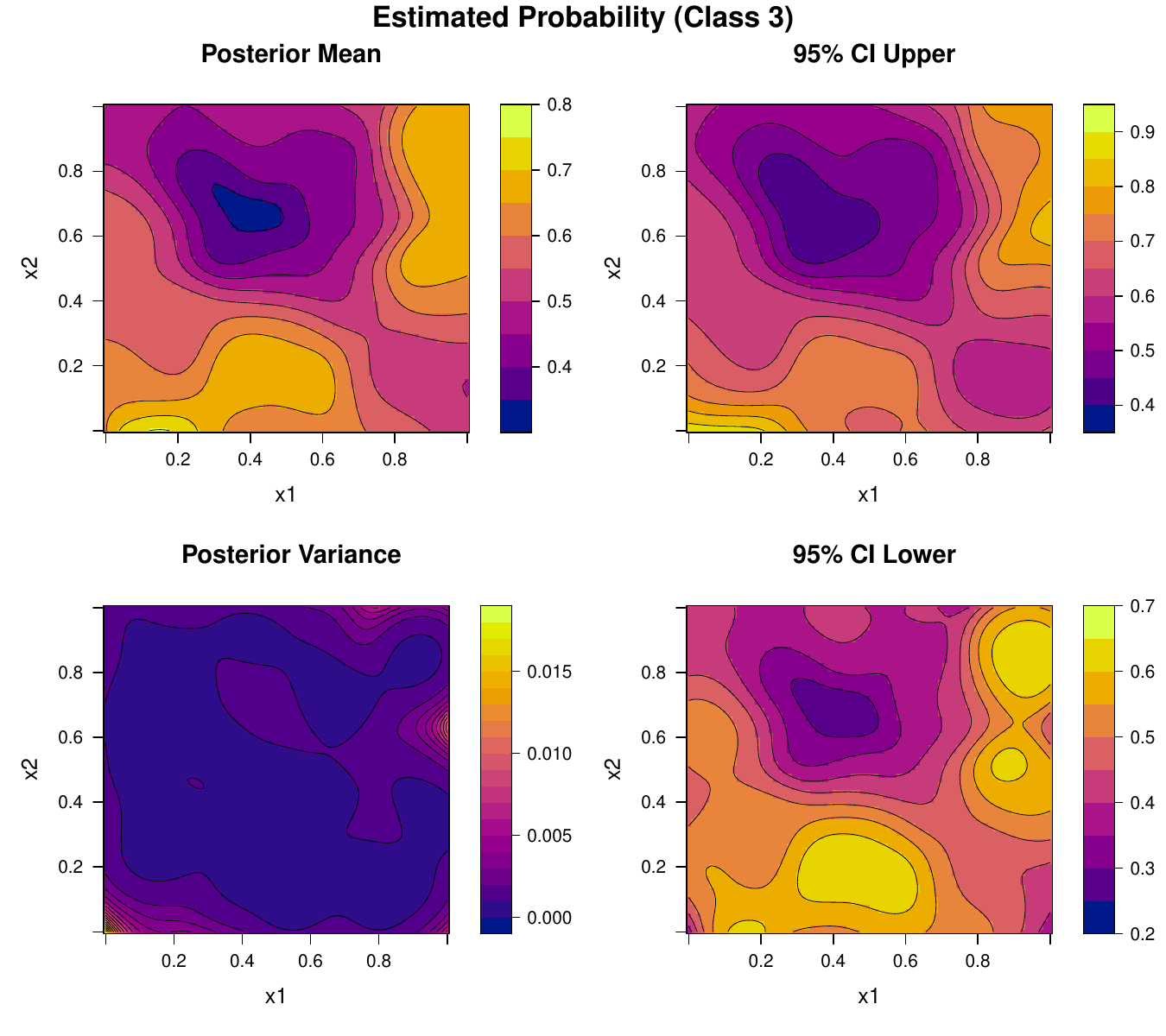}  
	\end{subfigure} 
	\caption{Class 3: true distribution (top) and DKP model posterior summaries (bottom)} 
	\label{fig:ex6-class3}
\end{figure}

Figures~\ref{fig:ex6-class1}--\ref{fig:ex6-class3} display the ground truth surfaces and corresponding posterior inference for each of the three classes. In each figure, the top panel shows the true class probability surface, while the bottom row presents the posterior mean, variance, and uncertainty bounds produced by the DKP model. 

\paragraph{Example 7} 
Consider another three-class classification task based on the well-known \emph{Iris} dataset available in \proglang{R}. This classic benchmark dataset contains measurements of three iris species---\emph{setosa}, \emph{versicolor}, and \emph{virginica}---each described by four features: sepal length, sepal width, petal length, and petal width. Due to the overlap in feature space, particularly between \emph{versicolor} and \emph{virginica}, the class boundaries are not linearly separable, making this dataset a standard testbed for evaluating multi-class classification algorithms. 
For visualization purposes, we restrict our analysis to the first two features: sepal length and sepal width. We fit the DKP model using a fixed prior specification with \code{r0 = 0.01} and \code{p0 = rep(1/3, 3)}, assuming equal prior probability for each class.

\begin{CodeChunk}
\begin{CodeInput}
R> data(iris)
R> X <- as.matrix(iris[, 1:2])
R> Xbounds <- rbind(c(4.2, 8), c(1.9, 4.5))
R> labels <- iris$Species
R> Y <- model.matrix(~ labels - 1)  # expand factors to a set of dummy variables
R> train_indices <- sample(1:nrow(iris), 0.7 * nrow(iris))
R> X_train <- X[train_indices, ]
R> Y_train <- Y[train_indices, ]
R> DKP_model_Class <- fit_DKP(
+    X_train, Y_train, Xbounds = Xbounds, loss = "log_loss",
+    prior = "fixed", r0 = 0.01, p0 = rep(1/3, 3))
R> X_test <- X[-train_indices, ]
R> Y_test <- Y[-train_indices, ]
R> dkp_pred_probs <- predict(DKP_model_Class, X_test)$mean
\end{CodeInput}
\end{CodeChunk} 

\begin{figure}[!t]
	\centering 
	\begin{subfigure}{\textwidth}
		\includegraphics[width=\linewidth]{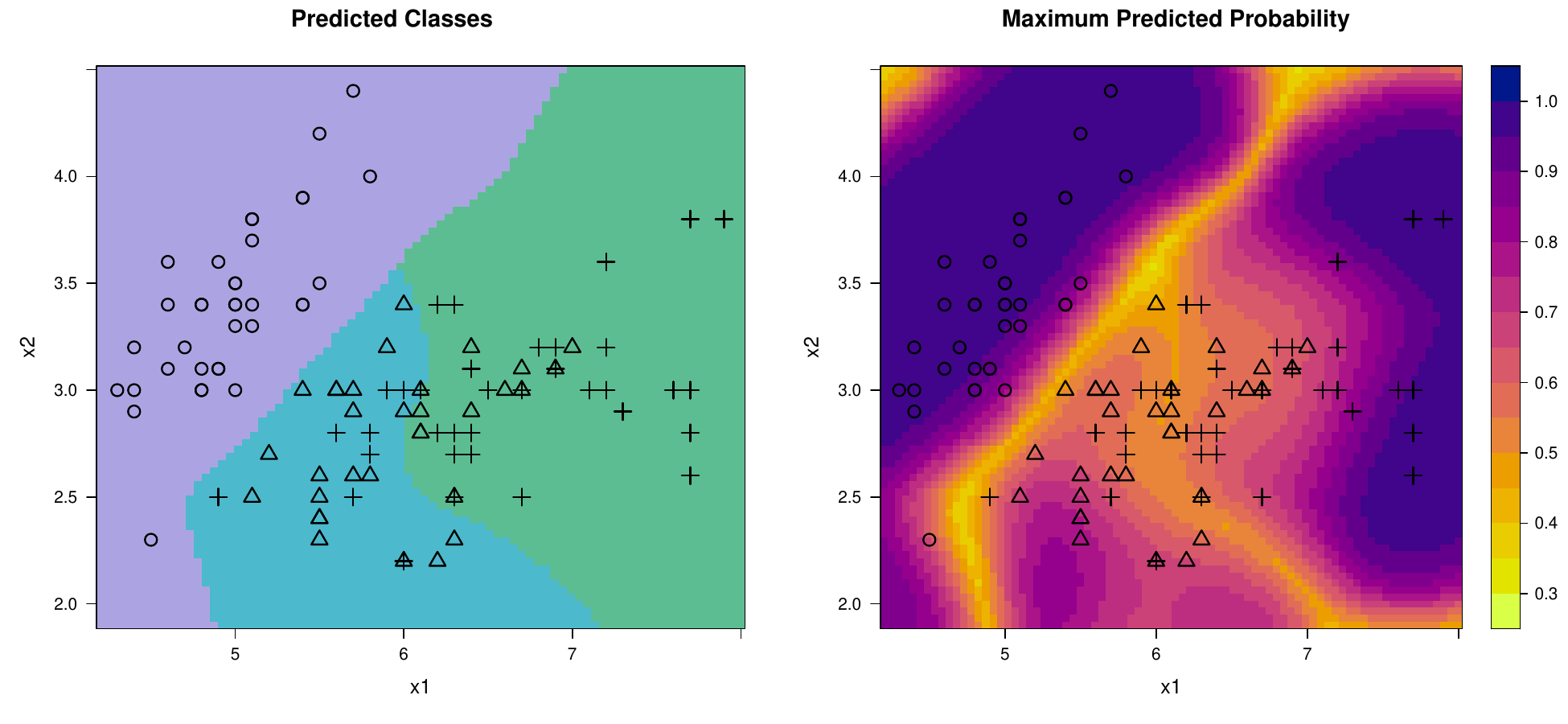} 
		\caption{DKP }
	\end{subfigure} 
	\begin{subfigure}{\textwidth}
		\includegraphics[width=\linewidth]{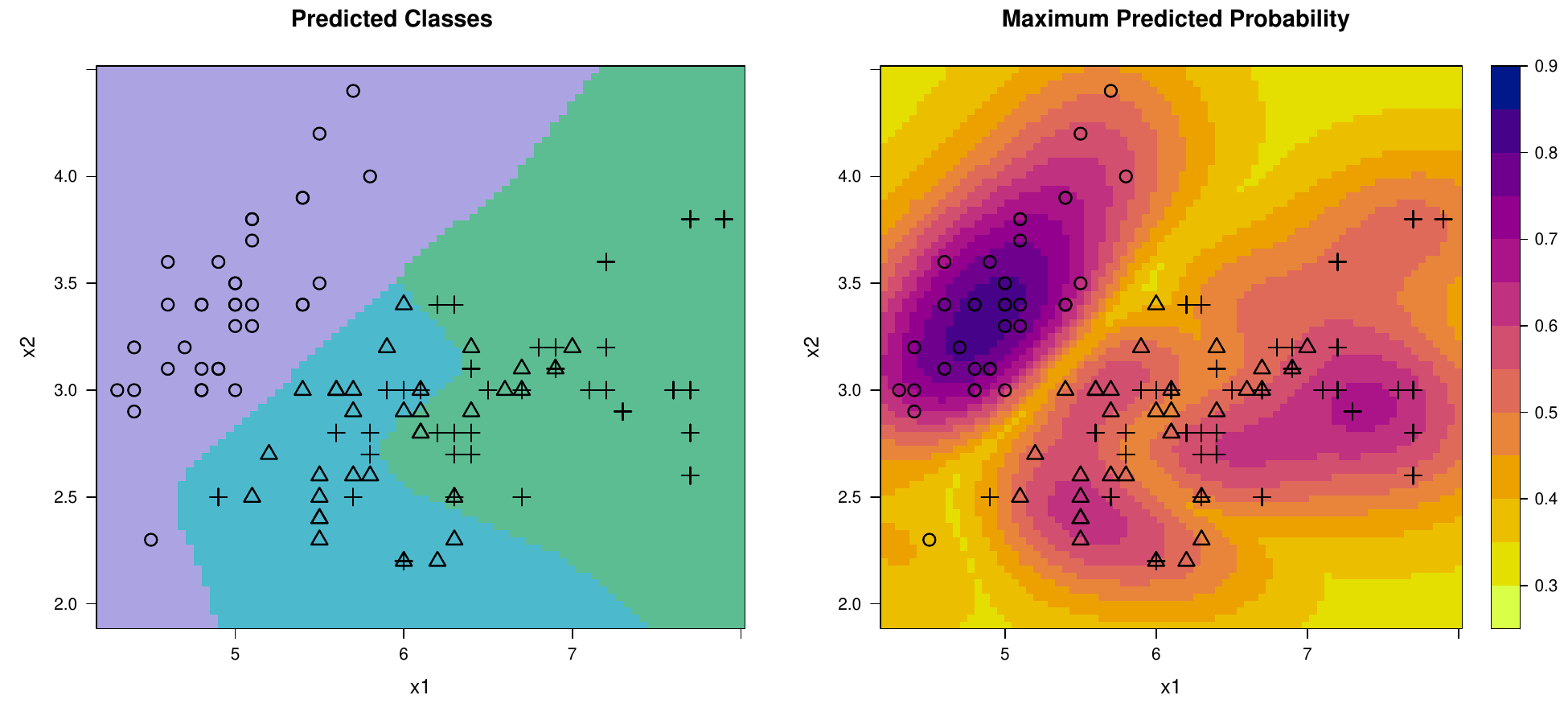} 
		\caption{LGP }
	\end{subfigure}  
	\caption{Classification and confidence visualization for the DKP and multiclass LGP fits on the \emph{Iris} dataset. The left panels show the predicted classification regions, with the training observations overlaid using class-specific symbols. The right panels show the maximum predicted class probability as a model-specific summary of classification confidence. Lower values indicate less decisive class-probability assignments, which typically occur near class boundaries.}
	\label{fig:ex7}
\end{figure}

Figure~\ref{fig:ex7} provides a visual comparison of the fitted DKP and multiclass LGP models. 
The upper-left panel shows that the DKP model clearly separates \emph{setosa} from the other two species, while producing a smooth transition between the overlapping \emph{versicolor} and \emph{virginica} regions. 
The upper-right panel displays the corresponding maximum posterior class probability $\max_j \pi_j(\vec{x})$. Values near 1 indicate decisive class assignments, whereas values closer to $1/3$ indicate greater ambiguity among the three classes. The lower-confidence regions are concentrated primarily near the fitted class boundaries, yielding a coherent representation of classification confidence.

For comparison, we fit a multiclass Gaussian-process classifier using the \pkg{kernlab} package \citep{Karatzoglou2004kernlab}. 
The \fct{gausspr} implementation fits pairwise binary Gaussian-process classifiers and combines their probability estimates through pairwise coupling. 
For evaluation, we compute one-vs-rest ROC curves for the three classes and report the corresponding macro-average AUC.

\begin{CodeChunk}
\begin{CodeInput}
R> iris_data <- data.frame(
+    Sepal.Length = iris$Sepal.Length,
+    Sepal.Width = iris$Sepal.Width,
+    Species = iris$Species
+  )
R> iris_train <- iris_data[train_indices, ]
R> iris_test <- iris_data[-train_indices, ] 
R> gausspr_model <- gausspr(Species ~ ., data = iris_train, 
+                           kernel = "rbfdot", kpar = "automatic") 
R> lgp_pred_probs <- predict(gausspr_model, newdata = iris_test, 
+                            type = "probabilities")
\end{CodeInput}
\end{CodeChunk} 

\begin{figure}[!t]
	\centering 
	\begin{subfigure}{0.48\textwidth}
		\includegraphics[width=\linewidth]{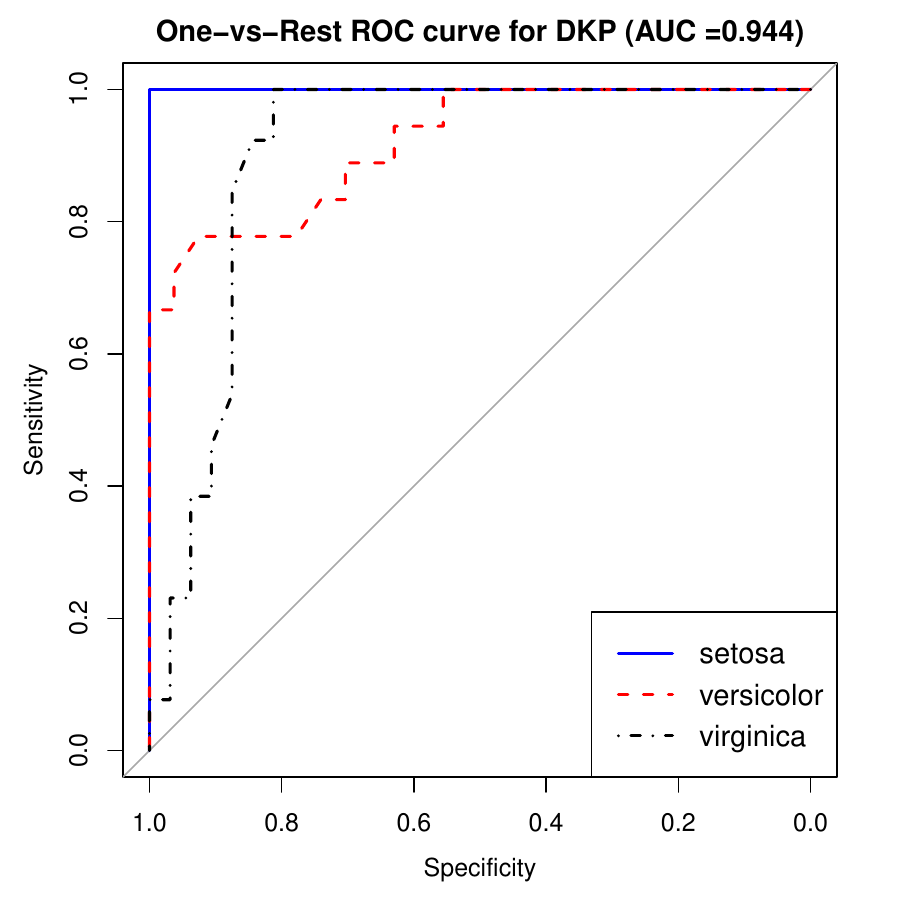}  
		\caption{DKP}
	\end{subfigure} 
	\begin{subfigure}{0.48\textwidth}
		\includegraphics[width=\linewidth]{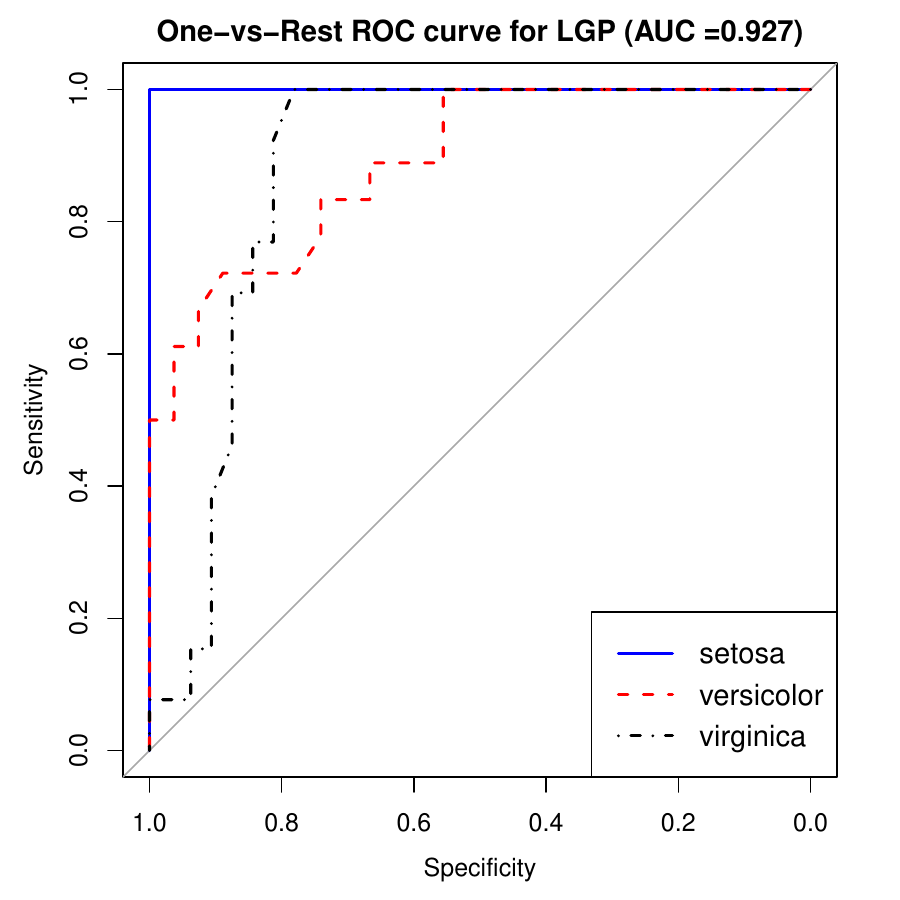}  
		\caption{LGP}
	\end{subfigure}    
	\caption{One-vs-Rest ROC curves and their respective macro-average AUCs for DKP and LGP models}
	\label{fig:ex7-roc}
\end{figure}

Compared with the DKP fit, the LGP model produces less smooth and more fragmented decision regions, particularly between the clusters corresponding to \emph{versicolor} and \emph{virginica}, as shown in the lower-left panel of Figure~\ref{fig:ex7}. 
Its maximum predicted probability surface in the lower-right panel is also less spatially coherent, with isolated pockets of relatively high or low confidence appearing within otherwise homogeneous class regions. 
In contrast, the DKP confidence surface varies more systematically with the fitted decision boundaries and provides a more stable model-based representation of classification confidence. 
The corresponding ROC curves in Figure~\ref{fig:ex7-roc} yield a macro-average AUC of 0.944 for DKP and 0.927 for LGP. 
Taken together, the smoother decision structure, more coherent confidence surface, and higher macro-average AUC indicate that DKP provides better predictive discrimination and a more stable representation of classification confidence than LGP in this example. 

\subsection{TwinBKP and TwinDKP Models}\label{sec:example_twin}

This subsection illustrates the TwinBKP and TwinDKP interfaces. 
The examples reuse the settings from Examples~2 and~5, but increase the sample size and replace the full-kernel fitting functions by their global-local counterparts. 
The purpose is to show that the scalable variants retain the same response formats and downstream S3 workflow, while replacing full kernel aggregation with twinning-selected global subsets and prediction-specific local neighbours.

\paragraph{Example 8} 
We first revisit the nonlinear binomial response setting in Example~2. 
The true probability function in \eqref{eq:bkp-1d-generate-function-2} is unchanged, but the number of input locations is increased from 30 to 500. 
The full BKP call is then replaced by \fct{fit\_TwinBKP}, which uses a twinning-selected global subset together with prediction-specific local neighbours.

\begin{CodeChunk}
\begin{CodeInput} 
R> n <- 500
R> Xbounds <- matrix(c(-2, 2), nrow = 1)
R> X <- lhs(n = n, rect = Xbounds)
R> true_pi <- true_pi_fun(X)
R> m <- sample(100, n, replace = TRUE)
R> y <- rbinom(n, size = m, prob = true_pi)
R> TwinBKP_model_1D_2 <- fit_TwinBKP(X, y, m, Xbounds = Xbounds)
R> print(TwinBKP_model_1D_2)
R> plot(TwinBKP_model_1D_2)
\end{CodeInput}
\end{CodeChunk}

\begin{CodeChunk}
\begin{CodeOutput}
       Twin Beta Kernel Process (TwinBKP) Model

Number of observations (n): 500
Input dimension (d):        1
Global kernel:              gaussian
Local kernel:               wendland
Isotropic:                  TRUE
theta_g:                    0.05
theta_l:                    0.0545
Loss function:              brier
Loss minimum:               0.03694
Prior:                      noninformative
Global subset size (g):     22 (target 22)
Local neighbours (l):       25
Twins runs:                 5
\end{CodeOutput}
\end{CodeChunk}

The first three arguments have the same meaning as in \fct{fit\_BKP}: \code{X} is the input matrix, \code{y} contains the observed success counts, and \code{m} contains the binomial trial sizes. 
The additional global-local structure is handled internally. 
In this run, the default rule selects 22 global points and 25 local neighbours per prediction location. 
The fitted \class{TwinBKP} object supports the same downstream workflow as a full \class{BKP} object. 

\begin{figure}[!t]
	\centering
	\includegraphics[width=0.8\linewidth]{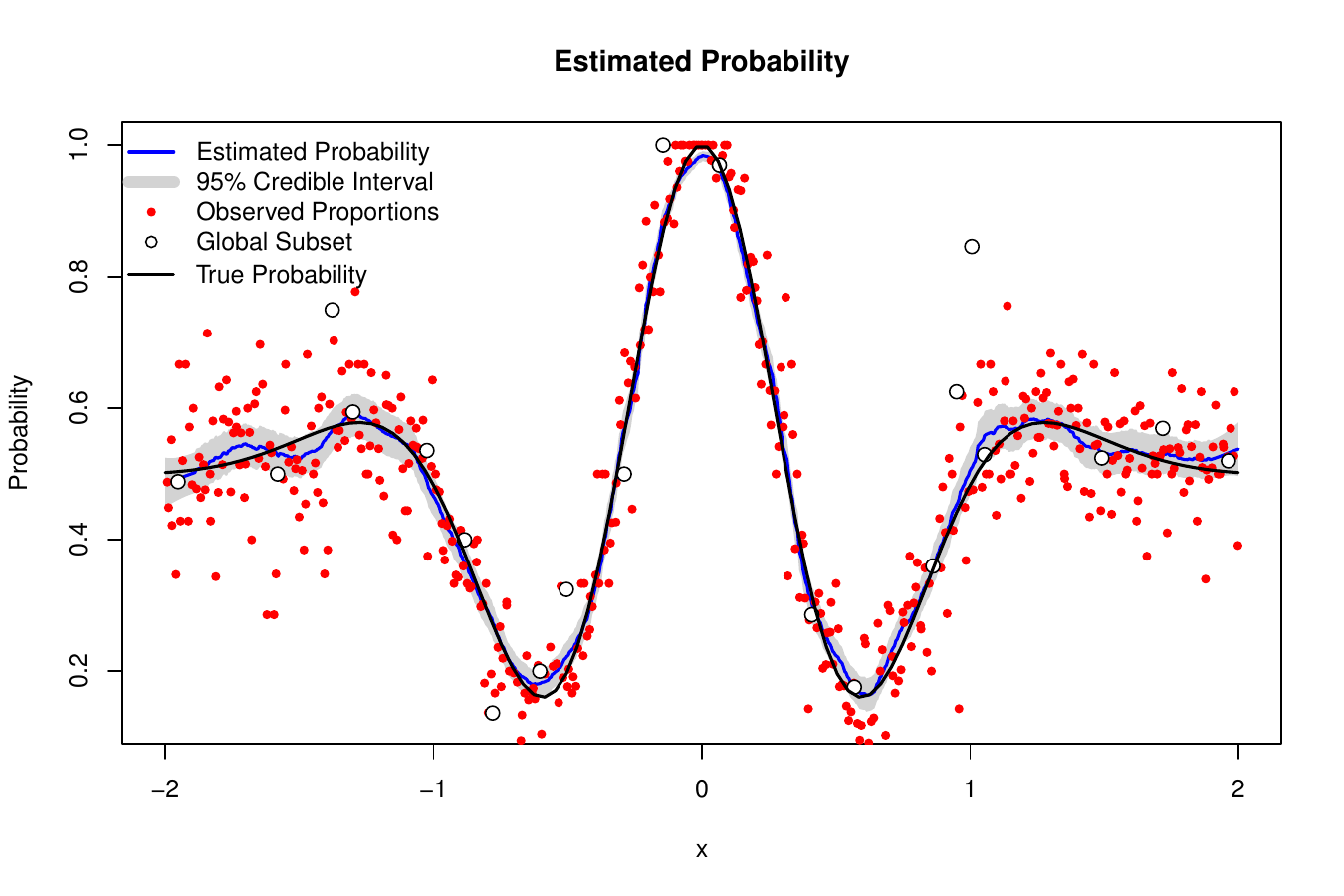}
	\caption{Posterior inference from the fitted TwinBKP model for the one-dimensional nonlinear binomial example. The blue curve is the posterior mean, the gray band is the pointwise 95\% credible interval, the red points are observed proportions, and the open circles indicate the twinning-selected global subset.}
	\label{fig:TwinBKP_1D}
\end{figure}

Figure~\ref{fig:TwinBKP_1D} shows that the TwinBKP fit follows the main nonlinear pattern of the probability function while using a compact global subset and local nearest-neighbour refinements. 
The posterior summaries retain the beta-binomial interpretation of the full BKP model, but the pseudo-count aggregation is performed over the union of the global subset and the prediction-specific local subset rather than over all training observations.


\paragraph{Example 9}
We next revisit the one-dimensional three-class multinomial response setting in Example~5 to illustrate the TwinDKP interface. 
The class-probability functions are unchanged, but the number of input locations is increased from 30 to 500. 
The full DKP call is replaced by \fct{fit\_TwinDKP}. 
As in the full DKP model, the response is supplied as an \code{n} by \code{q} matrix of multinomial counts.

\begin{CodeChunk}
\begin{CodeInput}
R> n <- 500
R> Xbounds <- matrix(c(-2, 2), nrow = 1)
R> X <- lhs(n = n, rect = Xbounds)
R> true_pi <- true_pi_fun(X)
R> m <- sample(150, n, replace = TRUE)
R> Y <- t(sapply(1:n, function(i) rmultinom(1, size = m[i], prob = true_pi[i, ])))
R> TwinDKP_model_1D <- fit_TwinDKP(X, Y, Xbounds = Xbounds)
R> plot(TwinDKP_model_1D)
\end{CodeInput}
\end{CodeChunk}

\begin{figure}[!t]
	\centering
	\includegraphics[width=\linewidth]{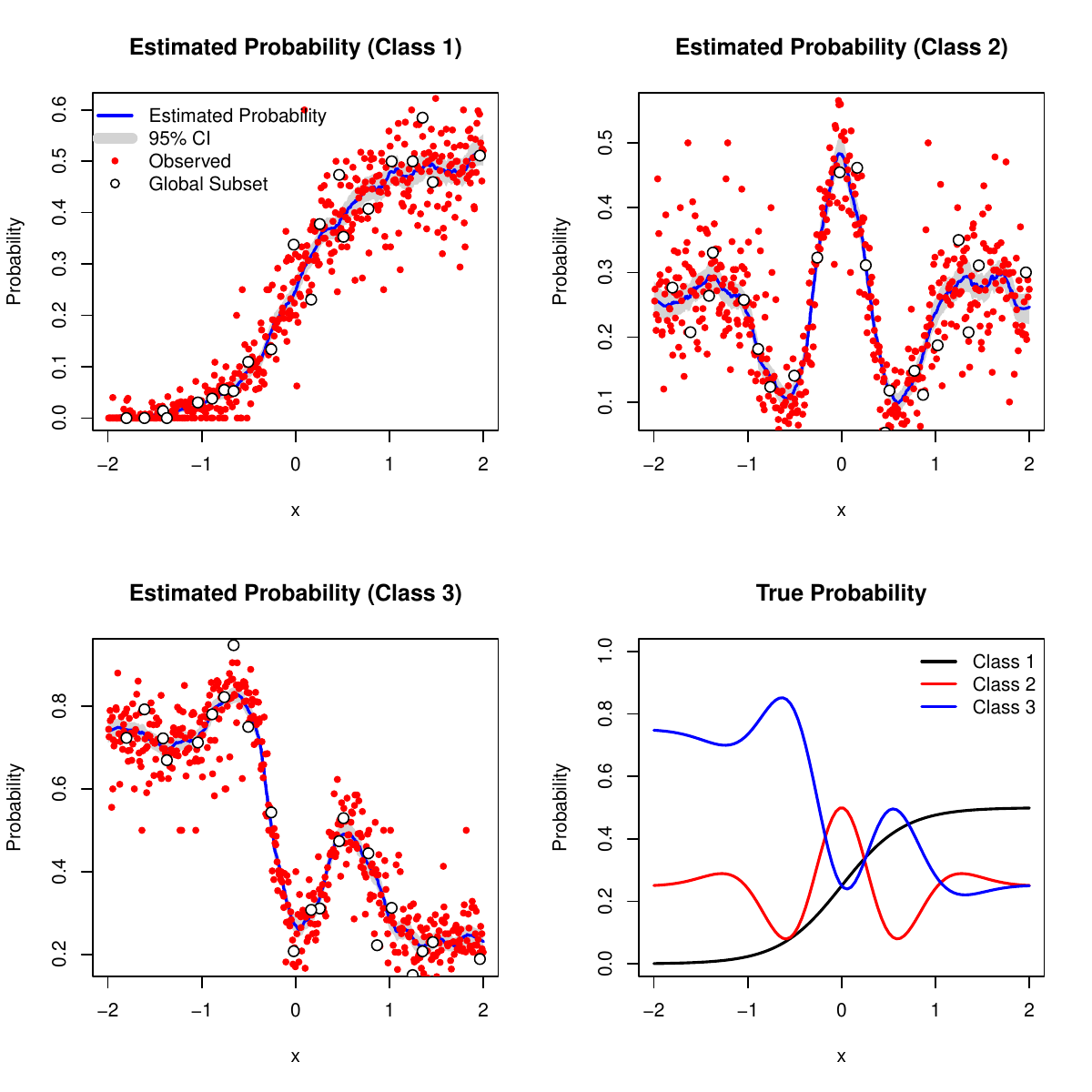}
	\caption{Posterior inference from the fitted TwinDKP model for the one-dimensional three-class multinomial example. The first three panels show the posterior mean class probabilities, pointwise 95\% credible intervals, observed class proportions, and the twinning-selected global subset. The fourth panel shows the true class-probability functions.}
	\label{fig:TwinDKP_1D}
\end{figure}

The fitted \class{TwinDKP} object can be visualized using the same \code{plot()} method as a full \class{DKP} object. 
Figure~\ref{fig:TwinDKP_1D} reports the fitted TwinDKP posterior summaries, with the true class-probability curves added as a fourth panel for comparison.

The TwinDKP fit follows the main features of the three class-probability functions while using the same global-local approximation as TwinBKP. 
The posterior summaries retain the Dirichlet-multinomial interpretation of the full DKP model, but the pseudo-count vector at each prediction location is aggregated over the union of the twinning-selected global subset and the prediction-specific local subset.


\FloatBarrier

\section{Real-data Applications}\label{sec:real_data}

This section illustrates the use of \pkg{BKP} in two real-data applications: spatial prevalence mapping of \emph{Loa loa} infection using aggregated binomial counts, and species distribution modeling for the Mourning Warbler using binary presence--absence data and high-dimensional environmental covariates. 
The data sources and preprocessing procedures are described in the corresponding subsections.

\subsection{Loa loa infection prevalence mapping}\label{sec:loaloa}

\subsubsection{Data Preparation}  

The \code{loaloa} dataset, available in the \pkg{RiskMap} package \citep{Giorgi2025RiskMap}, contains observations from 197 survey locations in North Cameroon collected between 1991 and 2001 \citep{Diggle2007Loa}.  
For each location, the dataset records the geographical coordinates (\code{LONGITUDE} and \code{LATITUDE}), the number of infected individuals (\code{NO_INF}), the total number examined (\code{NO_EXAM}), and additional environmental variables, including altitude and the maximum normalized difference vegetation index (NDVI).
 
In this analysis, geographical coordinates are used as the input variables \code{X}, the number of infected individuals as the binomial success count \code{y}, and the number examined as the trial count \code{m}.
For illustration, the locations are randomly divided into a training set containing 70\% of the observations and a testing set containing the remaining 30\%.

\begin{CodeChunk}
\begin{CodeInput}   
R> data("loaloa", package = "RiskMap")  
R> # Extract coordinates, infected counts, and numbers examined
R> X <- as.matrix(loaloa[, c("LONGITUDE", "LATITUDE")]) 
R> y <- loaloa$NO_INF
R> m <- loaloa$NO_EXAM 
R> # Randomly split into training (70%) and testing (30%) sets
R> train_idx <- sample(1:nrow(loaloa), 0.7 * nrow(loaloa))
R> X_train <- X[train_idx, ]
R> y_train <- y[train_idx]
R> m_train <- m[train_idx]
R> X_test <- X[-train_idx, ]
R> y_test <- y[-train_idx]
R> m_test <- m[-train_idx] 
\end{CodeInput} 
\end{CodeChunk}   

\begin{figure}[!t]
    \centering 
    \includegraphics[width=\linewidth]{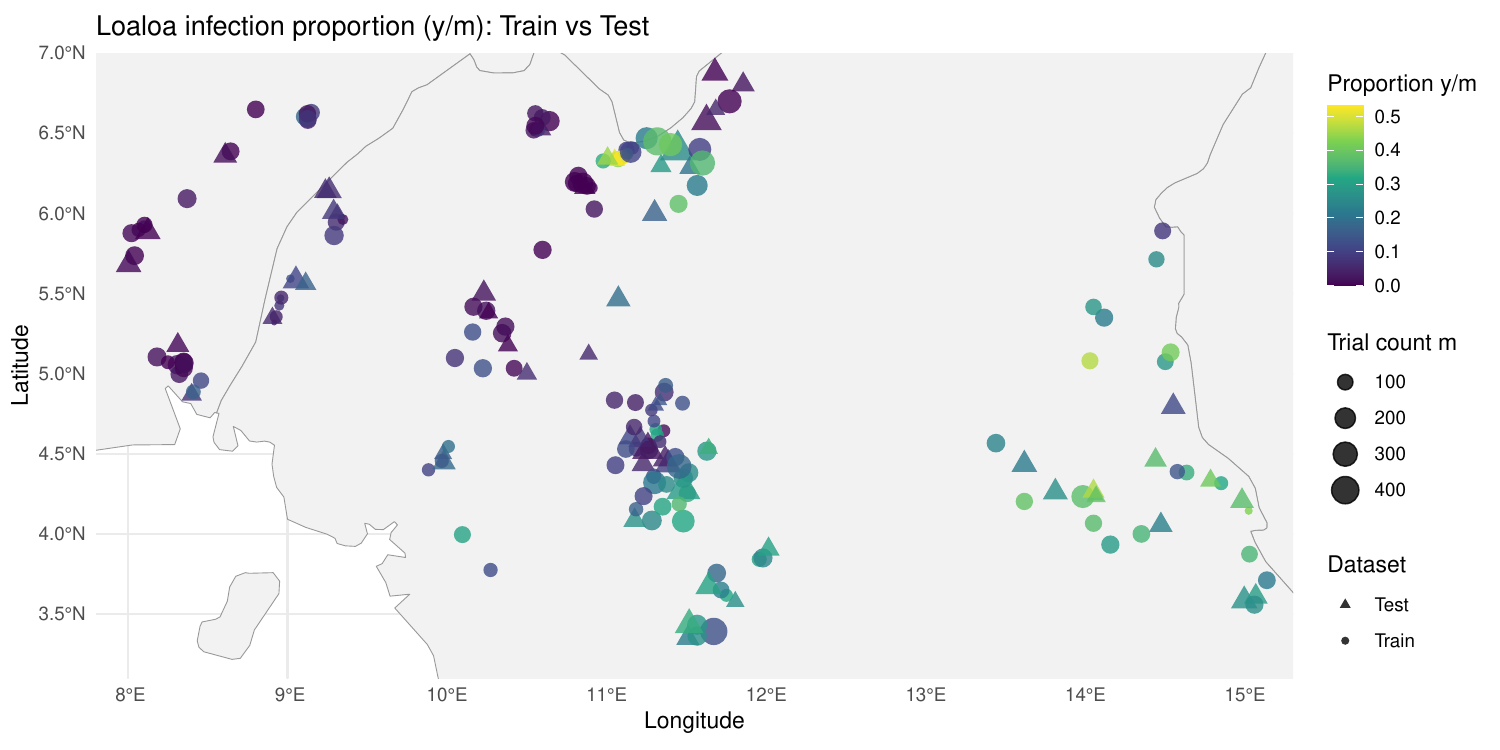}  
    \caption{Observed \emph{Loa loa} infection proportions \(y/m\) across North Cameroon. Point size is proportional to the number examined, color represents the observed infection proportion, and shape distinguishes the randomly selected training locations (circles; 70\%) from the testing locations (triangles; 30\%).}
    \label{fig:Loaloa_map}
\end{figure}

Figure~\ref{fig:Loaloa_map} displays the observed infection proportion $y/m$ at each survey location.
Point size represents the number examined and therefore indicates the nominal amount of binomial information available at each location, while color represents the observed infection proportion.
Point shape distinguishes the training and testing locations.
The observed proportions exhibit substantial spatial heterogeneity, including groups of locations with elevated prevalence between approximately $10^{\circ}\mathrm{E}$ and $12^{\circ}\mathrm{E}$, and another group near $14^{\circ}\mathrm{E}\)--\(15^{\circ}\mathrm{E}$.

\subsubsection{Model Fitting and Visualization}  

We fit a BKP model to the training data using \fct{fit\_BKP}.
The default specification uses the Brier score for LOOCV-based length-scale selection, a non-informative beta prior, and an isotropic Gaussian kernel.
The bounds supplied through \code{Xbounds} rescale longitude and latitude to the unit square before fitting.

\begin{CodeChunk}
\begin{CodeInput}
R> Xbounds <- matrix(c(7.8, 15.3, 3.1, 7.0), ncol = 2, byrow = TRUE)
R> bkp_fit <- fit_BKP(X_train, y_train, m_train, Xbounds = Xbounds)
R> summary(bkp_fit)
\end{CodeInput}
\begin{CodeOutput}

       Beta Kernel Process (BKP) Model   

Number of observations (n):  137
Input dimensionality (d):    2
Kernel type:                 (isotropic) gaussian
Optimized kernel parameters: 0.0486
Minimum achieved loss:       0.00943
Loss function:               brier
Prior type:                  noninformative

Posterior predictive summary (training points):
                      Mean Median     SD    Min    Max
Posterior means     0.1509 0.1273 0.1059 0.0077 0.4134
Posterior variances 0.0003 0.0001 0.0004 0.0000 0.0026
\end{CodeOutput}
\end{CodeChunk}  

For comparison, we also fit a logistic Gaussian process (LGP) model using \fct{gp\_init} and \fct{gp\_optim} from the \pkg{gplite} package \citep{Piironen2022gplite}.

\begin{CodeChunk}
\begin{CodeInput}   
R> gp <- gp_init(cf = cf_sexp(), lik = lik_binomial())
R> gp <- gp_optim(gp, X_train, y_train, trials = m_train, verbose = FALSE)
\end{CodeInput} 
\end{CodeChunk}  

\begin{figure}[!t]
    \centering 
    \includegraphics[width=\linewidth]{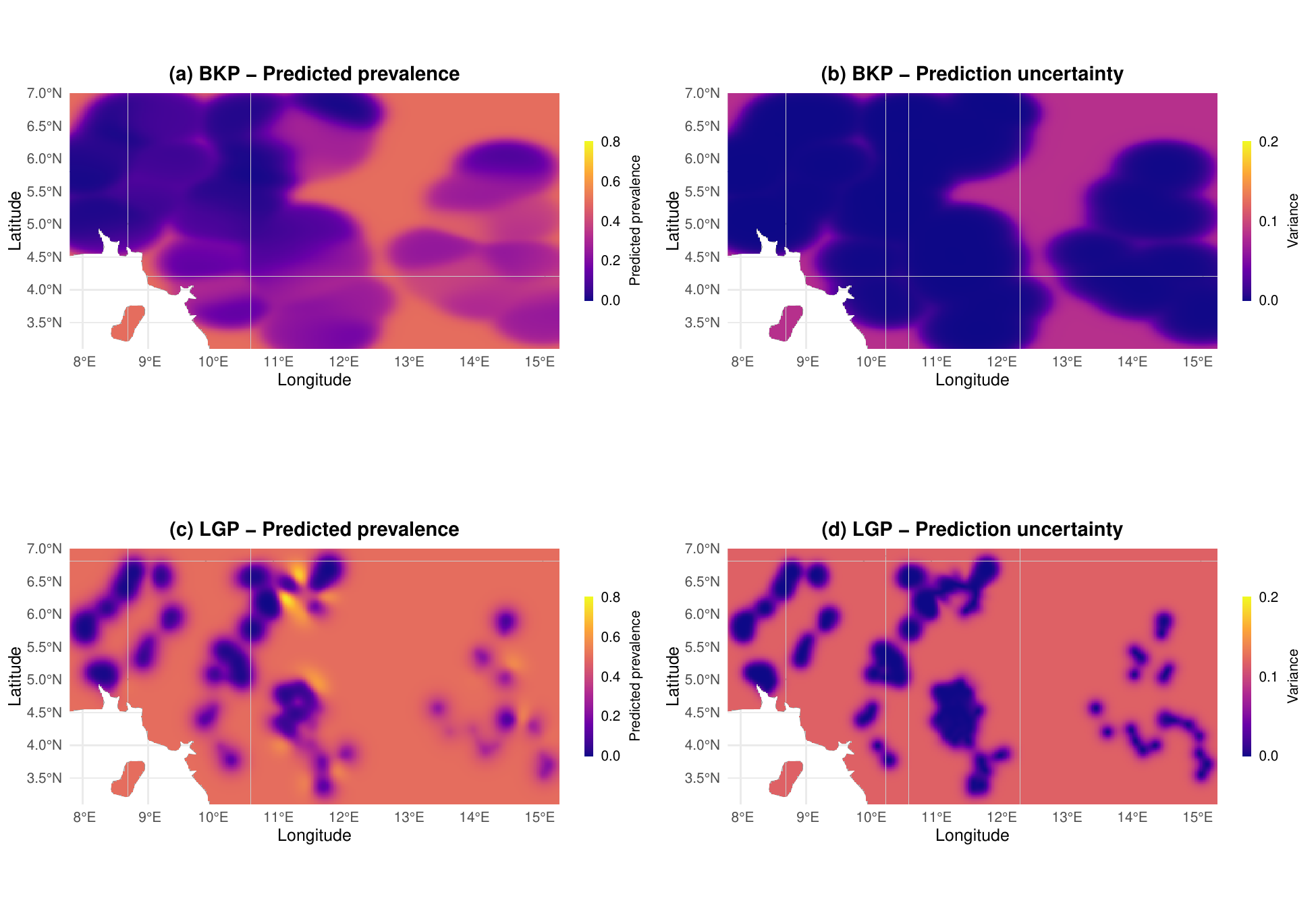}  
    \caption{Probability-scale prediction summaries for the BKP and LGP models fitted to the \emph{Loa loa} data. The left column shows fitted infection probabilities, and the right column shows the corresponding model-specific variance surfaces. Common color limits are used for the two probability panels and, separately, for the two variance panels.}
    \label{fig:Loaloa_combined}
\end{figure}

Figure~\ref{fig:Loaloa_combined} compares the fitted probability and variance surfaces from the two models.
The BKP probability surface contains several spatially connected regions of elevated predicted prevalence, including regions corresponding broadly to the higher observed proportions in Figure~\ref{fig:Loaloa_map}.
The LGP surface is more diffuse over parts of the study region and contains a larger number of small-scale local features.
These differences reflect the distinct smoothing constructions of the two methods:
BKP aggregates kernel-weighted success and failure pseudo-counts directly on the probability scale,
whereas LGP performs inference through a latent Gaussian process and a nonlinear link function.

The variance surfaces also exhibit different spatial patterns.
For BKP, larger posterior variances are concentrated mainly around transitions between lower- and higher-prevalence regions and in locations with weaker nearby data support.
The LGP variance is more broadly distributed over the study region, with local reductions around some observed locations.
Because BKP and LGP construct probability-scale uncertainty differently,
these variance maps should be interpreted as model-specific uncertainty summaries rather than as directly comparable frequentist coverage measures.
A separate simulation study of empirical pointwise interval coverage is reported in Appendix~\ref{app:coverage}.

\subsubsection{Predictive Performance on the Test Data} 

We next evaluate the fitted probabilities at the randomly withheld testing locations using the Brier score.

\begin{CodeChunk}
\begin{CodeInput}    
R> all_preds <- list()
R> all_preds$BKP <- predict(bkp_fit, Xnew = X_test)$mean
R> all_preds$LGP <- gp_pred(gp, X_test, transform = TRUE)$mean
R> # Calculate the empirical success rate for the test data
R> pi_test <- y_test / m_test 
R> # Calculate the Mean Squared Error (Brier Score)
R> mse_bkp <- mean((all_preds$BKP - pi_test)^2)
R> mse_gp <- mean((all_preds$LGP - pi_test)^2)   
\end{CodeInput}
\end{CodeChunk} 

For this random train--test split, the BKP model achieves a Brier score of 0.0073, compared with 0.0188 for the LGP model.
The lower score indicates that the BKP predictions agree more closely, on average, with the observed infection proportions at the withheld locations.
Together with the spatial patterns in Figure~\ref{fig:Loaloa_combined}, these results show that BKP provides an accurate and spatially coherent fit for this \emph{Loa loa} prevalence-mapping application.

\subsection{Mourning Warbler distribution modeling}\label{sec:warbler}

We next illustrate the use of BKP and TwinBKP for species distribution modeling using presence--absence observations of the Mourning Warbler in North America.
Unlike the previous application, the models are fitted in an eight-dimensional environmental covariate space rather than directly in geographic coordinates.
The example compares BKP, TwinBKP, and LGP using a predefined spatially structured test set and subsequently projects the fitted environmental relationships back into geographic space.

\subsubsection{Data Preparation} 

The occurrence data are derived from the 2011 North American Breeding Bird Survey (BBS) \citep{sauer2014north}, using a compiled subset provided by \citet{harris2015generating} and subsequently analyzed by \citet{golding2016fast}.
The dataset contains geographic coordinates, a binary response $y$ indicating presence ($y=1$) or absence ($y=0$), and eight continuous bioclimatic covariates.

Following \citet{golding2016fast}, longitude and latitude are used only to define the spatial partition and visualize the results; they are not included as model predictors.
Instead, all three models use the same eight bioclimatic variables.
The predefined disc-based spatial partition contains 1,613 training observations and 301 testing observations, with the testing observations forming geographically contiguous withheld regions.
This partition provides a more demanding assessment of spatial transfer than a conventional random train--test split.

\begin{CodeChunk}
\begin{CodeInput}   
R> # Load observation data and define training/testing splits
R> data <- read.csv("code/data/Mourning_Warbler.csv", stringsAsFactors = FALSE)
R> train_idx <- which(data$split == "training")
R> test_idx  <- which(data$split == "testing")
R> bio_cols <- grep("^wc2", names(data), value = TRUE)
R> # Extract 8 bioclimatic covariates
R> X_all <- as.matrix(data[, bio_cols])
R> X_train <- X_all[train_idx, ]; y_train <- data$y[train_idx]
R> m_train <- rep(1L, length(y_train))
R> X_test  <- X_all[test_idx, ];  y_test  <- data$y[test_idx]
R> Xbounds <- t(apply(X_train, 2, range))
R> # Load and crop the environmental raster for spatial projection
R> library(terra)
R> tif_files <- file.path("code/data/climate/wc2.1_10m", paste0(bio_cols, ".tif"))
R> na_clim <- crop(rast(tif_files), ext(-170.15, -46.70, 22.23, 71.90))
R> names(na_clim) <- bio_cols
\end{CodeInput} 
\end{CodeChunk}

\begin{figure}[!t]
	\centering 
	\includegraphics[width=0.8\linewidth]{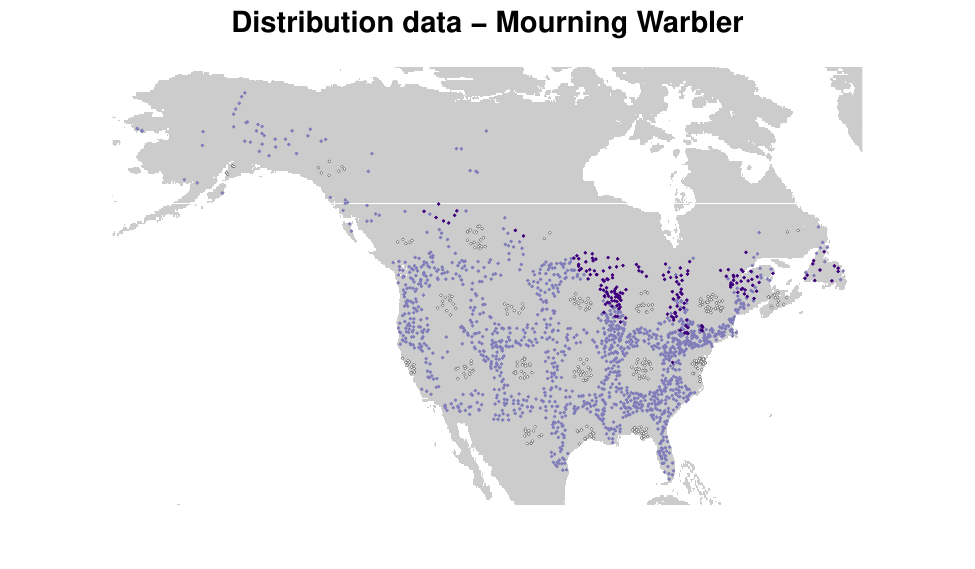} 
	\caption{Spatial distribution of the Mourning Warbler observations.
	Dark purple points indicate training presences, light purple points indicate training absences, and white points indicate the testing observations withheld under the disc-based spatial partition.}
	\label{fig:warbler_dist}
\end{figure}

Figure~\ref{fig:warbler_dist} shows the spatial distribution of the training and testing observations.
The withheld observations occur in geographically contiguous regions created by the disc-based partition, rather than being randomly interspersed among the training locations.

\subsubsection{Model Fitting and Predictive Evaluation}

We fit the full BKP model using \fct{fit\_BKP} with a fixed prior precision $r_0=0.1$, the log-loss criterion, and an isotropic Gaussian kernel.
Because \code{p0} is not supplied, the prior mean is set to the empirical presence proportion in the training data.
We then fit TwinBKP using the same global kernel and prior specification, together with the default compactly supported Wendland kernel for the local update.
For comparison, an LGP model with a Bernoulli likelihood and squared-exponential covariance function is fitted using the \pkg{gplite} package.

The elapsed times reported below include model fitting, hyperparameter optimization, and probability prediction at the 301 testing locations.

\begin{CodeChunk}
\begin{CodeInput}
R> # 1. Fit full BKP
R> bkp_fit <- fit_BKP(X_train, y_train, m_train, Xbounds = Xbounds,
+                     prior = "fixed", r0 = 0.1, loss = "log_loss")
R> bkp_pred <- predict(bkp_fit, Xnew = X_test)
R>
R> # 2. Fit TwinBKP
R> twin_fit <- fit_TwinBKP(X_train, y_train, m_train, Xbounds = Xbounds,
+                          prior = "fixed", r0 = 0.1, loss = "log_loss")
R> twin_pred <- predict(twin_fit, Xnew = X_test)
R>
R> # 3. Fit LGP (via gplite)
R> gp <- gp_init(cf = cf_sexp(), lik = lik_bernoulli())
R> gp <- gp_optim(gp, X_train, y_train, verbose = FALSE)
R> gp_pred <- gp_pred(gp, X_test, transform = TRUE)
\end{CodeInput}
\end{CodeChunk}

\begin{table}[!t]
	\centering
	\begin{tabular}{lccc}
		\toprule
		Model & AUC & Brier score & Elapsed time (s) \\
		\midrule
		BKP     & 0.888 & 0.0691 & 9.04 \\
		TwinBKP & 0.913 & 0.0674 & 0.05 \\
		LGP     & 0.852 & 0.0798 & 541.82 \\
		\bottomrule
	\end{tabular}
	\caption{Predictive performance and elapsed computation time for the Mourning Warbler application.
	AUC and Brier scores are evaluated on the predefined spatial test set containing 301 observations.
	Elapsed time includes model fitting, hyperparameter optimization, and prediction at the testing locations and is reported for a single replication run on the hardware described in the computational details.}
	\label{tab:model_comparison}
\end{table}

Predictive performance on the spatially withheld testing set is summarized using the area under the receiver operating characteristic curve (AUC) and the Brier score.
Table~\ref{tab:model_comparison} reports the predictive performance and elapsed computation times for the three fitted models.

On this predefined spatial test set, TwinBKP achieves the highest AUC of 0.913 and the lowest Brier score of 0.0674.
The full BKP model obtains an AUC of 0.888 and a Brier score of 0.0691, while the corresponding LGP values are 0.852 and 0.0798.
Thus, both BKP-based fits provide better discrimination and smaller probability errors than the fitted LGP model in this application.

TwinBKP also has the shortest observed elapsed time at 0.05 seconds, compared with 9.04 seconds for full BKP and 541.82 seconds for LGP.
These results illustrate the computational benefit of replacing full-data kernel aggregation with the twinning-based global-local approximation in this eight-dimensional example.

\subsubsection{Spatial Projection}

To visualize the fitted environmental relationships geographically, each model is evaluated at raster cells having complete values for all eight bioclimatic covariates.
The resulting occurrence probabilities and variance summaries are then mapped to the corresponding geographic locations.
Because longitude and latitude are not model predictors, these maps represent geographic projections of relationships learned in environmental covariate space rather than direct spatial smoothing of the occurrence data.

\begin{figure}[!t]
	\centering
	\includegraphics[width=\linewidth]{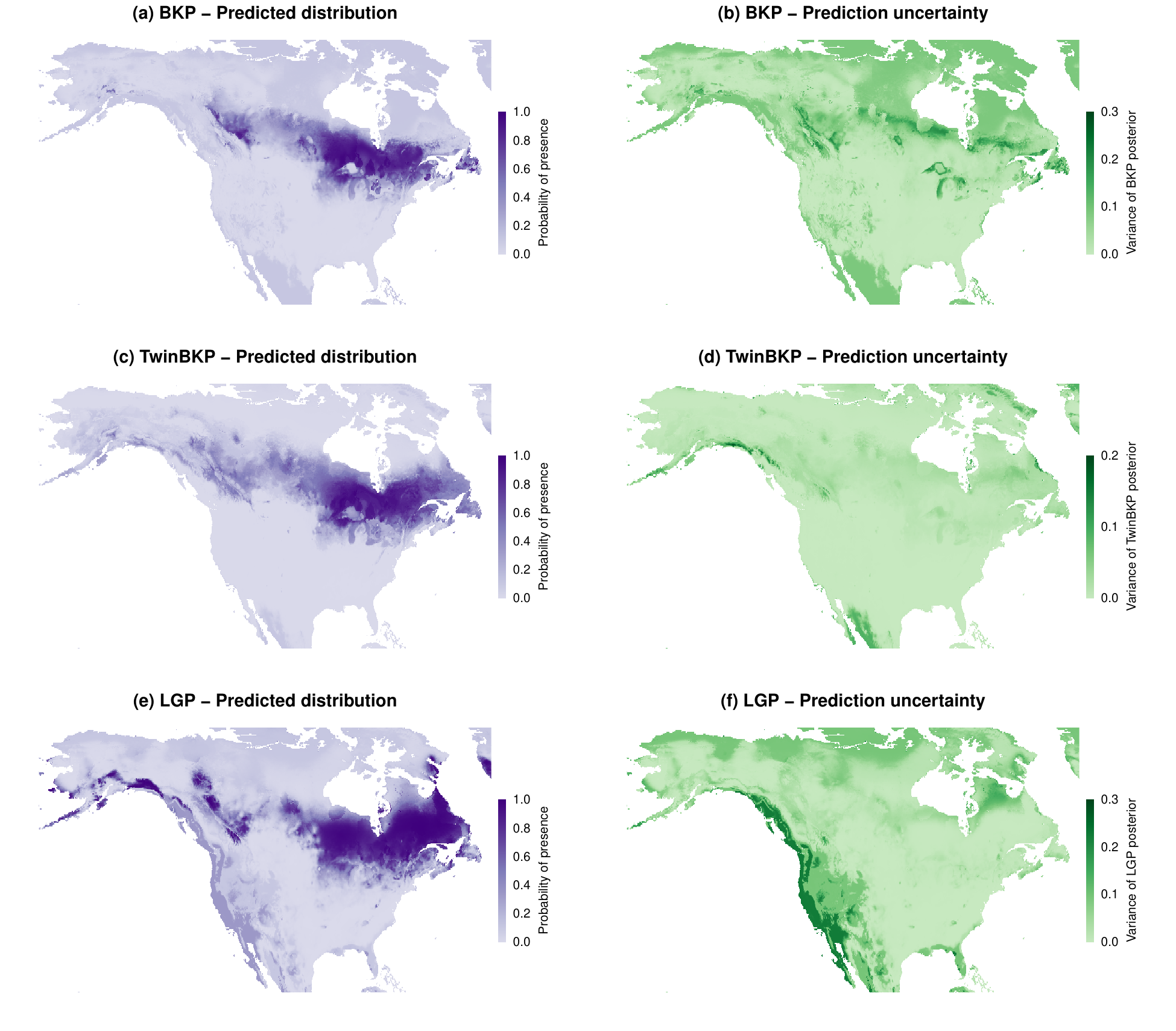}
	\caption{Spatial projections from the BKP, TwinBKP, and LGP models for the Mourning Warbler.
	The left column shows predicted probabilities of presence on a common scale from 0 to 1.
	The right column shows model-specific posterior variance summaries, with separate color scales for the three models.
	The top, middle, and bottom rows correspond to BKP, TwinBKP, and LGP, respectively.}
	\label{fig:warbler_maps_combined}
\end{figure}

Figure~\ref{fig:warbler_maps_combined} compares the projected probability and variance surfaces.
The probability panels use a common scale from 0 to 1, whereas the variance panels use model-specific scales.

All three models produce geographically structured occurrence-probability surfaces and identify broadly overlapping regions of elevated predicted presence.
The fitted maps nevertheless differ in the spatial extent and fine-scale variation of these regions, reflecting differences between the full probability-scale kernel update, the TwinBKP global-local approximation, and the latent-GP construction.

The variance maps also show substantial spatial heterogeneity within each fitted model.
Because the three methods use different posterior constructions and the variance panels have model-specific scales, these surfaces are most appropriately interpreted as within-model uncertainty summaries rather than as a direct ranking of uncertainty calibration across methods.
Predictions for raster cells containing environmental combinations that are weakly represented by the training data should additionally be interpreted as covariate-space extrapolations.

\FloatBarrier

\section{Summary and discussion} \label{sec:summary}

This article has presented \pkg{BKP}, a user-friendly \proglang{R} package for Beta Kernel Process (BKP) modeling.
The package implements a Bayesian-inspired, probability-scale kernel smoothing framework for input-dependent binomial probabilities, based on kernel-weighted pseudo-count aggregation and beta-binomial conjugate updating.
It provides a flexible and modular interface for fitting BKP models to both binary and aggregated binomial data, extends the same kernel-smoothed conjugate updating strategy to the Dirichlet Kernel Process (DKP) for categorical and multinomial responses, and includes TwinBKP and TwinDKP variants for twinning-based global-local approximation.
To our knowledge, \pkg{BKP} is the first publicly available software dedicated to BKP and DKP methodology, thereby filling an important gap in the toolkit for modeling input-dependent binomial and multinomial probability surfaces.

The framework is intended to complement, rather than replace, latent Gaussian-process classifiers and other nonparametric probabilistic models.
Its main advantages are that inference is performed directly on the probability scale, posterior summaries are available in closed form under the fitted local conjugate update, and the resulting workflow is transparent and easy to implement for binomial and multinomial response data.
At the same time, BKP should be interpreted as a Bayesian-inspired local-likelihood smoothing framework, not as a fully specified Bayesian stochastic-process model in the Gaussian-process sense.
Consequently, its credible intervals are pointwise posterior-style summaries under the fitted kernel-weighted update and do not carry frequentist coverage guarantees. Appendix~\ref{app:coverage} provides a simulation-based assessment of their empirical coverage and illustrates the effect of Shepard ESS calibration. Developing conformalized BKP procedures with finite-sample marginal coverage guarantees for future responses, together with calibrated pointwise intervals or simultaneous confidence bands for the latent probability surface, remains an important direction for future research.
In addition, the full BKP and DKP implementations remain quadratic full-kernel methods in memory, whereas TwinBKP and TwinDKP provide scalable global-local approximations for larger datasets.

Future development directions include extending the BKP framework to support more complex data structures, such as multivariate responses, functional data, time series, and combinations of qualitative and quantitative covariates.
Another promising avenue is to generalize the probability-scale kernel updating idea to likelihoods beyond the binomial and multinomial families.
For example, negative-binomial-type likelihoods may be useful for over-dispersed count data, where the variance exceeds the mean, a common feature in ecological surveys, RNA-seq gene expression counts, and epidemiological incidence data.
Geometric likelihoods, as special cases of negative-binomial models, may also be useful for modeling waiting times or the number of trials until the first success in reliability, survival, and event-process applications.
Developing such extensions would require likelihood-specific conjugate or approximate updating schemes, together with appropriate uncertainty calibration and model-selection criteria.

Finally, we welcome contributions from the community and invite developers to participate in the ongoing maintenance and extension of the package by submitting issues or pull requests through the \href{https://github.com/Jiangyan-Zhao/BKP}{\pkg{BKP} GitHub repository}.


\section*{Computational details  and reproducibility}
All analyses were conducted using \proglang{R} version 4.6.1 and \pkg{BKP} version 0.3.1. The LGP comparisons were implemented using \pkg{gplite} version 0.13.0 and \pkg{kernlab} version 0.9-33, while the Two Spirals benchmark and the \code{loaloa} dataset were obtained from \pkg{mlbench} version 2.1-8 and \pkg{RiskMap} version 1.0.0, respectively. \proglang{R} and the \proglang{R} packages used in this article are available from the Comprehensive \proglang{R} Archive Network (CRAN) at \url{https://CRAN.R-project.org/}. The TwinBKP and TwinDKP implementations use the header-only \code{nanoflann} library for kd-tree nearest-neighbour search and adapt the twinning subset-selection code from the \pkg{twingp} package. Copyright and license information for these third-party components is provided in \code{inst/COPYRIGHTS} of the \pkg{BKP} package. The multi-start length-scale tuning procedure follows the log-scale parameterization and search-region construction used in \pkg{GPfit}. 
All timing experiments reported in Figure~\ref{fig:elapsed_time} and Table~\ref{tab:model_comparison} were conducted on a machine equipped
with an Apple M4 chip (10-core CPU) and 16 GB of unified memory.
GPU acceleration was not used, and all methods were run using their default single-threaded settings. 
The computation times in Figure~\ref{fig:elapsed_time} are averages over 20 independent repetitions, whereas those in Table~\ref{tab:model_comparison} are reported from a single run.

Complete reproducibility materials for this article are available from the \href{https://github.com/Jiangyan-Zhao/BKP-paper}
{\pkg{BKP}-paper reproducibility repository}.
The repository contains the analysis scripts, data-processing code, generated figures, and numerical results for the illustrative examples, real-data applications, and appendix simulation.
The computational environment is recorded in \code{renv.lock} and can be restored using \code{renv::restore()}.
The main analysis workflow can be run from the repository root using \code{source("code/run_all.R")}; instructions for rerunning the computationally intensive timing experiments are provided in the repository README.

\section*{Acknowledgments}

The authors thank the editorial team and the anonymous reviewers for their constructive comments and suggestions, which helped improve the clarity and reproducibility of the article.


\bibliography{refs}

\newpage
\appendix
\section{Pointwise Interval-Coverage Simulation} \label{app:coverage}

We evaluate the empirical pointwise coverage of the uncertainty intervals produced by \pkg{BKP} using the one-dimensional oscillatory probability function from Example~2 in Equation~\ref{eq:bkp-1d-generate-function-2}. The comparison includes standard BKP with \code{ess = "none"}, Shepard effective-sample-size calibrated BKP with \code{ess = "shepard"}, hereafter denoted BKP-ESS, and a logistic Gaussian process (LGP).

For an evaluation set $\mathcal{E}=\{x_1,\ldots,x_N\}$, the empirical pointwise coverage rate in one simulation replication is defined as
\[
\widehat{C}(\mathcal{E})=
\frac{1}{N}\sum_{i=1}^{N}
\mathds{1}\left\{L(x_i)\leq\pi(x_i)\leq U(x_i)\right\},
\]
where $L(x_i)$ and $U(x_i)$ are the lower and upper bounds of a nominal $95\%$ pointwise interval for the latent probability $\pi(x_i)$.

We consider training sample sizes $n=30$ and $n=100$, with 100 independently generated datasets for each setting. In each replication, the training inputs are generated by Latin hypercube sampling over $[-2,2]$, the binomial trial sizes are sampled from $\{1,\ldots,100\}$, and the responses are generated from the true probability function in Equation~\ref{eq:bkp-1d-generate-function-2}. All three models, including their hyperparameter optimization procedures, are refitted independently in every replication. The BKP and BKP-ESS intervals are obtained from the quantiles of the fitted beta distributions, whereas the LGP intervals are obtained from the transformed posterior quantiles returned by \pkg{gplite}.

Coverage is evaluated on two sets of input locations. The evaluation grid consists of the same 2,000 equally spaced points over $[-2,2]$ in every replication and is used to assess coverage throughout the input domain. The training-location coverage is evaluated at the $n$ randomly generated training inputs from the corresponding replication. The evaluation grid contains no additional observed responses and should therefore not be interpreted as a conventional held-out testing dataset.

The nominal $95\%$ interval level is not an automatic finite-sample frequentist coverage guarantee for either BKP or LGP. Both methods produce model-based posterior intervals whose empirical coverage may depend on the kernel specification, likelihood, prior, and hyperparameter estimation. Conventional Gaussian-process intervals may similarly exhibit substantial under-coverage under model misspecification \citep{Papadopoulos2024GPRCP}.

\begin{figure}[!t]
	\centering
	\includegraphics[width=0.8\linewidth]{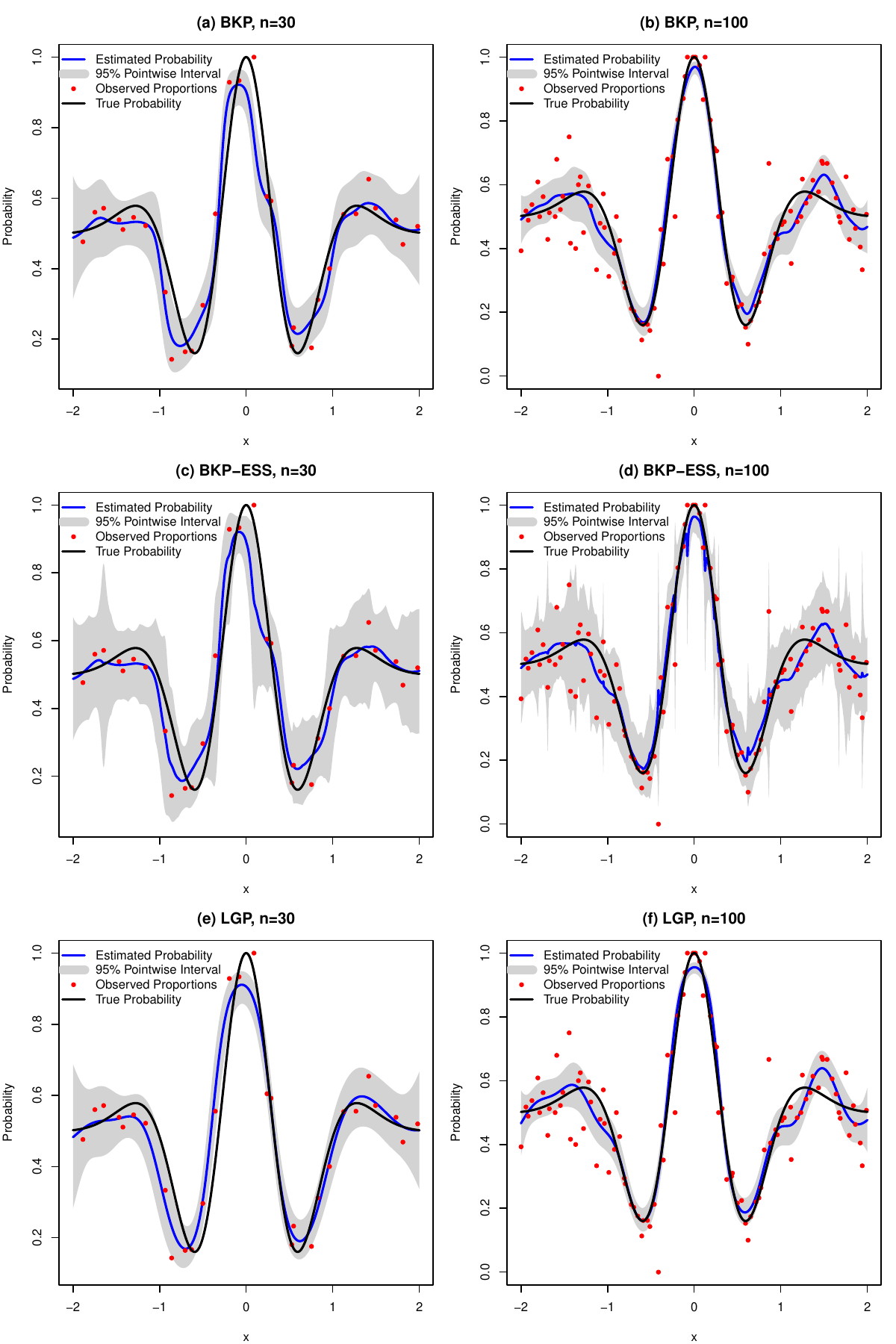}
	\caption{Representative fitted probability surfaces and pointwise $95\%$ intervals for Example~2. The rows correspond to BKP, BKP-ESS, and LGP, respectively, while the columns correspond to training sample sizes $n=30$ and $n=100$. Each panel is based on one simulation replication and is evaluated on the fixed evaluation grid of 2,000 points.}
	\label{fig:ex2_combined}
\end{figure}

\begin{table}[!t]
	\centering
	\begin{tabular}{lllc}
		\toprule
		Evaluation set & Sample size & Model & Coverage (\%) \\
		\midrule
		\multirow{6}{*}{Evaluation grid}
		& \multirow{3}{*}{$n=30$}
		& BKP     & 87.1 (6.4) \\
		& & BKP-ESS & 94.1 (3.1) \\
		& & LGP     & 91.4 (4.0) \\
		\cmidrule{2-4}
		& \multirow{3}{*}{$n=100$}
		& BKP     & 90.9 (4.1) \\
		& & BKP-ESS & 97.6 (1.0) \\
		& & LGP     & 89.5 (3.6) \\
		\midrule
		\multirow{6}{*}{Training locations}
		& \multirow{3}{*}{$n=30$}
		& BKP     & 88.1 (6.7) \\
		& & BKP-ESS & 94.6 (3.3) \\
		& & LGP     & 91.1 (5.0) \\
		\cmidrule{2-4}
		& \multirow{3}{*}{$n=100$}
		& BKP     & 90.8 (4.1) \\
		& & BKP-ESS & 97.6 (1.0) \\
		& & LGP     & 89.8 (3.7) \\
		\bottomrule
	\end{tabular}
	\caption{Empirical coverage of nominal $95\%$ pointwise intervals. Entries are mean coverage percentages across 100 independent replications, with standard deviations of the replication-specific coverage rates in parentheses. The evaluation-grid results use the same 2,000 equally spaced points over $[-2,2]$ in every replication, whereas the training-location results use the randomly generated training inputs from each replication.}
	\label{tab:coverage_results}
\end{table}

Figure~\ref{fig:ex2_combined} shows representative fitted probability surfaces and pointwise $95\%$ intervals from one simulation replication. Table~\ref{tab:coverage_results} summarizes the mean and standard deviation of the replication-specific coverage rates across the 100 replications.

The standard BKP and LGP intervals exhibit under-coverage in all four settings. BKP-ESS produces coverage close to the nominal level for $n=30$, with mean coverage of $94.1\%$ on the evaluation grid and $94.6\%$ at the training locations. For $n=100$, its mean coverage increases to $97.6\%$ for both evaluation sets, indicating moderately conservative intervals. The standard deviations are also smaller for BKP-ESS, particularly when $n=100$.

These results show that Shepard ESS calibration substantially changes the concentration of the BKP intervals and reduces under-coverage under the present simulation design. They do not establish a general frequentist coverage guarantee, because coverage may depend on the data-generating mechanism and model specification.

A natural direction for future research is to develop conformalized BKP procedures that provide finite-sample marginal coverage guarantees for future Bernoulli, binomial, or multinomial responses under exchangeability. BKP posterior summaries or beta-binomial and Dirichlet-multinomial predictive distributions could be used to construct suitable nonconformity scores. Coverage guarantees for the latent probability surface $\pi(x)$ are distinct from response-level conformal coverage and require separate development of calibrated pointwise intervals or simultaneous confidence bands.

\end{document}